\documentclass[aps,prl,reprint,groupedaddress,longbibliography]{revtex4-1}

\usepackage{graphicx, subfigure, scalefnt,amsmath,lipsum}
\usepackage{bibunits}
\DeclareGraphicsExtensions{.pdf}
\begin{document}
\begin{bibunit}


\title{Bianisotropic Metasurfaces: Ultra-thin Surfaces for Complete Control of Electromagnetic Wavefronts}


\author{Carl Pfeiffer and Anthony Grbic}

\email[]{To whom correspondence should be addressed. \\Email: agrbic@umich.edu}
\affiliation{Department of Electrical Engineering and Computer Science, University of Michigan,\\
Ann Arbor, MI, 48109-2122, USA}


\date{\today}

\begin{abstract}
Complete control of electromagnetic fields requires particles that exhibit bianisotropic constituent parameters (i.e. permittivity, permeability, and chirality). Here, methods to analyze and synthesize two-dimensional, bianisotropic metamaterials (metasurfaces) are presented. First, closed-form expressions are derived relating the reflection and transmission coefficients of a general bianisotropic metasurface to its constituent surface parameters. Next, a systematic method to design bianisotropic metasurfaces is presented. It is analytically shown that cascading anisotropic, patterned metallic sheets (electric sheet admittances) can provide electric, magnetic, and chiral responses. To demonstrate the utility of the design procedure, four devices exhibiting exotic polarization transformations are presented: a polarization rotator, an asymmetric circular polarizer, an asymmetric linear polarizer, and a symmetric circular polarizer. The optimal performance at centimeter, millimeter, and micrometer wavelengths highlights the versatility of the design process.
\end{abstract}


\maketitle


Metasurfaces that exhibit a purely electric response have shown extraordinary capabilities in controlling an electromagnetic wavefront \cite{yu2014flat,kildishev2013planar,silva2014performing}. Some of the most exciting examples have been subwavelength focusing and revising Snell's laws of refraction \cite{grbic2008near,yu2011light}. Recently, it was shown that adding a magnetic response to metasurfaces can remove reflection losses, and dramatically increase their efficiency \cite{pfeiffer2013metamaterial}. Here, anisotropy and chirality are also systematically incorporated into metasurface design to allow for complete control of an electromagnetic wavefront's polarization. To date, many structures have been reported that exhibit novel polarization effects such as asymmetric transmission \cite{fedotov2006asymmetric,wu2013giant,khanikaev2010one}, and giant optical activity \cite{niemi2013synthesis,ye201090}. However, the performance of most devices has been suboptimal since a systematic design methodology for realizing these low symmetry structures has not been established. Designs typically employ a resonant geometry that exhibits the necessary mirror and rotational symmetry for a desired bianisotropic response, but there is no guarantee that the performance is optimal \cite{menzel2010advanced}. Additionally, the principle of operation is often device specific, and its generalization to other designs is not straightforward. The physical models for bianisotropic metasurfaces reported to date are complex and valid only when the metasurface consists of infinitesimally small dipoles \cite{zhang2013Interference,niemi2013synthesis}.

Here, a simplified method to analyze arbitrary bianisotropic metasurfaces is presented. Closed-form expressions are derived that relate the reflection and transmission coefficients (scattering parameters) to the constituent surface parameters. In addition, a method to systematically design bianisotropic structures is introduced. It is shown that cascading anisotropic, patterned metallic sheets can provide significant control over the constituent surface parameters. A transfer matrix approach is used to analytically solve for the scattering parameters (S-parameters) of the structure, enabling devices with optimal performance. The ability to realize a wide range of constituent surface parameters is demonstrated with four different devices: a polarization rotator, an asymmetric circular polarizer, an asymmetric linear polarizer, and a symmetric circular polarizer. For brevity, the asymmetric linear polarizer and symmetric circular polarizer are presented in the supplementary information.

\begin{figure}
\centering
    \subfigure[]{
    \includegraphics[width=2in]{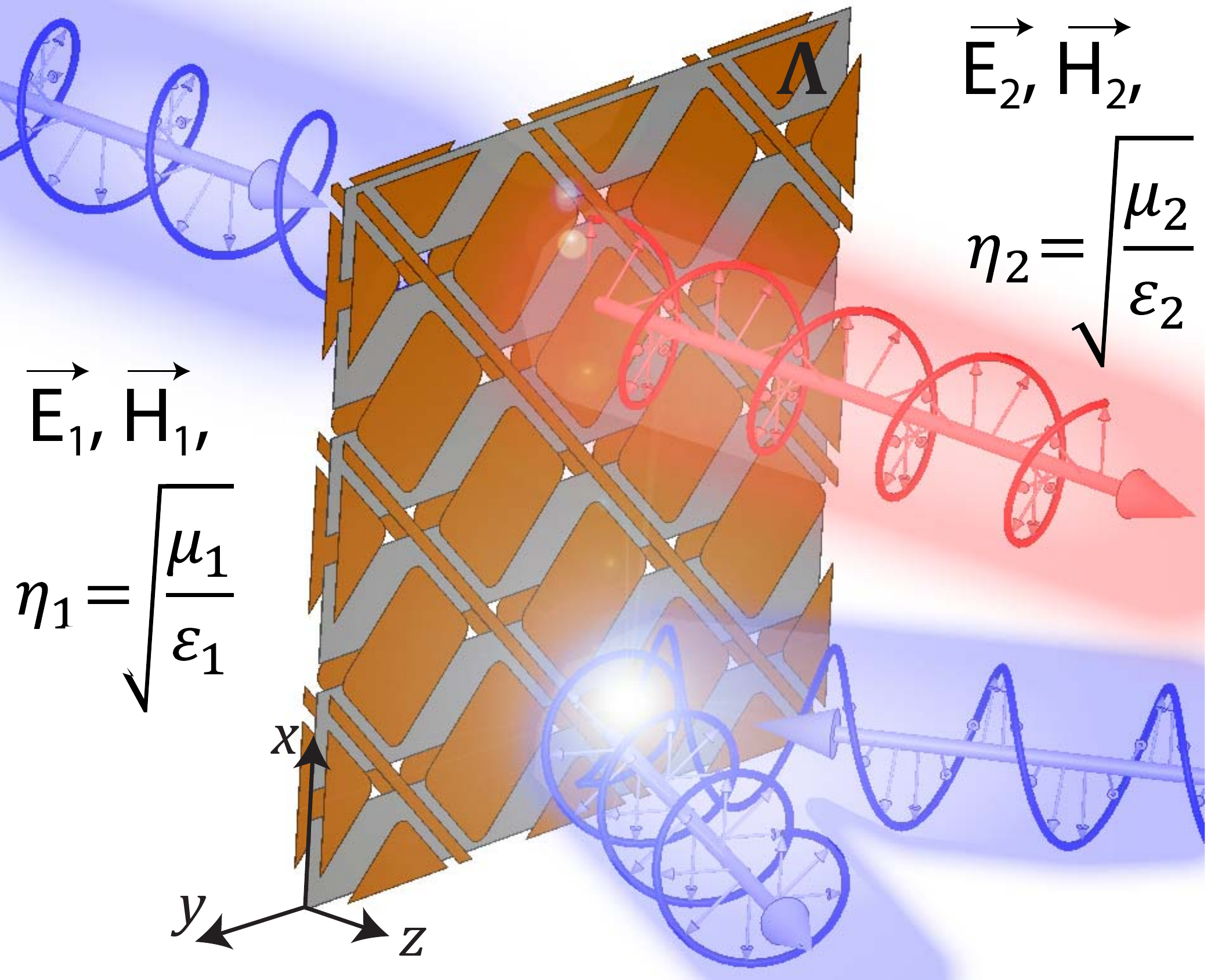}
    \label{fig:IdealMetasurface}}
        \subfigure[]{
    \includegraphics[width=2in]{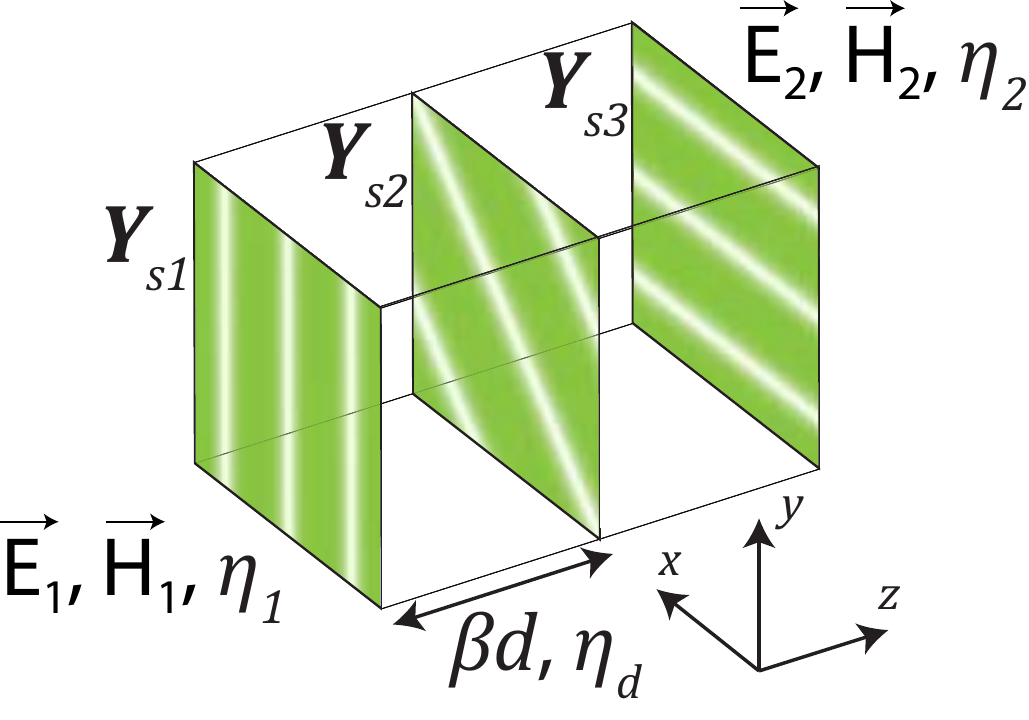}
    \label{fig:CascadedSheets}}
  \caption{\textbf{(a)} Bianisotropic metasurfaces exhibiting electric, magnetic, and chiral responses can achieve complete control of the polarization of an electromagnetic wavefront. This artistic rendering shows the example of an asymmetric circular polarizer converting right-handed-circularly polarized light from Region 1 to left-handed-circularly-polarized light in Region 2. However, right-handed-circularly polarized light is completely reflected when incident from Region 2. \textbf{(b)} Anisotropic sheet admittances cascaded in the direction of propagation can realize a wide range of constituent surface parameters. Provided the overall thickness of the cascaded sheets is subwavelength, they can be modeled as a single bianisotropic metasurface.}\label{fig:BianisotropicMetasurface}
\end{figure}

To begin, consider two regions of space (Regions 1 and 2) with wave impedances given by $\eta_1=\sqrt{\mu_1/\epsilon_1}$ and $\eta_2=\sqrt{\mu_2/\epsilon_2}$, respectively. The two regions of space are separated by an arbitrary metasurface along the $z=0$ plane, as shown in Fig. \ref{fig:IdealMetasurface}. The metasurface is illuminated by plane waves, which are assumed to be normally incident. The scattering parameters (S-parameters) are equal to the ratio of the scattered electric field to the incident electric field. In general, $\textbf{S}_{nm}=\begin{pmatrix}S_{nm}^{xx} & S_{nm}^{xy}\\S_{nm}^{yx} & S_{nm}^{yy}\end{pmatrix}$ is a 2x2 matrix relating the field scattered into Region $n$ when a plane wave is normally incident from Region $m$. For example, $S_{21}^{yx}$ represents the $y$-polarized field transmitted into Region 2 when an $x$-polarized plane wave is incident from Region 1. The parameters $\textbf{S}_{11}$ and $\textbf{S}_{22}$ are the reflection coefficients when viewed from Regions 1 and 2 respectively, and $\textbf{S}_{21}$ and $\textbf{S}_{12}$ are the transmission coefficients when viewed from Regions 1 and 2, respectively. The transmission coefficient is often referred to as the Jones matrix \cite{menzel2010advanced}.

An arbitrary metasurface can be modeled as a two-dimensional array of polarizable particles \cite{kuester2003averaged}. Each particle is characterized by its quasi-static electric and magnetic polarizabilities $(\boldsymbol{\alpha}_{e,m})$, defined as the ratio of the dipole moment to the local field. When these particles are closely spaced across a two-dimensional surface, a surface polarizability $(\boldsymbol{\alpha}^{s}_{e,m})$ that accounts for coupling between particles can be defined \cite{kuester2003averaged}. It is interpreted as the effective polarizability density per unit area,
\begin{equation}\label{eqn:polarizability}
 \begin{pmatrix}
      \textbf{p}^{s} \\
      \textbf{m}^{s} \\
   \end{pmatrix} =
 \begin{pmatrix}
      \boldsymbol{\alpha}^{s}_{ee} & \boldsymbol{\alpha}^{s}_{em} \\
      \boldsymbol{\alpha}^{s}_{me} & \boldsymbol{\alpha}^{s}_{mm}
   \end{pmatrix}\begin{pmatrix}
       \textbf{E}\\
       \textbf{H}
   \end{pmatrix}
\end{equation}
Here, $\textbf{p}^{s}=[p^{s}_x\quad p^{s}_y]^T$ and $\textbf{m}^{s}=[m^{s}_x\quad m^{s}_y]^T$ represent the electric and magnetic dipole moments, while $\textbf{E}=[E_x\quad E_y]^T$ and $\textbf{H}=[H_x\quad H_y]^T$ represent the average field tangential to the surface.

A time-harmonic progression of $e^{j\omega t}$ is assumed, where $\omega$ is the radial frequency and $t$ is time. We then define an electric sheet admittance $(\textbf{Y}=j\omega\boldsymbol{\alpha}^{s}_{ee})$, magnetic sheet impedance $(\textbf{Z}=j\omega\boldsymbol{\alpha}^{s}_{mm})$, and dimensionless chirality tensors $(\boldsymbol{\chi}=j\omega\boldsymbol{\alpha}^{s}_{em}, \boldsymbol{\Upsilon}=j\omega\boldsymbol{\alpha}^{s}_{me})$ in terms of the surface polarizabilities. Multiplying both sides of (\ref{eqn:polarizability}) by $j\omega$ and noting that a time varying dipole moment can be equated to a surface current, the electric and magnetic surface current established on the metasurface is related to the average, tangential electric and magnetic fields,
\begin{equation}\label{eqn:Lambda}
 \begin{pmatrix}
      \textbf{J}^{s} \\
      \textbf{M}^{s} \\
   \end{pmatrix} =
 \begin{pmatrix}
      \textbf{Y} & \boldsymbol{\chi} \\[0.3em]
      \boldsymbol{\Upsilon} & \textbf{Z}
   \end{pmatrix}\begin{pmatrix}
       \textbf{E} \\
       \textbf{H}
   \end{pmatrix}
   =\boldsymbol{\Lambda}\begin{pmatrix}
       \textbf{E} \\
       \textbf{H}
   \end{pmatrix}
\end{equation}
The variables $\textbf{Y}$, $\boldsymbol{\chi}$, $\boldsymbol{\Upsilon}$, and $\textbf{Z}$ are all 2x2 tensors that relate the $x$ and $y$ field components to the $x$ and $y$ current components: $\textbf{Y}=\begin{pmatrix}Y_{xx} & Y_{xy}\\
Y_{yx} & Y_{yy}\end{pmatrix}$, $\boldsymbol{\chi}=\begin{pmatrix}\chi_{xx} & \chi_{xy}\\ \chi_{yx} & \chi_{yy}\end{pmatrix}$, $\boldsymbol{\Upsilon}=\begin{pmatrix}\Upsilon_{xx} & \Upsilon_{xy}\\ \Upsilon_{yx} & \Upsilon_{yy}\end{pmatrix}$, $\textbf{Z}=\begin{pmatrix}Z_{xx} & Z_{xy}\\ Z_{yx} & Z_{yy}\end{pmatrix}$. Intuitively, $\textbf{Y}$ and $\textbf{Z}$ are the two dimensional equivalent of electric and magnetic material susceptibilities, respectively \cite{holloway2012overview}. Similarly, $\boldsymbol{\chi}$ and $\boldsymbol{\Upsilon}$ are the two dimensional equivalent of the chirality parameters. It should be noted that if reciprocal materials are used, $\textbf{Y}=\textbf{Y}^T$, $\boldsymbol{\Upsilon}=-\boldsymbol{\chi}^T$, and $\textbf{Z}=\textbf{Z}^T$ \cite{kong1972theorems}. In addition, if lossless materials are used, $\textbf{Y}$ and $\textbf{Z}$ are purely imaginary, whereas $\boldsymbol{\Upsilon}$ and $\boldsymbol{\chi}$ are purely real \cite{kong1972theorems}. For now, the analysis is kept as general as possible, and no assumption is made on reciprocity or loss.

Once a physical model is derived, the S-parameters can be found by enforcing the boundary condition of (\ref{eqn:Lambda}) (see supplementary information),
\begin{align}\label{eqn:Sparam}
\begin{pmatrix}\textbf{S}_{11} & \textbf{S}_{12}\\[0.7em]
\textbf{S}_{21} & \textbf{S}_{22} \end{pmatrix} &=
\begin{pmatrix}\frac{\textbf{Y}}{2}-\frac{\boldsymbol{\chi}\textbf{n}}{2\eta_1} +\frac{\textbf{I}}{\eta_1} & \frac{\textbf{Y}}{2}+\frac{\boldsymbol{\chi}\textbf{n}}{2\eta_2} +\frac{\textbf{I}}{\eta_2}\\[0.7em]
-\frac{\textbf{Z}\textbf{n}}{2\eta_1}+\frac{\boldsymbol{\Upsilon}}{2}-\textbf{n} & \frac{\textbf{Z}\textbf{n}}{2\eta_2}+\frac{\boldsymbol{\Upsilon}}{2}+\textbf{n}
\end{pmatrix}^{-1}\nonumber\\
&\cdot
\begin{pmatrix}-\frac{\textbf{Y}}{2} -\frac{\boldsymbol{\chi}\textbf{n}}{2\eta_1}+ \frac{\textbf{I}}{\eta_1} & -\frac{\textbf{Y}}{2} +\frac{\boldsymbol{\chi}\textbf{n}}{2\eta_2}+ \frac{\textbf{I}}{\eta_2}\\[0.7em]
-\frac{\textbf{Z}\textbf{n}}{2\eta_1}-\frac{\boldsymbol{\Upsilon}}{2} +\textbf{n} & \frac{\textbf{Z}\textbf{n}}{2\eta_2}-\frac{\boldsymbol{\Upsilon}}{2}-\textbf{n}
\end{pmatrix}
\end{align}
where $\textbf{I}=\begin{pmatrix}1&0\\0&1\end{pmatrix}$ is the identity matrix and $\textbf{n}=\begin{pmatrix}0&-1\\1&0\end{pmatrix}$ is the 90$^{\circ}$ rotation matrix.

Alternatively, the constituent surface parameters $(\boldsymbol{\Lambda})$ can be written in terms of the S-parameters (see supplementary information). This allows for the synthesis of metasurfaces from the desired S-parameters,
\begin{align}\label{eqn:SheetImpedance}
\begin{pmatrix}\textbf{Y} & \boldsymbol{\chi} \\[0.3em]
      \boldsymbol{\Upsilon} & \textbf{Z}\end{pmatrix}&=
      2\begin{pmatrix}\frac{\textbf{I}}{\eta_1}-\frac{\textbf{S}_{11}}{\eta_1} -\frac{\textbf{S}_{21}}{\eta_2} & \frac{\textbf{I}}{\eta_2}-\frac{\textbf{S}_{12}}{\eta_1} -\frac{\textbf{S}_{22}}{\eta_2}\\[0.3em]
      \textbf{n}+\textbf{n}\textbf{S}_{11}-\textbf{n}\textbf{S}_{21} & -\textbf{n}+\textbf{n}\textbf{S}_{12}-\textbf{n}\textbf{S}_{22}\end{pmatrix} \nonumber\\
      \cdot& \begin{pmatrix}\textbf{I}+\textbf{S}_{11}+\textbf{S}_{21} & \textbf{I}+\textbf{S}_{12}+\textbf{S}_{22}\\[0.3em]
      \frac{\textbf{n}}{\eta_1}-\frac{\textbf{n}\textbf{S}_{11}}{\eta_1} +\frac{\textbf{n}\textbf{S}_{21}}{\eta_2} & -\frac{\textbf{n}}{\eta_2}-\frac{\textbf{n}\textbf{S}_{12}}{\eta_1} +\frac{\textbf{n}\textbf{S}_{22}}{\eta_2}\end{pmatrix}^{-1}
\end{align}
Similar to homogenization procedures for bulk metamaterials \cite{smith2002determination}, (\ref{eqn:Sparam}) and (\ref{eqn:SheetImpedance}) provide a powerful framework to design and analyze metasurfaces that realize arbitrary polarization, phase, and amplitude transformations.

Next, a geometry is proposed that can achieve a wide range of constituent surface parameters. The geometry consists of cascaded metallic sheets (electric sheet admittances) as shown in Fig. \ref{fig:CascadedSheets}. This cascaded structure can be modeled as a single bianisotropic metasurface provided that its overall thickness is subwavelength. This geometry is attractive because it allows for straightforward design and fabrication from microwave to optical wavelengths \cite{al2011wideband,liu2007three}. Inspiration for this geometry is derived from recent work showing that the diagonal elements of the electric and magnetic surface susceptibility tensors can be completely controlled with cascaded sheets \cite{monticone2013full}. In addition, if the sheets are anisotropic, polarization controlling devices can be achieved, such as quarter-wave plates, half-wave plates, and circular polarizers \cite{pfeiffer2013cascaded,pfeiffer2013millimeter,zhao2012twisted}. However, in all these previous examples simplifying assumptions were made, which introduced limitations. If these assumptions are removed, far greater control over the achievable constituent surface parameters is possible.

Next, a transfer matrix ($\textbf{ABCD}$ matrix) approach is employed to develop a relation between the cascaded sheet admittances and the S-parameters \cite{zhao2012twisted}. This approach relates the total field in Regions 1 and 2 by the $\textbf{ABCD}$ matrix,
\begin{equation}
\begin{pmatrix}\textbf{E}_1\\ \textbf{H}_1\end{pmatrix} =
\begin{pmatrix} \textbf{A} & \textbf{B}\\ \textbf{C} & \textbf{D}\end{pmatrix}
\begin{pmatrix}\textbf{E}_2\\ \textbf{H}_2\end{pmatrix}
\end{equation}
where \textbf{A}, \textbf{B}, \textbf{C}, \textbf{D} are each 2x2 matrices relating the $x$ and $y$ field components. For example, three cascaded sheet admittances have the transfer matrix,

\small
\begin{align}\label{eqn:ABCD}
&\begin{pmatrix} \textbf{A} & \textbf{B}\\ \textbf{C} & \textbf{D}\end{pmatrix}
=\left[\begin{pmatrix}
      \textbf{I} & \textbf{0}\\
      \textbf{n}\textbf{Y}_{s1} & \textbf{I}
   \end{pmatrix} \begin{pmatrix}
      \textrm{cos}(\beta d)\textbf{I} & -j\textrm{sin}(\beta d)\eta_d\textbf{n}\\[0.3em]
j\textrm{sin}(\beta d)\eta_d^{-1}\textbf{n} & \textrm{cos}(\beta d)\textbf{I}
   \end{pmatrix}
   \right.\nonumber\\
   \cdot& \left.\begin{pmatrix}
      \textbf{I} & \textbf{0}\\
      \textbf{n}\textbf{Y}_{s2} & \textbf{I}
   \end{pmatrix} \begin{pmatrix}
      \textrm{cos}(\beta d)\textbf{I} & -j\textrm{sin}(\beta d)\eta_d\textbf{n}\\[0.3em]
j\textrm{sin}(\beta d)\eta_d^{-1}\textbf{n} & \textrm{cos}(\beta d)\textbf{I}
   \end{pmatrix}  \begin{pmatrix}
      \textbf{I} & \textbf{0}\\
      \textbf{n}\textbf{Y}_{s3} & \textbf{I}
   \end{pmatrix}\right]
\end{align}
\normalsize
Here, $\eta_d$ is the substrate impedance, $\beta d$ is the interlayer electrical spacing, and $\textbf{Y}_{sn}$ is the admittance of the $n^{\textrm{th}}$ sheet (see Fig. \ref{fig:CascadedSheets}).

The reflection and transmission coefficients of the structure can then be related to the $\textbf{ABCD}$ matrix of the cascaded sheet admittances (see supplementary information),
\begin{equation}\label{eqn:SparamsFromABCD1}
\begin{pmatrix} \textbf{S}_{11} & \textbf{S}_{12}\\[0.3em] \textbf{S}_{21} & \textbf{S}_{22}\end{pmatrix}
=\begin{pmatrix}
      -\textbf{I} & \frac{\textbf{B}\textbf{n}}{\eta_2}+\textbf{A}\\[0.3em]
      \frac{\textbf{n}}{\eta_1} & \frac{\textbf{D}\textbf{n}}{\eta_2}+\textbf{C}
   \end{pmatrix}^{-1}\begin{pmatrix}
      \textbf{I} & \frac{\textbf{B}\textbf{n}}{\eta_2}-\textbf{A}\\[0.3em]
      \frac{\textbf{n}}{\eta_1} & \frac{\textbf{D}\textbf{n}}{\eta_2}-\textbf{C}
   \end{pmatrix}
\end{equation}
Now that a relation between cascaded sheet admittances and the S-parameters is established, the sheets can be systematically designed. First, the necessary sheet admittances that realize a desired S-parameter distribution for a given substrate impedance $(\eta_d)$ and interlayer electrical spacing $(\beta d)$ are numerically found (see supporting information). Once the required sheet admittances are known, their physical realization is straightforward. Typically, each sheet is independently designed by patterning metal on a dielectric substrate. Frequency-selective surface theory has provided extensive literature on realizing arbitrary electric sheet admittances \cite{munk2005frequency}. At optical frequencies, dielectric patterning also becomes an attractive option \cite{monticone2013full,wu2013silicon}.

To demonstrate the versatility of this design process, four devices exhibiting novel polarization transformations are presented: a polarization rotator, an asymmetric circular polarizer, an asymmetric linear polarizer, and a symmetric circular polarizer. Each structure requires significantly different constituent surface parameters. For brevity, the asymmetric linear polarizer and symmetric circular polarizer are detailed in the supplementary information. For additional details on the design process, simulations, analysis, characterization, and comparisons to prior art please see the supplementary information.

Polarization rotation (chirality) is commonly used in analytical chemistry, biology, and crystallography for identifying the spatial structure of molecules \cite{rogacheva2006giant}. In addition, it provides an alternative route to achieve negative refraction \cite{pendry2004chiral}. A polarization rotator with a reflection coefficient equal to zero and transmission coefficient equal to,
\begin{equation}\label{eqn:polRotator}
\textbf{S}_{21}=e^{j\phi}\begin{pmatrix}0 & -1\\1&0\end{pmatrix}
\end{equation}
is considered \cite{niemi2013synthesis,ye201090}. In other words, any incident plane wave traveling in the $+z$ direction that is linearly polarized will experience a clockwise polarization rotation of $90^{\circ}$ upon transmission, when viewed from Region 1. By inserting (\ref{eqn:polRotator}) into (\ref{eqn:SheetImpedance}), the ideal constituent parameters of this device can be derived,

\small
\begin{equation}\label{eqn:LambdaPolRotator}
 \boldsymbol{\Lambda}=\begin{pmatrix}
     \frac{-2j\textrm{ tan}(\phi)}{\eta_{\circ}} & 0 & -2\textrm{ sec}(\phi) & 0 \\[0.3em]
       0 & \frac{-2j\textrm{ tan}(\phi)}{\eta_{\circ}} & 0 & -2\textrm{ sec}(\phi) \\[0.3em]
       2\textrm{ sec}(\phi) & 0 & -2j\eta_{\circ}\textrm{ tan}(\phi) & 0 \\
       0 & 2\textrm{ sec}(\phi) & 0 & -2j\eta_{\circ}\textrm{ tan}(\phi)
   \end{pmatrix}
\end{equation}
\normalsize
The metasurface is isotropic and chiral.

When realizing polarization transformations, the absolute phase delay $(\phi)$ generated by the metasurface is typically not important for most applications. Therefore, the phase delay can be viewed as a free parameter that can be adjusted to increase the bandwidth and reduce the loss of the metasurface.
\begin{figure}[ht]
      \centering
    \subfigure[]{
    \includegraphics[width=1.7in]{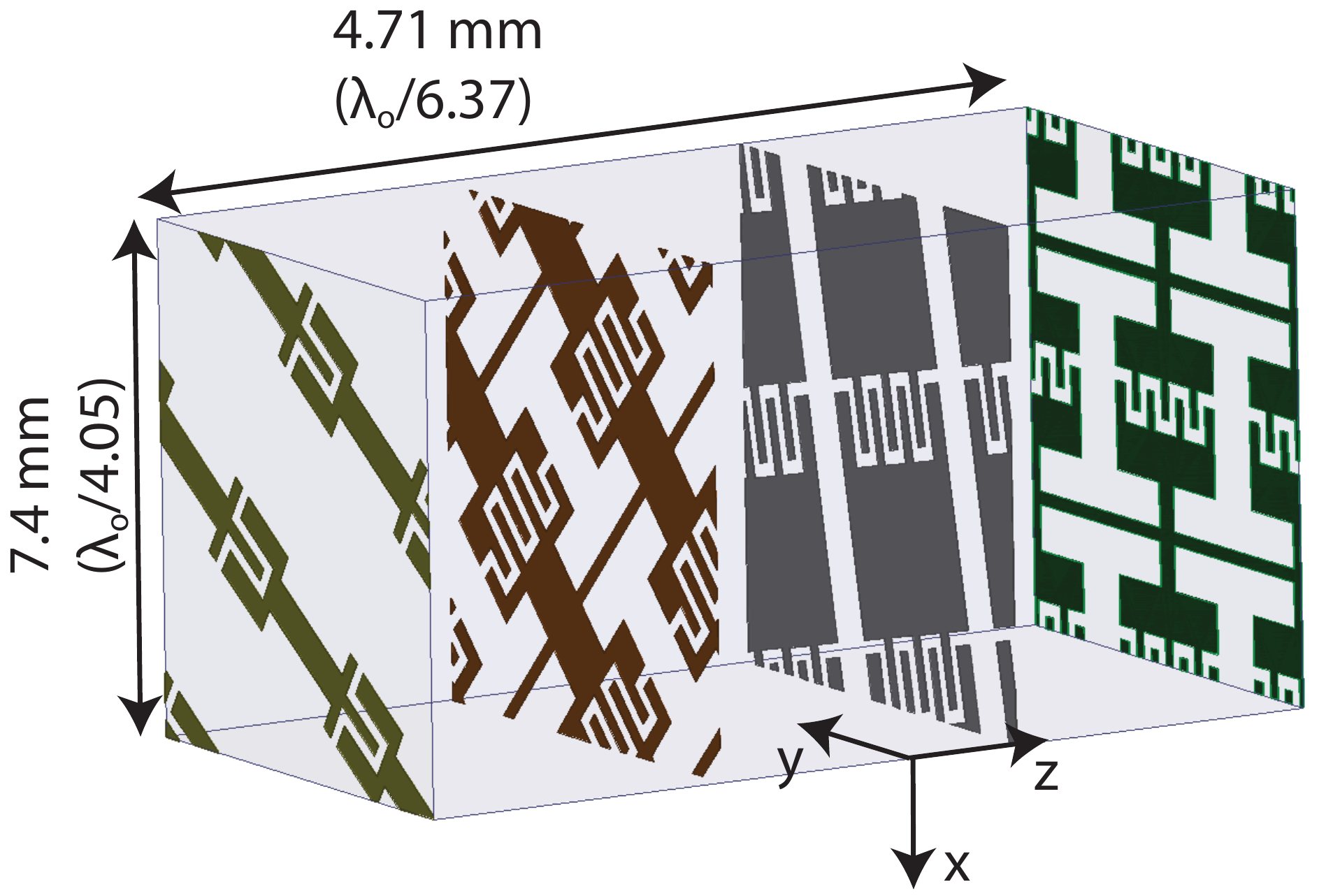}
    \label{fig:PolRotatorPic}}
     \subfigure[]{
    \includegraphics[width=1in]{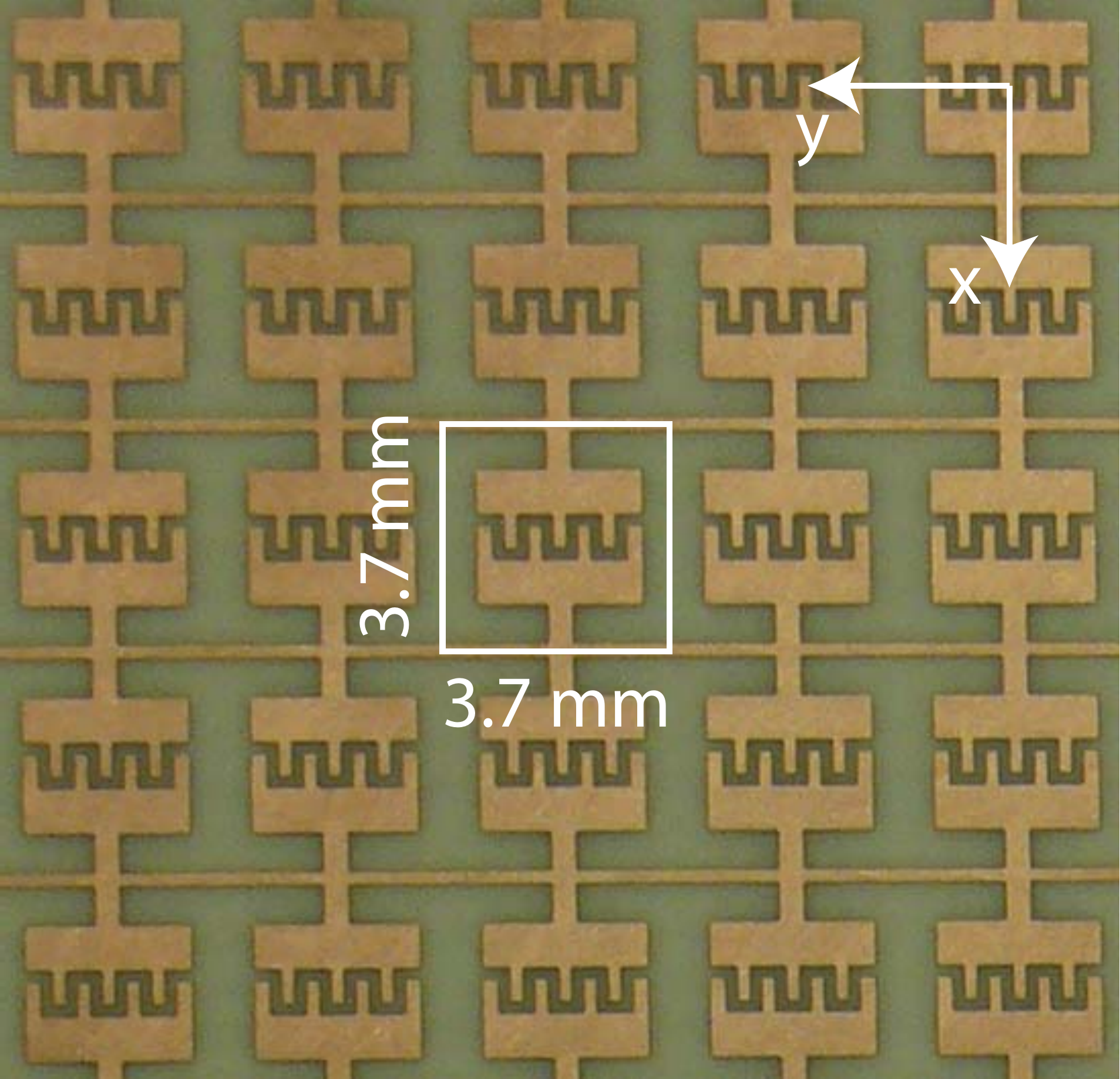}
    \label{fig:PolRotatorFabPic}}
    \subfigure[]{
    \includegraphics[width=1.55in]{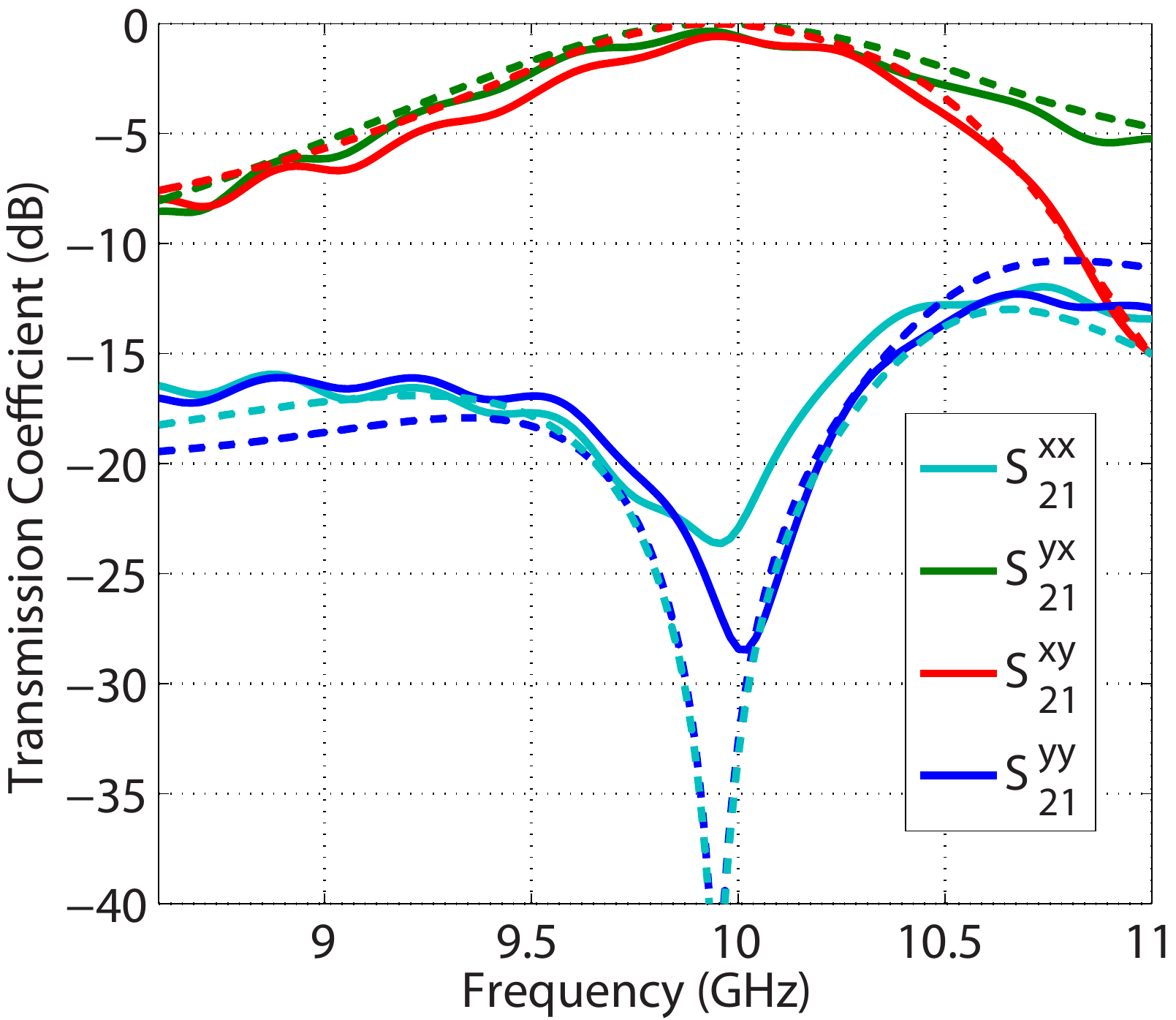}
    \label{fig:PolRotatorPerformance}}
    \subfigure[]{
    \includegraphics[width=1.55in]{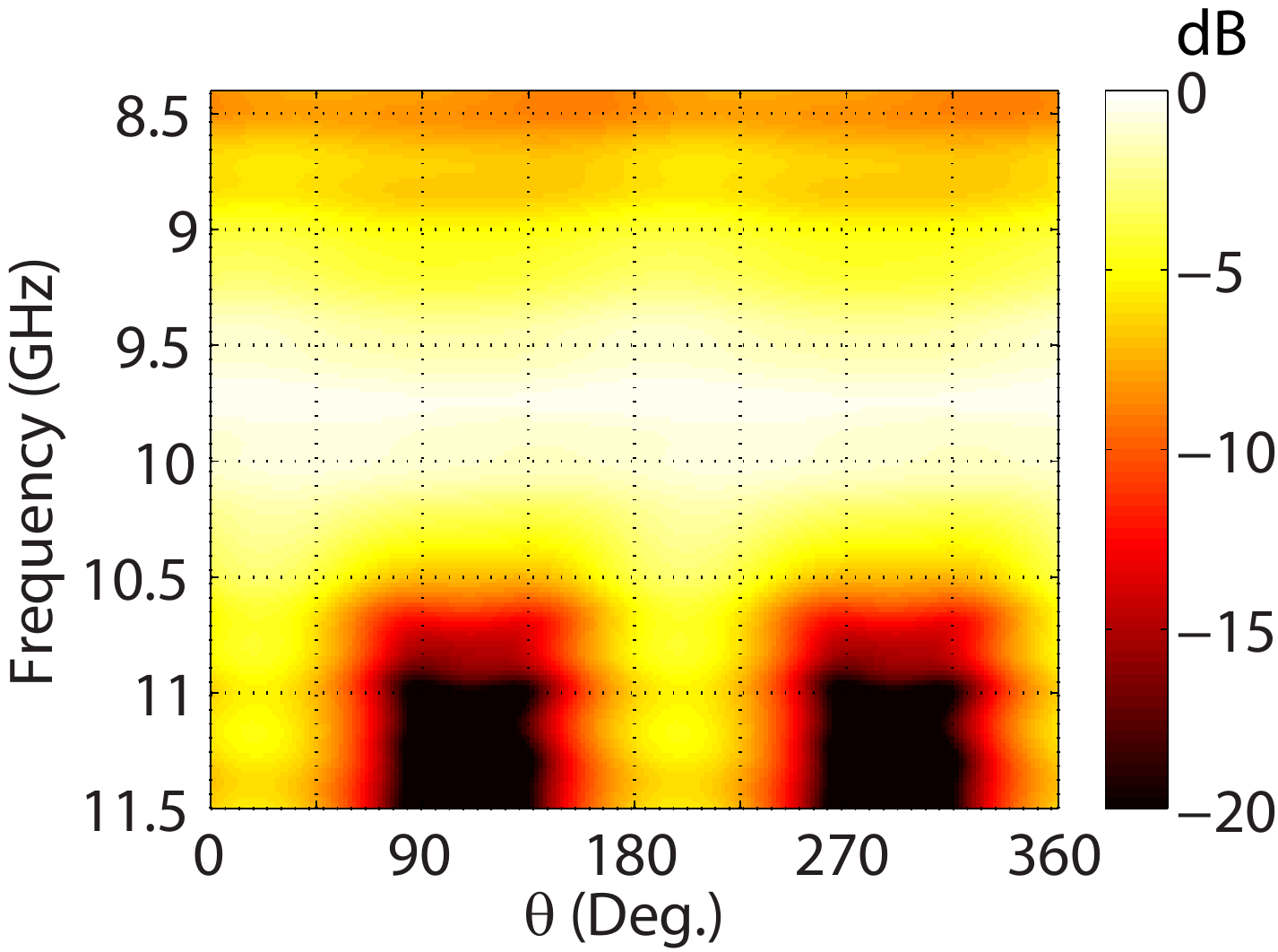}
    \label{fig:PolRotatorPerformance2}}
  \caption{Metasurface exhibiting polarization rotation near 10 GHz. \textbf{(a)} Schematic of the unit cell. For clarity, the $z$-axis is scaled by a factor of 3 so that all four sheets are visible. \textbf{(b)} Bottom sheet $(\textbf{Y}_{s4})$ of the fabricated polarization rotator. \textbf{(c)} Transmission coefficient for an incident plane wave traveling in the +$z$ direction. Measured data is denoted by solid lines, whereas simulated is denoted by dashed lines. For clarity, the measured data is frequency shifted by +0.20 GHz in the plot. \textbf{(d)} Measured cross-polarized transmission coefficient as a function of frequency and incident linear polarization. The angle $\theta$ refers to the angle between the $x$ and $y$ axes of the incident linear polarization. It can be seen that the cross-polarized transmission coefficient is near 0 dB, independent of $\theta$.}\label{fig:PolRotator}
\end{figure}

Next, the bianisotropic metasurface is realized by cascading four patterned metallic sheets on a low loss substrate (see supplementary information). A section of the designed polarization rotator is shown in Fig. \ref{fig:PolRotatorPic}, and the bottom sheet of the fabricated structure is shown in Fig. \ref{fig:PolRotatorFabPic}. The simulated and measured performance is shown in Fig. \ref{fig:PolRotatorPerformance}. There is a 2\% frequency shift between measurement and simulation due to fabrication tolerances. To highlight the similarities between measurement and simulation, Fig. \ref{fig:PolRotatorPerformance} adds a 0.2 GHz frequency shift to the measured data. It is also important to note that (\ref{eqn:LambdaPolRotator}) dictates that a polarization rotator must be isotropic. The isotropic response of the fabricated structure can be verified by rotating the incident linear polarization by an angle $\theta$ about the $z$-axis. As shown in Fig. \ref{fig:PolRotatorPerformance2}, the measured cross-polarized transmission coefficient is high and independent of the incident linear polarization at the operating frequency of 9.8 GHz. This is in contrast to the more common half-wave plate, which only achieves a high cross-polarization when the incident field is polarized at $45^{\circ}$ relative to its crystal axis. The fractional bandwidth of this structure was measured to be 8.7\%. The bandwidth is defined here to be the frequency range over which the cross-polarized transmission coefficient is greater than -3 dB and a co-polarized transmission coefficient is less than -10 dB, independent of the incident linear polarization.

Another interesting example of polarization control by a metasurface is asymmetric transmission for circularly polarized light \cite{fedotov2006asymmetric,wu2013giant}. This metasurface converts right-handed-circular to left-handed-circular when traveling in the $+z$ direction. It has the following transmission coefficient,
\begin{equation}\label{eqn:asymmetriccircular}
\textbf{S}_{21}=\frac{e^{j\phi}}{2}\begin{pmatrix}1 & j\\j&-1\end{pmatrix}
\end{equation}
However, when propagating in the $-z$ direction, the same metasurface converts left-handed-circular to right-handed-circular. Therefore it exhibits asymmetric transmission for circular polarization. It should be noted that this does not violate reciprocity since $\textbf{S}_{21}=\textbf{S}_{12}^T$, and hence the performance of the structure can be analyzed by only considering plane waves incident from Region 1.  The constituent surface parameters are given by,

\small
\begin{equation}\label{eqn:LambdaAsymmetricCircularTransmission}
\boldsymbol{\Lambda}= \begin{pmatrix}
      \frac{-2j\textrm{ tan}(\phi/2)}{\eta_{\circ}} & 0 & 0 & 0 \\
       0 & \frac{2j\textrm{ cot}(\phi/2)}{\eta_{\circ}} & 0 & 0 \\
       0 & 0 & -2j\eta_{\circ}\textrm{ tan}(\phi) & 2j\eta_{\circ}\textrm{ sec}(\phi) \\
       0 & 0 & 2j\eta_{\circ}\textrm{ sec}(\phi) & -2j\eta_{\circ}\textrm{ tan}(\phi)
   \end{pmatrix}
\end{equation}
\normalsize

As was previously noted, asymmetric transmission does not require three-dimensional chirality $(\boldsymbol{\chi}=\boldsymbol{\Upsilon}=0)$ \cite{fedotov2006asymmetric,menzel2010advanced}. However, the principle axes of the electric and magnetic responses should be rotated with respect to each other since $Y_{xy}=0$ and $Z_{xy}\ne0$. For an operating frequency of 77 GHz, the designed unit cell (see supplementary information) is shown in Fig. \ref{fig:AsymmetricCircularPic}, and the top sheet of the fabricated structure is shown in Fig. \ref{fig:AsymmetricCircularFabPic}. The simulated and measured transmission coefficients are shown in Fig. \ref{fig:AsymmetricCircularPerformance}. The surface exhibits near perfect conversion of right-handed-circular into left handed circular when propagating in the $+z$ direction. In addition the measured asymmetric response is broadband: S$_{21}^{++}$, S$_{21}^{+-}$, and S$_{21}^{--}$ are below -10 dB and S$_{21}^{-+}$ is above -0.8 dB over a bandwidth of 20\%. The superscript `+' denotes right-handed-circular polarization and `-' denotes left-handed-circular polarization.
\begin{figure}[ht]
      \centering
    \subfigure[]{
    \includegraphics[width=1.55in]{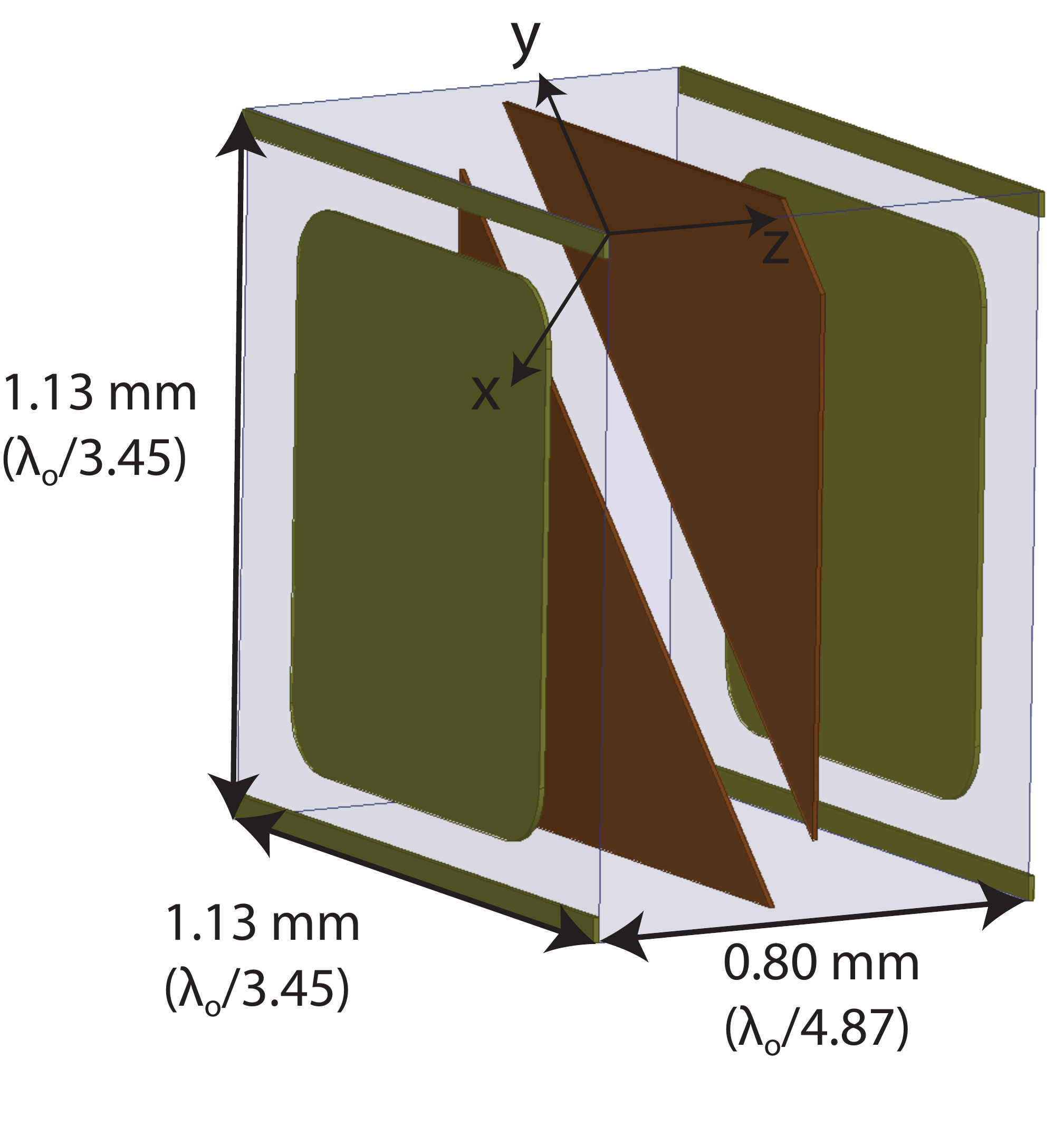}
    \label{fig:AsymmetricCircularPic}}
     \subfigure[]{
    \includegraphics[width=1.55in]{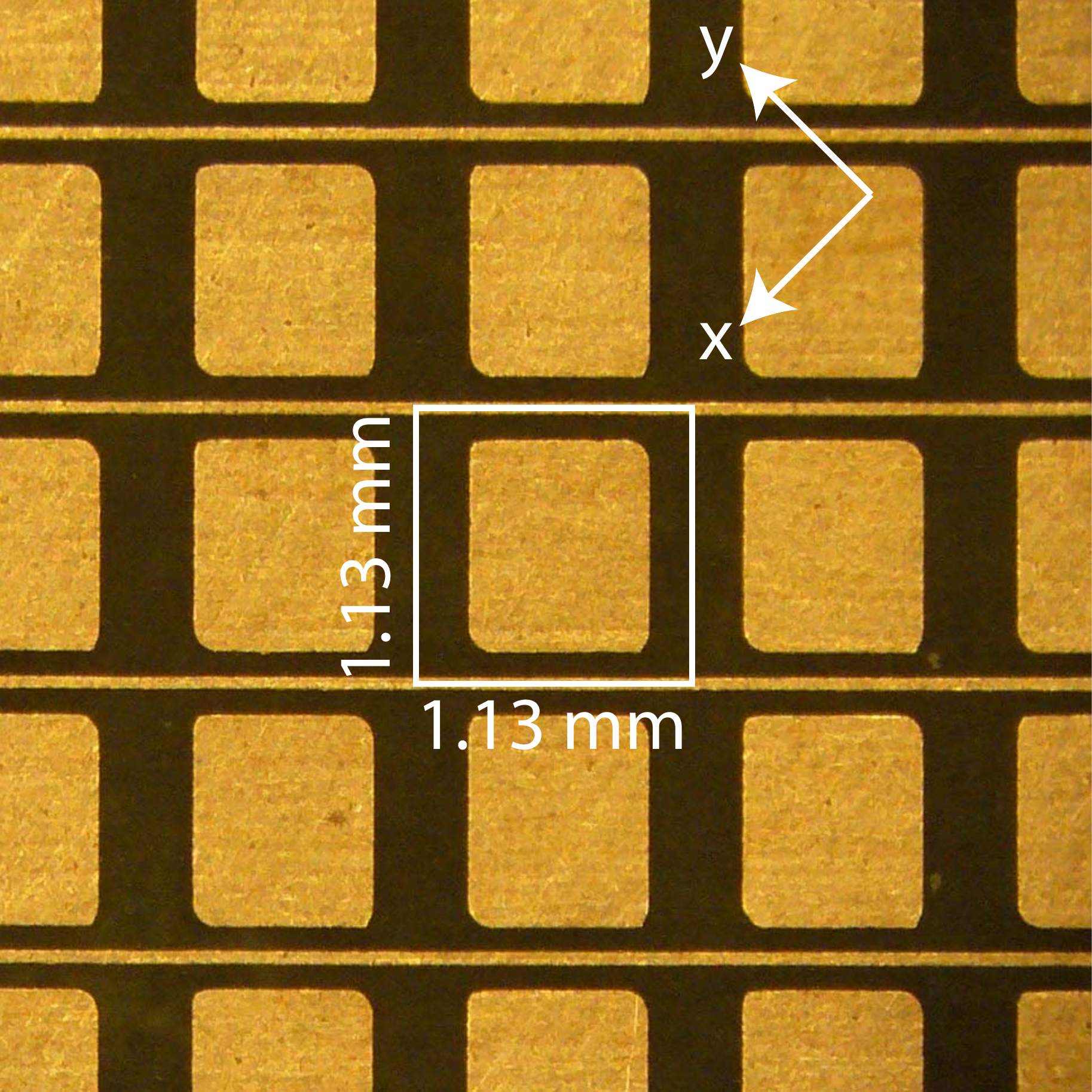}
    \label{fig:AsymmetricCircularFabPic}}
    \subfigure[]{
    \includegraphics[width=1.9in]{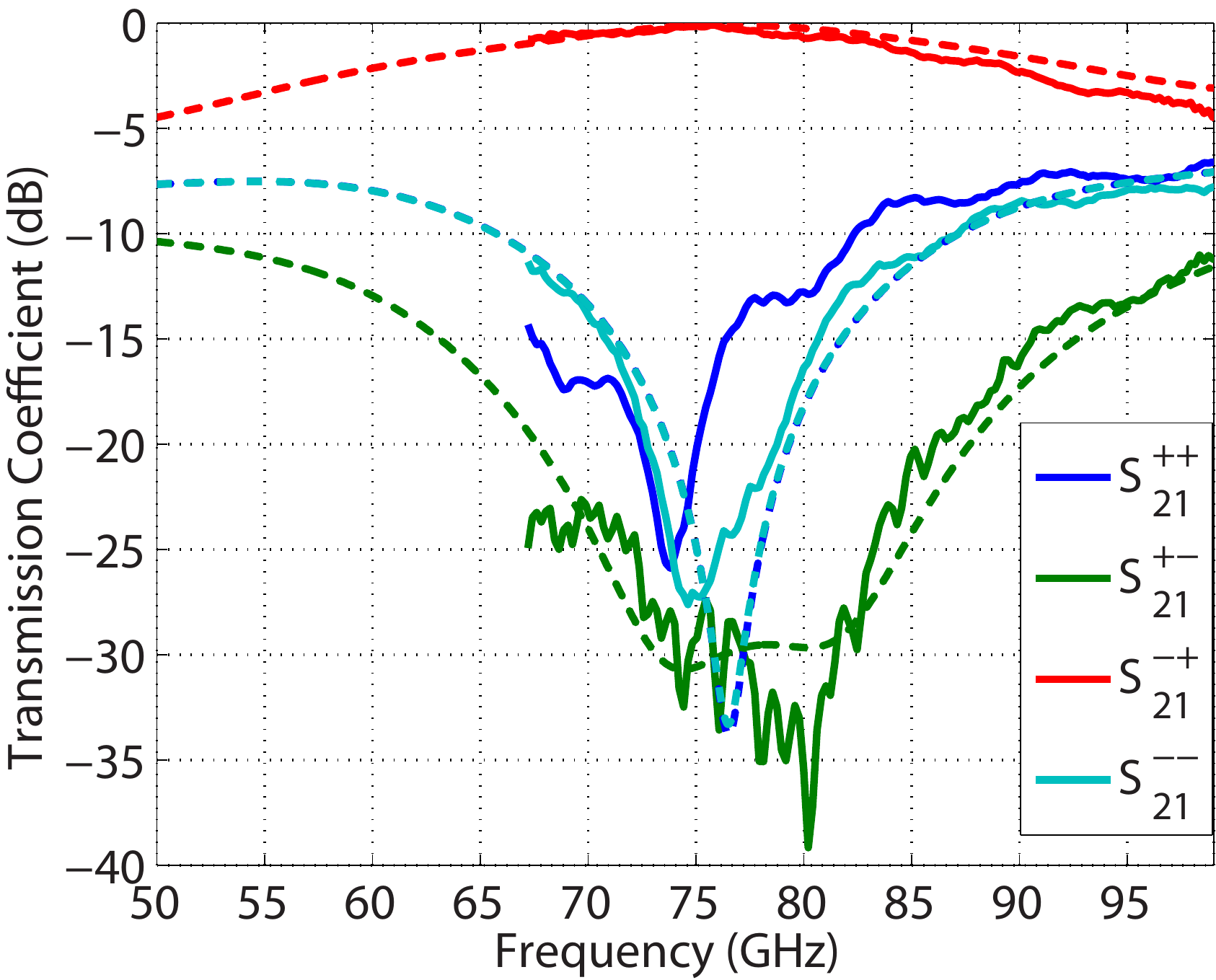}
    \label{fig:AsymmetricCircularPerformance}}
  \caption{Metasurface exhibiting asymmetric circular transmission at millimeter-wave frequencies. \textbf{(a)} Schematic of the unit cell. \textbf{(b)} Top sheet $(\textbf{Y}_{s1})$ of the fabricated asymmetric circular polarizer. \textbf{(c)} Transmission coefficient for an incident plane wave traveling in the +$z$ direction, where the superscript `+' denotes right-handed-circular and `-' denotes left-handed-circular. Measured data is denoted by solid lines, whereas simulated is denoted by dashed lines. }\label{fig:AsymmetricCircular}
\end{figure}

It should be emphasized that both of these structures have fundamentally different operating principles than devices that achieve polarization control through cascading Jones matrices. The structures presented here rely on the interference of multiple reflections between the sheets to achieve novel polarization effects, while also maintaining a subwavelength profile. In contrast, simply cascading the Jones matrices of wave-plates and linear polarizers does not take advantage of the interference between sheets, and therefore are significantly bulkier \cite{Roy1996Reciprocal}.

In future work, inhomogeneous metasurfaces can be designed that enable both wavefront and polarization control \cite{pfeiffer2013millimeter}. If lossy materials are used, this work can find applications in perfect absorbers and stealth technologies \cite{serdyukov2001electromagnetics}. Further, three-dimensional metamaterials can also benefit from this work \cite{pendry2004chiral}. By cascading the unit cell in the propagation direction ($z$-direction), a bulk bianisotropic response is attainable \cite{fietz2010homogenization}.

This work was supported by a US Air Force grant (FA4600-06-D003) and the National Science Foundation Materials Research Science and Engineering Center program DMR 1120923 (Center for Photonics and Multiscale Nanomaterials at the University of Michigan).

\end{bibunit}
\begin{bibunit}

\setcounter{figure}{0}
\setcounter{equation}{0}
\renewcommand{\thefigure}{S\arabic{figure}}
\renewcommand{\theequation}{s\arabic{equation}}
\newpage
{\large\textbf{Supplemental Materials}}
\section{Fabrication and measurement procedures}

Metasurfaces that provide polarization rotation and asymmetric circular transmission were experimentally demonstrated in the main text. Both metasurfaces were fabricated with commercial printed-circuit-board techniques (photolithography, chemical etching, and substrate bonding). The substrates used for the polarization rotator and asymmetric circular polarizer were Rogers 4003 $(\epsilon_r=3.55$, tan $\delta=0.0027)$ and Rogers 5880 Duroid $(\epsilon_r=2.2$, tan $\delta=0.0009)$, respectively. The patterned substrates of the polarization rotator were bonded together with 100 $\mu$m thickness, Rogers 4450B Bondply $(\epsilon_r=3.54$, tan $\delta=0.004)$. The asymmetric circular polarizer was bonded together with 38 $\mu$m thickness, Rogers 3001 Bondply $(\epsilon_r=2.28$, tan $\delta=0.003)$. The copper cladding was 18 $\mu$m thick for the polarization rotator and 9 $\mu$m thick for the asymmetric circular polarizer. The polarization rotator was shaped as a 12-sided regular polygon (dodecagon) with a maximum dimension of 19.5 cm. The asymmetric circular polarizer was square, with length and width equal to 8.5 cm.

\begin{figure}[ht]
\centering
\centerline{\includegraphics[width=3.1in] {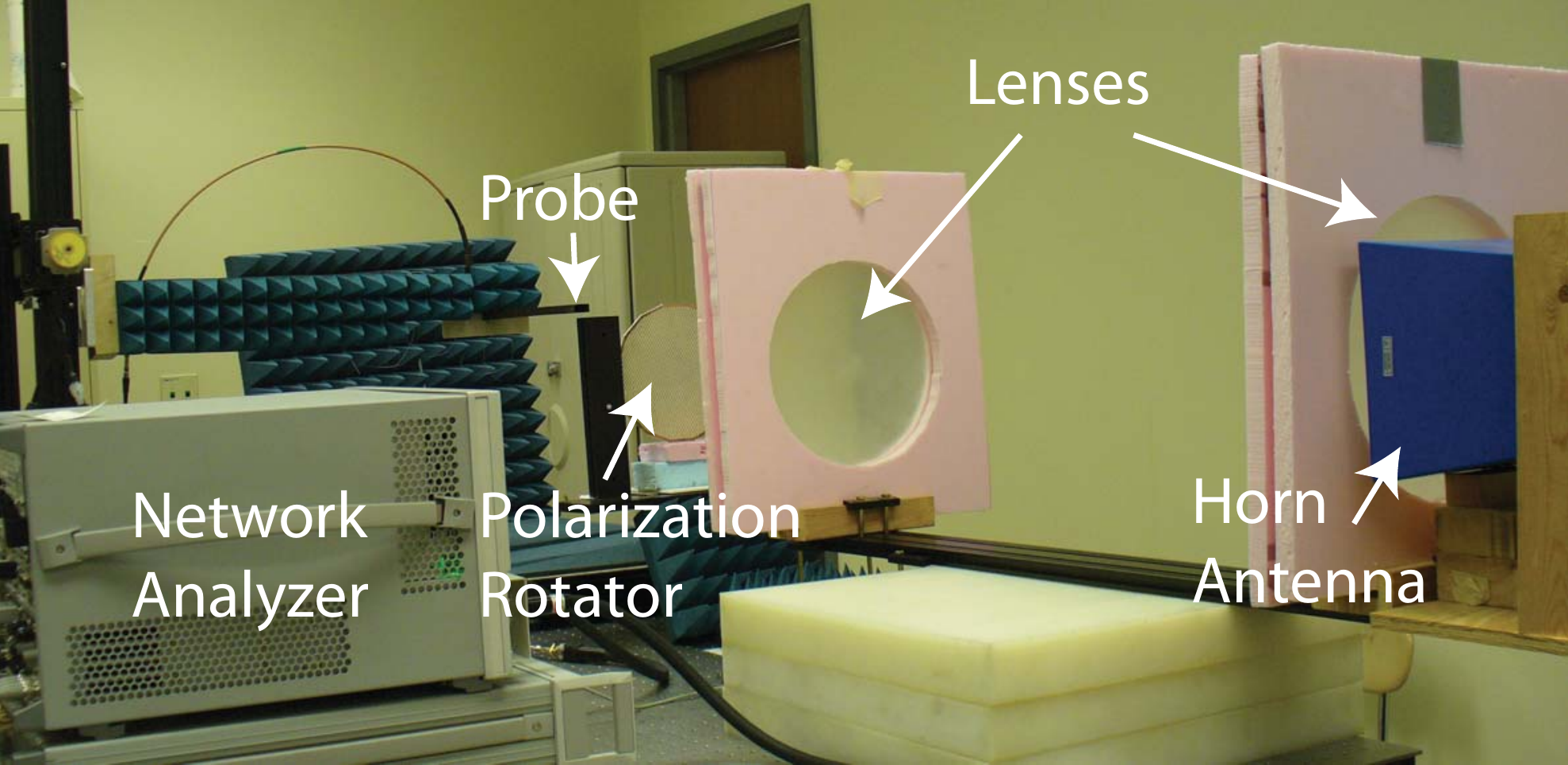}}
\caption{Experimental setup of the X-band near-field scanning system used to characterize the performance of the polarization rotator}
\label{fig:XBandNearFieldSystem}
\end{figure}

The polarization rotator was experimentally measured using the X-band near-field scanning system described in \cite{pfeiffer2013metamaterial}, and shown in Fig. \ref{fig:XBandNearFieldSystem}. To approximate the plane wave excitation, a quasi-optical Gaussian beam telescope was used. The telescope consisted of a rectangular horn antenna (Dorado GH-90-25) with a gain of 25 dBi, and a pair of lenses separated by the sum of their focal distances. The horn antenna produced a quasi-Gaussian beam with 88\% of its power coupled to the fundamental Gaussian mode. The pair of lenses focused the Gaussian beam to a beam waist with size and location independent of frequency \cite{goldsmith1992quasi}. The two lenses were identical, each made of Rexolite $(n=1.59)$ and bi-hyperbolic in shape. The diameter of the lenses was 32.5 cm with input and output focal distances of 45 cm. The focused Gaussian beam was measured to have a 114 mm beam waist diameter, 1.8 m from the phase center of the horn antenna. The rectangular horn antenna generating the Gaussian beam was connected to the transmitting port of a vector network analyzer (Agilent E8361A). An open-ended, WR-90 waveguide probe was connected to the receive port (see Fig. \ref{fig:XBandNearFieldSystem}). The ratio of the received voltage to the transmitted voltage provided the magnitude and phase of the field at the waveguide probe's position. Using a two-dimensional translation stage, the field was sampled over a 200 mm x 200 mm area $(6.7\lambda_{\circ}$ x $6.7\lambda_{\circ})$, 100 mm $(3.33\lambda)$ behind the metasurface. The field was sampled every 20 mm x 20 mm $(0.67\lambda_{\circ}$ x $0.67\lambda_{\circ})$, which is a higher sampling rate than necessary since the beam is approximately paraxial.

\begin{figure}[ht]
\centering
\centerline{\includegraphics[width=3.1in]{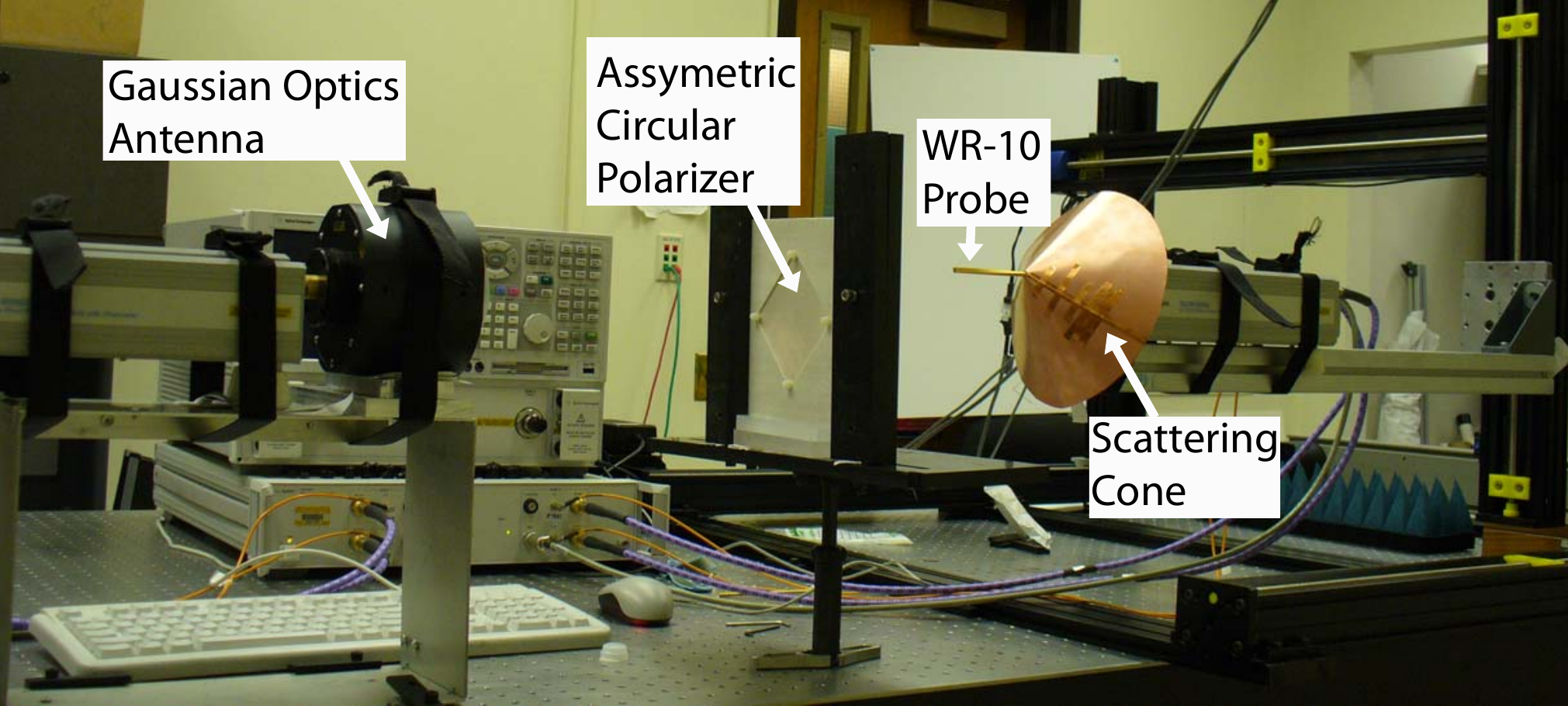}}
\caption{Experimental setup of the mm-wave near-field scanning system used to characterize the performance of the asymmetric circular polarizer.}
\label{fig:mmWaveNearFieldSystem}
\end{figure}

The asymmetric circular polarizer was measured using the mm-wave near-field scanning system described in \cite{pfeiffer2013millimeter}, and shown in Fig. \ref{fig:mmWaveNearFieldSystem}. This system operates from 67 GHz to 100 GHz. A Gaussian-Optics-Antenna (Millitech GOA-10-R00004F) was connected to the transmitting port of a vector network analyzer (Agilent E8361A). The antenna illuminated the transmitarray at normal incidence with a focused Gaussian beam whose measured beam waist was 40 mm in diameter \cite{goldsmith1992quasi}. An open-ended WR-10 waveguide probe was connected to the receive port of the network analyzer to measure the transmitted electric field. A scattering cone minimized backscatter from the metallic structure supporting the waveguide probe \cite{gregson2007principles}. Using a two-dimensional translation stage with 5 $\mu$m accuracy, the field was sampled over an 82 mm x 82 mm area $(21\lambda_{\circ}$ x $21\lambda_{\circ})$, 17 mm $(4.4\lambda_{\circ})$ behind the metasurface. The field was sampled every 4.1 mm x 4.1 mm ($1.05\lambda_{\circ}$ x $1.05\lambda_{\circ}$), which is a higher sampling rate than necessary since the beam is paraxial.

To ensure that the measurements were accurate, a number of precautions were taken. The diameters of the metasurfaces were at least 1.7 times the measured beam waist diameter of the incident Gaussian beam. This limited diffraction effects \cite{goldsmith1992quasi}. The field at the edge of the sampled area was approximately 25 dB below the peak value for the X-band system, and 30 dB below the peak value for the mm-wave system. This ensured that the majority of the power was sampled. To reduce the effects of multiple reflections between the metasurface, lenses, and antennas, time domain gating techniques were employed. The far-field was used to characterize both metasurfaces by appropriately Fourier transforming the measured near-field. To properly extract the far-field using a waveguide probe, the probe's radiation pattern was determined using full-wave electromagnetic simulations, and probe correction was applied \cite{yaghjian1986overview}. The system was calibrated by first measuring the far-field of the incident Gaussian beam, without the metasurface present. The metasurface was then placed at the beam waist of the Gaussian beam, and again the far-field was measured. The far-field of the metasurface was then normalized by the peak amplitude of the incident beam's far-field. The amplitude of the transmission coefficient of the metasurface was determined by taking the square root of the ratio of the transmitted to incident far-field power. The phase was determined by noting the phase of the transmitted electric field in the direction of the main beam.

For both structures, the transmitting antennas illuminated the metasurface with a vertical polarization. The transmitted vertical polarizations were first measured using standard open-ended waveguide probes. The horizontal polarization of the X-band near-field scanning system was then measured by rotating the waveguide probe by $90^{\circ}$ and scanning the field again. The horizontal polarization of the mm-wave near-field scanning system was measured with a waveguide probe that has a $90^{\circ}$ twist. To measure the transmission coefficient for a horizontally polarized plane wave, the metasurface was rotated by $90^{\circ}$, and again the vertical and horizontal polarizations were measured. This procedure provided the entire transmission coefficient matrix, $\textbf{S}_{21}=\begin{pmatrix}S_{21}^{xx} & S_{21}^{xy}\\S_{21}^{yx} & S_{21}^{yy}\end{pmatrix}$. Once the transmission coefficient was determined in the linear polarization basis vectors, it could be rewritten in terms of any desired polarization basis, such as circular \cite{menzel2010advanced}.

\section{Relating S-parameters to constituent surface parameters}
In the main text, we defined the constituent surface parameters that model an arbitrary bianisotropic metasurface. Here, relations between the S-parameters and the constituent parameters are derived using this well-defined boundary condition. In contrast, previously reported analyses modeled bianisotropic metasurfaces as coupled electric and magnetic dipoles \cite{zhang2013Interference}. If the interaction from neighboring particles is also considered, the polarizabilities needed to achieve an arbitrary polarization transformation can be solved for, as in \cite{niemi2013synthesis}. However, this model is rather complex, and valid only when the metasurface consists of infinitesimally small dipoles \cite{niemi2013synthesis}. This is a significant limitation since the bandwidth of periodic structures is typically enhanced by evanescent coupling between neighboring unit cells \cite{rudolph2012broadband}. In addition, the design process to realize a physical structure is not straightforward. Alternatively, it is possible to model a metasurface as a thin bianisotropic slab \cite{shaltout2013homogenization}. However, the scattering parameters cannot be solved in closed form, and the interpretation is not strictly appropriate since the thickness of a metasurface is ambiguous \cite{holloway2011characterizing}.

To begin, the boundary condition given by (2) of the main text is explicitly written in terms of the tangential field in Regions 1 and 2,
\begin{eqnarray}
\textbf{Y}\left(\frac{\textbf{E}_1+\textbf{E}_2}{2}\right)+ \boldsymbol{\chi}\left(\frac{\textbf{H}_1+\textbf{H}_2}{2}\right)&=& \hat{n}\times\left(\textbf{H}_2-\textbf{H}_1\right)\nonumber\\
&=&\textbf{n} (\textbf{H}_2-\textbf{H}_1)\label{eqn:BC1}\\
\boldsymbol{\Upsilon}\left(\frac{\textbf{E}_1+\textbf{E}_2}{2}\right)+ \textbf{Z}\left(\frac{\textbf{H}_1+\textbf{H}_2}{2}\right)&=&-\hat{n}\times (\textbf{E}_2-\textbf{E}_1)\nonumber\\
&=&-\textbf{n} (\textbf{E}_2-\textbf{E}_1)\label{eqn:BC2}
\end{eqnarray}

Consider an x-polarized plane wave, normally incident on the bianisotropic metasurface from Region 1. The field in Region 1 is expressed as $\textbf{E}_1^+=\textbf{I}_x+\textbf{S}_{11}^x$ and $\textbf{H}_1^+=\eta_1^{-1}\textbf{n}(\textbf{I}_x-\textbf{S}_{11}^x)$. The field in Region 2 is written as $\textbf{E}_2^+=\textbf{S}_{21}^x$ and $\textbf{H}_2^+=\eta_2^{-1}\textbf{n}\textbf{S}_{21}^x$. Here, the $+$ sign indicates the excitation is incident from Region 1 (traveling in the $+z$ direction), $\textbf{I}_x=(1\quad 0)^T$, $\textbf{S}_{11}^x=(S_{11}^{xx} \quad S_{11}^{yx})^T$, $\textbf{S}_{21}^x=(S_{21}^{xx} \quad S_{21}^{yx})^T$, and $\textbf{n}=\begin{pmatrix}0&-1\\1&0\end{pmatrix}$. Substituting these expressions for $\textbf{E}^+$ and $\textbf{H}^+$ into (\ref{eqn:BC1}) and (\ref{eqn:BC2}), the S-parameters are related to the constituent surface parameters,
\begin{eqnarray}
\frac{\textbf{Y}}{2}(\textbf{I}_x+\textbf{S}_{11}^x+\textbf{S}_{21}^x)&+& \frac{\boldsymbol{\chi}\textbf{n}}{2}\left(\frac{\textbf{I}_x}{\eta_1} -\frac{\textbf{S}_{11}^x}{\eta_1}+\frac{\textbf{S}_{21}^x}{\eta_2}\right)\nonumber\\
&=&-\left(\frac{-\textbf{I}_x}{\eta_1} +\frac{\textbf{S}_{11}^x}{\eta_1}+\frac{\textbf{S}_{21}^x}{\eta_2}\right) \label{eqn:BC1b}\\
\frac{\boldsymbol{\Upsilon}}{2}(\textbf{I}_x+\textbf{S}_{11}^x+\textbf{S}_{21}^x)&+& \frac{\textbf{Z}\textbf{n}}{2}\left(\frac{\textbf{I}_x}{\eta_1} -\frac{\textbf{S}_{11}^x}{\eta_1}+\frac{\textbf{S}_{21}^x}{\eta_2}\right)\nonumber\\
&=&-\textbf{n}(-\textbf{I}_x -\textbf{S}_{11}^x+\textbf{S}_{21}^x) \label{eqn:BC2b}
\end{eqnarray}

This linear system of equations is solved in closed form,
\begin{eqnarray}
\begin{pmatrix}\textbf{S}_{11}^x\\[0.7em]
\textbf{S}_{21}^x \end{pmatrix} &=&
\begin{pmatrix}\frac{\textbf{Y}}{2}-\frac{\boldsymbol{\chi}\textbf{n}}{2\eta_1} +\frac{\textbf{I}}{\eta_1} & \frac{\textbf{Y}}{2}+\frac{\boldsymbol{\chi}\textbf{n}}{2\eta_2} +\frac{\textbf{I}}{\eta_2}\\[0.7em]
-\frac{\textbf{Z}\textbf{n}}{2\eta_1}+\frac{\boldsymbol{\Upsilon}}{2}-\textbf{n} & \frac{\textbf{Z}\textbf{n}}{2\eta_2}+\frac{\boldsymbol{\Upsilon}}{2}+\textbf{n}
\end{pmatrix}^{-1}\nonumber\\
&&\cdot
\begin{pmatrix}-\frac{\textbf{Y}\textbf{I}_x}{2} -\frac{\boldsymbol{\chi}\textbf{n}\textbf{I}_x}{2\eta_1}+ \frac{\textbf{I}_x}{\eta_1}\\[0.7em]
-\frac{\textbf{Z}\textbf{n}\textbf{I}_x}{2\eta_1}-\frac{\boldsymbol{\Upsilon} \textbf{I}_x}{2}+\textbf{n}\textbf{I}_x
\end{pmatrix}
\end{eqnarray}
where $\textbf{I}=\begin{pmatrix}1&0\\0&1\end{pmatrix}$ is the identity matrix. Similarly, $\textbf{S}_{11}^y$ and $\textbf{S}_{21}^y$ are solved by replacing $\textbf{I}_x=(1\quad 0)^T$ with $\textbf{I}_y=(0\quad 1)^T$.

The variables, $\textbf{S}_{12}$ and $\textbf{S}_{22}$ are also solved using the boundary condition from (\ref{eqn:BC1}) and (\ref{eqn:BC2}). A normally incident x-polarized plane wave excitation is stipulated in Region 2. The field in Region 2 is then expressed as $\textbf{E}_2^-=\textbf{I}_x+\textbf{S}_{22}^x$, and $\textbf{H}_2^-=\eta_2^{-1}\textbf{n}(-\textbf{I}_x+\textbf{S}_{22}^x)$. The field in Region 1 is written as $\textbf{E}_1^-=\textbf{S}_{12}^x$ and $\textbf{H}_1^-=-\eta_1^{-1}\textbf{n}\textbf{S}_{12}^x$. Analogous to before, the $-$ sign indicates the excitation is incident from Region 2, $\textbf{S}_{12}^x=(S_{12}^{xx} \quad S_{12}^{yx})^T$, and $\textbf{S}_{22}^x=(S_{22}^{xx} \quad S_{22}^{yx})^T$. The expressions for $\textbf{E}^-$ and $\textbf{H}^-$ are then substituted into (\ref{eqn:BC1}) and (\ref{eqn:BC2}),
\begin{eqnarray}
\frac{\textbf{Y}}{2}(\textbf{I}_x+\textbf{S}_{12}^x+\textbf{S}_{22}^x)&+& \frac{\boldsymbol{\chi}\textbf{n}}{2}\left(-\frac{\textbf{I}_x}{\eta_2} -\frac{\textbf{S}_{12}^x}{\eta_1}+\frac{\textbf{S}_{22}^x}{\eta_2}\right)\nonumber\\
&=&-\left(\frac{-\textbf{I}_x}{\eta_2} +\frac{\textbf{S}_{12}^x}{\eta_1}+\frac{\textbf{S}_{22}^x}{\eta_2}\right)\\
\frac{\boldsymbol{\Upsilon}}{2}(\textbf{I}_x+\textbf{S}_{12}^x+\textbf{S}_{22}^x)&+& \frac{\textbf{Z}\textbf{n}}{2}\left(-\frac{\textbf{I}_x}{\eta_2} -\frac{\textbf{S}_{12}^x}{\eta_1}+\frac{\textbf{S}_{22}^x}{\eta_2}\right)\nonumber\\
&=&-\textbf{n}(\textbf{I}_x -\textbf{S}_{12}^x+\textbf{S}_{22}^x)
\end{eqnarray}
Again, there are two equations and two unknowns that can be solved,
\begin{eqnarray}
\begin{pmatrix}\textbf{S}_{12}^x\\[0.7em]
\textbf{S}_{22}^x \end{pmatrix} &=&
\begin{pmatrix}\frac{\textbf{Y}}{2}-\frac{\boldsymbol{\chi}\textbf{n}}{2\eta_1} +\frac{\textbf{I}}{\eta_1} & \frac{\textbf{Y}}{2}+\frac{\boldsymbol{\chi}\textbf{n}}{2\eta_2} +\frac{\textbf{I}}{\eta_2}\\[0.7em]
-\frac{\textbf{Z}\textbf{n}}{2\eta_1}+\frac{\boldsymbol{\Upsilon}}{2}-\textbf{n} & \frac{\textbf{Z}\textbf{n}}{2\eta_2}+\frac{\boldsymbol{\Upsilon}}{2}+\textbf{n}
\end{pmatrix}^{-1}\nonumber\\
&&\cdot
\begin{pmatrix}-\frac{\textbf{Y}\textbf{I}_x}{2} +\frac{\boldsymbol{\chi}\textbf{n}\textbf{I}_x}{2\eta_2}+ \frac{\textbf{I}_x}{\eta_2}\\[0.7em]
\frac{\textbf{Z}\textbf{n}\textbf{I}_x}{2\eta_2}-\frac{\boldsymbol{\Upsilon} \textbf{I}_x}{2}-\textbf{n}\textbf{I}_x
\end{pmatrix}
\end{eqnarray}

The expressions $\textbf{S}_{12}^y$ and $\textbf{S}_{22}^y$ are solved by replacing $\textbf{I}_x$ with $\textbf{I}_y$. Therefore, all the S-parameters are written concisely as,
\begin{align}\label{eqn:SparamfromLambda}
\begin{pmatrix}\textbf{S}_{11} & \textbf{S}_{12}\\[0.7em]
\textbf{S}_{21} & \textbf{S}_{22} \end{pmatrix} &=
\begin{pmatrix}\frac{\textbf{Y}}{2}-\frac{\boldsymbol{\chi}\textbf{n}}{2\eta_1} +\frac{\textbf{I}}{\eta_1} & \frac{\textbf{Y}}{2}+\frac{\boldsymbol{\chi}\textbf{n}}{2\eta_2} +\frac{\textbf{I}}{\eta_2}\\[0.7em]
-\frac{\textbf{Z}\textbf{n}}{2\eta_1}+\frac{\boldsymbol{\Upsilon}}{2}-\textbf{n} & \frac{\textbf{Z}\textbf{n}}{2\eta_2}+\frac{\boldsymbol{\Upsilon}}{2}+\textbf{n}
\end{pmatrix}^{-1}\nonumber\\
&\cdot
\begin{pmatrix}-\frac{\textbf{Y}}{2} -\frac{\boldsymbol{\chi}\textbf{n}}{2\eta_1}+ \frac{\textbf{I}}{\eta_1} & -\frac{\textbf{Y}}{2} +\frac{\boldsymbol{\chi}\textbf{n}}{2\eta_2}+ \frac{\textbf{I}}{\eta_2}\\[0.7em]
-\frac{\textbf{Z}\textbf{n}}{2\eta_1}-\frac{\boldsymbol{\Upsilon}}{2} +\textbf{n} & \frac{\textbf{Z}\textbf{n}}{2\eta_2}-\frac{\boldsymbol{\Upsilon}}{2}-\textbf{n}
\end{pmatrix}
\end{align}
which is identical to (3) of the main text.

Alternatively, the constituent surface parameters can be written in terms of the S-parameters. In total there are four illuminations (x-polarized and y-polarized from the front and back of the metasurface). These illuminations are inserted into (\ref{eqn:BC1}) and (\ref{eqn:BC2}),
\begin{eqnarray}
\begin{pmatrix}\textbf{Y} & \boldsymbol{\chi} \\[0.3em]
      \boldsymbol{\Upsilon} & \textbf{Z}\end{pmatrix}&&
      \begin{pmatrix}\frac{\textbf{E}_1^++\textbf{E}_2^+}{2} & \frac{\textbf{E}_1^-+\textbf{E}_2^-}{2}\\[0.3em]
      \frac{\textbf{H}_1^++\textbf{H}_2^+}{2} & \frac{\textbf{H}_1^-+\textbf{H}_2^-}{2}\end{pmatrix}\nonumber\\
      &&=\begin{pmatrix}\textbf{n} (\textbf{H}_2^+-\textbf{H}_1^+) & \textbf{n} (\textbf{H}_2^--\textbf{H}_1^-)\\[0.3em]
      -\textbf{n} (\textbf{E}_2^+-\textbf{E}_1^+) & -\textbf{n} (\textbf{E}_2^--\textbf{E}_1^-)\end{pmatrix}
\end{eqnarray}

Substituting the expressions for $\textbf{E}_{1,2}^{+,-}$ and $\textbf{H}_{1,2}^{+,-}$, and bringing the average field values to the right hand side of the equation, the constituent surface parameters are solved,
\begin{align}\label{eqn:LambdafromSparam}
\begin{pmatrix}\textbf{Y} & \boldsymbol{\chi} \\[0.3em]
      \boldsymbol{\Upsilon} & \textbf{Z}\end{pmatrix}&=
      2\begin{pmatrix}\frac{\textbf{I}}{\eta_1}-\frac{\textbf{S}_{11}}{\eta_1} -\frac{\textbf{S}_{21}}{\eta_2} & \frac{\textbf{I}}{\eta_2}-\frac{\textbf{S}_{12}}{\eta_1} -\frac{\textbf{S}_{22}}{\eta_2}\\[0.3em]
      \textbf{n}+\textbf{n}\textbf{S}_{11}-\textbf{n}\textbf{S}_{21} & -\textbf{n}+\textbf{n}\textbf{S}_{12}-\textbf{n}\textbf{S}_{22}\end{pmatrix} \nonumber\\
      &\cdot \begin{pmatrix}\textbf{I}+\textbf{S}_{11}+\textbf{S}_{21} & \textbf{I}+\textbf{S}_{12}+\textbf{S}_{22}\\[0.3em]
      \frac{\textbf{n}}{\eta_1}-\frac{\textbf{n}\textbf{S}_{11}}{\eta_1} +\frac{\textbf{n}\textbf{S}_{21}}{\eta_2} & -\frac{\textbf{n}}{\eta_2}-\frac{\textbf{n}\textbf{S}_{12}}{\eta_1} +\frac{\textbf{n}\textbf{S}_{22}}{\eta_2}\end{pmatrix}^{-1}
\end{align}
which is identical to (4) of the main text.

\section{Relating S-parameters to cascaded sheet admittances}

Following an approach similar to that in \cite{zhao2011homogenization}, the reflection and transmission properties of the cascaded sheet admittances are solved. The transfer matrix approach is taken ($\textbf{ABCD}$ matrix), which reduces the analysis to matrix multiplication once the transfer matrix of the sheet admittances and dielectric substrate are derived.

To begin, the transfer matrix ($\textbf{ABCD}$ matrix) of an arbitrary structure is defined by relating the total field in Regions 1 and 2,
\begin{equation}
\begin{pmatrix}\textbf{E}_1\\ \textbf{H}_1\end{pmatrix} =
\begin{pmatrix} \textbf{A} & \textbf{B}\\ \textbf{C} & \textbf{D}\end{pmatrix}
\begin{pmatrix}\textbf{E}_2\\ \textbf{H}_2\end{pmatrix}
\end{equation}
As before, we define, $\textbf{E}_{1,2}=[E_{1,2}^x\quad E_{1,2}^y]^T$, $\textbf{H}_{1,2}=[H_{1,2}^x \quad H_{1,2}^y]^T$, $\textbf{A}=\begin{pmatrix}A_{xx} & A_{xy}\\
A_{yx} & A_{yy}\end{pmatrix}$, $\textbf{B}=\begin{pmatrix}B_{xx} & B_{xy}\\
B_{yx} & B_{yy}\end{pmatrix}$, $\textbf{C}=\begin{pmatrix}C_{xx} & C_{xy}\\
C_{yx} & C_{yy}\end{pmatrix}$, and $\textbf{D}=\begin{pmatrix}D_{xx} & D_{xy}\\
D_{yx} & D_{yy}\end{pmatrix}$.

The $\textbf{ABCD}$ matrix of an electric sheet admittance $\textbf{Y}_{s}$ is then derived. First it is noted that the boundary condition of an electric sheet admittance can be written as,
\begin{equation}\label{eqn:SheetAdmittanceBC}
\hat{n}\times\left(\textbf{H}_2-\textbf{H}_1\right)=\textbf{n}\left(\textbf{H}_2-\textbf{H}_1\right)=\textbf{Y}_s\textbf{E}_1=\textbf{Y}_s\textbf{E}_2
\end{equation}
Two separate conditions are then stipulated: Condition A ($\textbf{E}_2=\textbf{I}$, and $\textbf{H}_2=\textbf{0}$) and Condition B ($\textbf{E}_2=\textbf{0}$, $\textbf{H}_2=\textbf{I}$). Thus we have,
\begin{equation}
\begin{pmatrix}\textbf{E}_1^A & \textbf{E}_1^B\\ \textbf{H}_1^A & \textbf{H}_1^B\end{pmatrix} =
\begin{pmatrix} \textbf{A} & \textbf{B}\\ \textbf{C} & \textbf{D}\end{pmatrix}
\begin{pmatrix}\textbf{E}_2^A & \textbf{E}_2^B\\ \textbf{H}_2^A & \textbf{H}_2^B\end{pmatrix}=\begin{pmatrix} \textbf{A} & \textbf{B}\\ \textbf{C} & \textbf{D}\end{pmatrix}
\end{equation}
Enforcing the boundary condition of an electric sheet admittance under the two separate conditions, the field in Region 1 is solved, thus providing the $\textbf{ABCD}$ matrix of an electric sheet admittance,
\begin{equation}
\begin{pmatrix} \textbf{A} & \textbf{B}\\ \textbf{C} & \textbf{D}\end{pmatrix}
=\begin{pmatrix}
      \textbf{I} & \textbf{0}\\
      \textbf{n}\textbf{Y}_{s} & \textbf{I}
   \end{pmatrix}
\end{equation}
It should be noted that $-\textbf{n}\textbf{n}=\textbf{I}$.

The $\textbf{ABCD}$ matrix of a dielectric substrate with wave impedance $\eta_d$ and thickness $\beta d$ is then derived. First consider Condition A ($\textbf{E}_2=\textbf{I}$, and $\textbf{H}_2=\textbf{0}$). This is equivalent to the case where a plane wave is incident from Region 1 with an incident electric field of $\textbf{I}/2$. We must also require Region 1 to have a wave impedance of $\eta_d$, and Region 2 have an infinite wave impedance (perfect magnetic conductor). Thus the field in Region 1 is written as,
\begin{eqnarray}
\textbf{E}_1^A&=&\textbf{I}\frac{e^{j\beta d}+e^{-j\beta d}}{2}=\textbf{I}\textrm{ cos}(\beta d)\nonumber\\
 \textbf{H}_1^A&=&\textbf{n}\frac{e^{j\beta d}-e^{-j\beta d}}{2\eta_d}=\textbf{n}j\eta_d^{-1}\textrm{ sin}(\beta d)
\end{eqnarray}
Similarly, Condition B ($\textbf{E}_2=\textbf{0}$, $\textbf{H}_2=\textbf{I}$) is equivalent to the case where a plane wave is incident from Region 1 with an incident magnetic field of $\textbf{I}/2$, Region 1 has a wave impedance of $\eta_d$, and Region 2 is replaced with a perfect electric conductor,
\begin{eqnarray}
\textbf{E}_1^B&=&-\textbf{n}\eta_d\frac{e^{j\beta d}-e^{-j\beta d}}{2}=-\textbf{n}j\eta_d\textrm{ sin}(\beta d)\nonumber\\
 \textbf{H}_1^B&=&\textbf{I}\frac{e^{j\beta d}+e^{-j\beta d}}{2}=\textbf{I}\textrm{ cos}(\beta d)
\end{eqnarray}
Thus the $\textbf{ABCD}$ matrix of a dielectric substrate is written as,
\begin{equation}
\begin{pmatrix} \textbf{A} & \textbf{B}\\ \textbf{C} & \textbf{D}\end{pmatrix}
=\begin{pmatrix}
      \textrm{cos}(\beta d)\textbf{I} & -j\textrm{sin}(\beta d)\eta_d\textbf{n}\\[0.3em]
j\textrm{sin}(\beta d)\eta_d^{-1}\textbf{n} & \textrm{cos}(\beta d)\textbf{I}
   \end{pmatrix}
\end{equation}

When three electric sheet admittances are separated by dielectric spacers, the $\textbf{ABCD}$ matrix of the entire structure is written as,
\small
\begin{align}\label{eqn:ABCD3layer}
&\begin{pmatrix} \textbf{A} & \textbf{B}\\ \textbf{C} & \textbf{D}\end{pmatrix}
=\left[\begin{pmatrix}
      \textbf{I} & \textbf{0}\\
      \textbf{n}\textbf{Y}_{s1} & \textbf{I}
   \end{pmatrix} \begin{pmatrix}
      \textrm{cos}(\beta d)\textbf{I} & -j\textrm{sin}(\beta d)\eta_d\textbf{n}\\[0.3em]
j\textrm{sin}(\beta d)\eta_d^{-1}\textbf{n} & \textrm{cos}(\beta d)\textbf{I}
   \end{pmatrix}
   \right.\nonumber\\
   \cdot& \left.\begin{pmatrix}
      \textbf{I} & \textbf{0}\\
      \textbf{n}\textbf{Y}_{s2} & \textbf{I}
   \end{pmatrix} \begin{pmatrix}
      \textrm{cos}(\beta d)\textbf{I} & -j\textrm{sin}(\beta d)\eta_d\textbf{n}\\[0.3em]
j\textrm{sin}(\beta d)\eta_d^{-1}\textbf{n} & \textrm{cos}(\beta d)\textbf{I}
   \end{pmatrix}  \begin{pmatrix}
      \textbf{I} & \textbf{0}\\
      \textbf{n}\textbf{Y}_{s3} & \textbf{I}
   \end{pmatrix}\right]
\end{align}
\normalsize
which is identical to (6) of the main text.

Next, the $\textbf{ABCD}$ matrix of an arbitrary structure is related to its S-parameters. The total field in Regions 1 and 2 can be written as,
\begin{align}\label{eqn:ABCDSparam}
\begin{pmatrix}
      \textbf{I}+\textbf{S}_{11} & \textbf{S}_{12}\\[0.3em]
      \frac{\textbf{n}}{\eta_1}(\textbf{I}-\textbf{S}_{11}) & -\frac{\textbf{n}}{\eta_1}\textbf{S}_{12}
   \end{pmatrix} &=\begin{pmatrix} \textbf{A} & \textbf{B}\\ \textbf{C} & \textbf{D}\end{pmatrix}
   \nonumber\\
   \cdot& \begin{pmatrix}
      \textbf{S}_{21} & \textbf{I}+\textbf{S}_{22}\\[0.3em]
      \frac{\textbf{n}}{\eta_2}\textbf{S}_{21} & \frac{\textbf{n}}{\eta_2}(-\textbf{I}+\textbf{S}_{22})
   \end{pmatrix}
\end{align}
Then, the $\textbf{ABCD}$ matrix of an arbitrary structure is written in terms of the S-parameters by solving (\ref{eqn:ABCDSparam}),
\begin{align}
\begin{pmatrix} \textbf{A} & \textbf{B}\\ \textbf{C} & \textbf{D}\end{pmatrix}
&=\begin{pmatrix}
      \textbf{I}+\textbf{S}_{11} & \textbf{S}_{12}\\[0.3em]
      \frac{\textbf{n}}{\eta_1}(\textbf{I}-\textbf{S}_{11}) & -\frac{\textbf{n}}{\eta_1}\textbf{S}_{12}
   \end{pmatrix}
   \nonumber\\
   \cdot& \begin{pmatrix}
      \textbf{S}_{21} & \textbf{I}+\textbf{S}_{22}\\[0.3em]
      \frac{\textbf{n}}{\eta_2}\textbf{S}_{21} & \frac{\textbf{n}}{\eta_2}(-\textbf{I}+\textbf{S}_{22})
   \end{pmatrix}^{-1}
\end{align}

Alternatively, the S-parameters can be written in terms of the $\textbf{ABCD}$ matrix by solving (\ref{eqn:ABCDSparam}),
\begin{equation}\label{eqn:SparamsFromABCD}
\begin{pmatrix} \textbf{S}_{11} & \textbf{S}_{12}\\[0.3em] \textbf{S}_{21} & \textbf{S}_{22}\end{pmatrix}
=\begin{pmatrix}
      -\textbf{I} & \frac{\textbf{B}\textbf{n}}{\eta_2}+\textbf{A}\\[0.3em]
      \frac{\textbf{n}}{\eta_1} & \frac{\textbf{D}\textbf{n}}{\eta_2}+\textbf{C}
   \end{pmatrix}^{-1}\begin{pmatrix}
      \textbf{I} & \frac{\textbf{B}\textbf{n}}{\eta_2}-\textbf{A}\\[0.3em]
      \frac{\textbf{n}}{\eta_1} & \frac{\textbf{D}\textbf{n}}{\eta_2}-\textbf{C}
   \end{pmatrix}
\end{equation}
which is identical to (7) of the main text.

\section{Finding the sheet admittances}
Analytically solving for the S-parameters of a given structure is straightforward. For example, the S-parameters of three cascaded sheet admittances can be found by inserting (\ref{eqn:ABCD3layer}) into (\ref{eqn:SparamsFromABCD}). However, we are looking to solve the inverse problem: the S-parameters are stipulated and the necessary sheet admittances are found. Since the necessary sheet admittances could not be solved analytically, numerical solvers were employed. The fmincon function provided by Matlab's optimization toolbox was used to perform a gradient descent method. The specific cost function that was minimized was $|S_{21}^{xx}-T^{xx}|^2+|S_{21}^{xy}-T^{xy}|^2+|S_{21}^{yx}-T^{yx}|^2+|S_{21}^{yy}-T^{yy}|^2$, where $\textbf{S}_{21}$ is the transmission coefficient of the cascaded sheet admittances, and $\textbf{T}$ is the desired transmission coefficient. Since this is a nonlinear problem, the gradient descent method may only return a local minimum rather than the global minimum, depending on the initial starting point. Nevertheless, the optimizer typically converged to the global minimum with less than 10 randomly seeded initial starting points.

\begin{figure}[ht]
    \centering
    \includegraphics[width=2.4in]{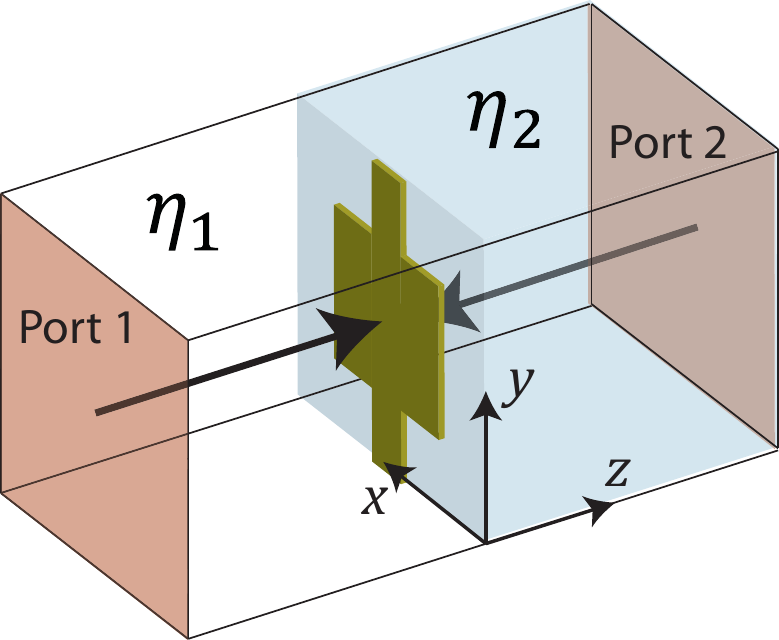}
  \caption{Simulation used to design each sheet admittance.}\label{fig:SheetAdmittanceExtraction}
\end{figure}

Once a desired sheet admittance is determined, it is realized by patterning metal on a dielectric substrate. The metallic pattern is designed through scattering simulations using the full-wave solver Ansys HFSS. Fig. \ref{fig:SheetAdmittanceExtraction} demonstrates how each sheet is simulated. Floquet ports excite normally incident plane waves, and are de-embedded to the $z=0$ plane. Infinite periodicity is assumed by stipulating periodic boundary conditions along the other four sides. As shown in Fig. \ref{fig:SheetAdmittanceExtraction}, the sheet admittance is in general located between two different media with wave impedances given by $\eta_1$ and $\eta_2$. It is important to note that, in addition to the metallic pattern, the sheet admittance itself is also a function of the media surrounding it.

An iterative approach is used to design the sheet admittance. First, the dimensions of a metallic pattern are stipulated, and the structure is simulated. The sheet admittance is extracted from simulation by noting the simulated reflection coefficients,
\begin{eqnarray}\label{eqn:SheetExtraction}
\textbf{Y}_s&=&\left(\frac{\textbf{I}-\textbf{S}_{11}}{\eta_1}- \frac{\textbf{I}+\textbf{S}_{11}}{\eta_2}\right)\left(\textbf{I}+\textbf{S}_{11}\right)^{-1}\nonumber\\ &=&\left(\frac{\textbf{I}-\textbf{S}_{22}}{\eta_2}- \frac{\textbf{I}+\textbf{S}_{22}}{\eta_1}\right)\left(\textbf{I}+\textbf{S}_{22}\right)^{-1}
\end{eqnarray}
Then considering the equivalent circuit of the sheet, the dimensions are adjusted while also attempting to maximize bandwidth and minimize loss. Once the sheet admittances are realized, their cascaded response is calculated using the transfer matrix approach described earlier. It should be noted that the transfer matrix approach only accounts for propagating modes within the structure, and all evanescent modes are neglected. Thus, the analysis only works when evanescent coupling between the sheets is negligible. The validity of this approximation is improved by reducing the cell size and increasing the interlayer spacing.

\section{Polarization rotator}
Chiral materials with a strong rotary power are commonly used in analytical chemistry, biology, and crystallography for identifying the spatial structure of molecules \cite{rogacheva2006giant}. Chirality has also received significant attention because it provides an alternative route to achieve negative refraction \cite{pendry2004chiral}. A particularly interesting structure that exhibits a strong chiral response is the polarization rotator, which rotates an incident linear polarization by $90^{\circ}$ upon transmission. Previously, polarization rotation was accomplished with an isotropic helical structure \cite{niemi2013synthesis}. However, the three-dimensional geometry requires metallized via holes, which become prohibitively difficult to fabricate at higher frequencies. In addition, the structure exhibited a large insertion loss (S$_{21}$=-5 dB). Alternatively, bilayered metamaterials that utilize two sheet admittances separated by an electrically thin thickness can also act as polarization rotators \cite{rogacheva2006giant,ye201090}. These results demonstrated that complex helical patterns are not required to achieve significant chirality. In \cite{rogacheva2006giant}, it was shown that bilayered metamaterials have orders of magnitude larger rotary powers than naturally occurring gyrotropic crystals in the visible spectrum. The rotary power was later increased by optimizing the patterns on each sheet \cite{ye201090}. However, the design process was not straightforward, which led to a narrow bandwidth and low transmission coefficient (-5 dB) at the $90^{\circ}$ rotation angle. Here, a systematic method for designing polarization rotators is presented, which leads to optimal performance.

The polarization rotator presented in the main text has four patterned metallic sheets. To analyze this structure, (\ref{eqn:ABCD3layer}) must be modified to account for the fourth sheet,

\small
\begin{align}\label{eqn:ABCD2}
&\begin{pmatrix} \textbf{A} & \textbf{B}\\ \textbf{C} & \textbf{D}\end{pmatrix}
=\left[\begin{pmatrix}
      \textbf{I} & \textbf{0}\\
      \textbf{n}\textbf{Y}_{s1} & \textbf{I}
   \end{pmatrix} \begin{pmatrix}
      \textrm{cos}(\beta d_1)\textbf{I} & -j\textrm{sin}(\beta d_1)\eta_d\textbf{n}\\[0.3em]
j\textrm{sin}(\beta d_1)\eta_d^{-1}\textbf{n} & \textrm{cos}(\beta d_1)\textbf{I}
   \end{pmatrix}
   \right.\nonumber\\
   \cdot& \left.\begin{pmatrix}
      \textbf{I} & \textbf{0}\\
      \textbf{n}\textbf{Y}_{s2} & \textbf{I}
   \end{pmatrix} \begin{pmatrix}
      \textrm{cos}(\beta d_2)\textbf{I} & -j\textrm{sin}(\beta d_2)\eta_d\textbf{n}\\[0.3em]
j\textrm{sin}(\beta d_2)\eta_d^{-1}\textbf{n} & \textrm{cos}(\beta d_2)\textbf{I}
   \end{pmatrix}  \begin{pmatrix}
      \textbf{I} & \textbf{0}\\
      \textbf{n}\textbf{Y}_{s3} & \textbf{I}
   \end{pmatrix}\right.\nonumber\\
   &\left. \begin{pmatrix}
      \textrm{cos}(\beta d_1)\textbf{I} & -j\textrm{sin}(\beta d_1)\eta_d\textbf{n}\\[0.3em]
j\textrm{sin}(\beta d_1)\eta_d^{-1}\textbf{n} & \textrm{cos}(\beta d_1)\textbf{I}
   \end{pmatrix}  \begin{pmatrix}
      \textbf{I} & \textbf{0}\\
      \textbf{n}\textbf{Y}_{s4} & \textbf{I}
   \end{pmatrix}\right]
\end{align}
\normalsize
It should be noted that due to the adhesive layers used in fabrication, we allowed the middle dielectric spacer $(d_2)$ to be a different thickness than the outer dielectric spacers $(d_1)$.

As mentioned in the main text (see Eqs. (8) and (9)), a polarization rotator with a reflection coefficient equal to zero and transmission coefficient equal to,
\begin{equation}\label{eqn:polRotatorS}
\textbf{S}_{21}=e^{j\phi}\begin{pmatrix}0 & -1\\1&0\end{pmatrix}
\end{equation}
is considered. This device has constituent surface parameters given by,

\small
\begin{equation}\label{eqn:LambdaPolRotatorS}
 \boldsymbol{\Lambda}=\begin{pmatrix}
     \frac{-2j\textrm{ tan}(\phi)}{\eta_{\circ}} & 0 & -2\textrm{ sec}(\phi) & 0 \\[0.3em]
       0 & \frac{-2j\textrm{ tan}(\phi)}{\eta_{\circ}} & 0 & -2\textrm{ sec}(\phi) \\[0.3em]
       2\textrm{ sec}(\phi) & 0 & -2j\eta_{\circ}\textrm{ tan}(\phi) & 0 \\
       0 & 2\textrm{ sec}(\phi) & 0 & -2j\eta_{\circ}\textrm{ tan}(\phi)
   \end{pmatrix}
\end{equation}
\normalsize

The necessary cascaded sheet admittances that realize a polarization rotator are numerically found by inserting (\ref{eqn:polRotatorS}) into (\ref{eqn:SparamsFromABCD}), and combining the result with (\ref{eqn:ABCD2}). If the operating frequency equals 10 GHz, $\phi=-40^{\circ}$, $\eta_1=\eta_2=\eta_{\circ}$, $\eta_d=\eta_{\circ}/1.88$, $\beta d_1=2\pi/10.48$, and $\beta d_2=2\pi/9.54$, the required sheet admittances are $\textbf{Y}_{s1}=\frac{j}{\eta_{\circ}}\begin{pmatrix}0.92 & -1.39\\-1.39 & 2.14\end{pmatrix}$, $\textbf{Y}_{s2}=\frac{j}{\eta_{\circ}}\begin{pmatrix}5.21 & -8.07\\-8.07 & 5.21\end{pmatrix}$, $\textbf{Y}_{s3}=\frac{j}{\eta_{\circ}}\begin{pmatrix}7.88 & -1.17\\-1.17 & 2.50\end{pmatrix}$, and $\textbf{Y}_{s4}=\frac{j}{\eta_{\circ}}\begin{pmatrix}5.67 & 0\\0 & -2.63\end{pmatrix}$. It was found that the bandwidth is maximized by stipulating the transmitted phase to be $\phi=-40^{\circ}$.

\begin{figure}[ht]
    \centering
    \subfigure[]{
    \includegraphics[width=1.55in]{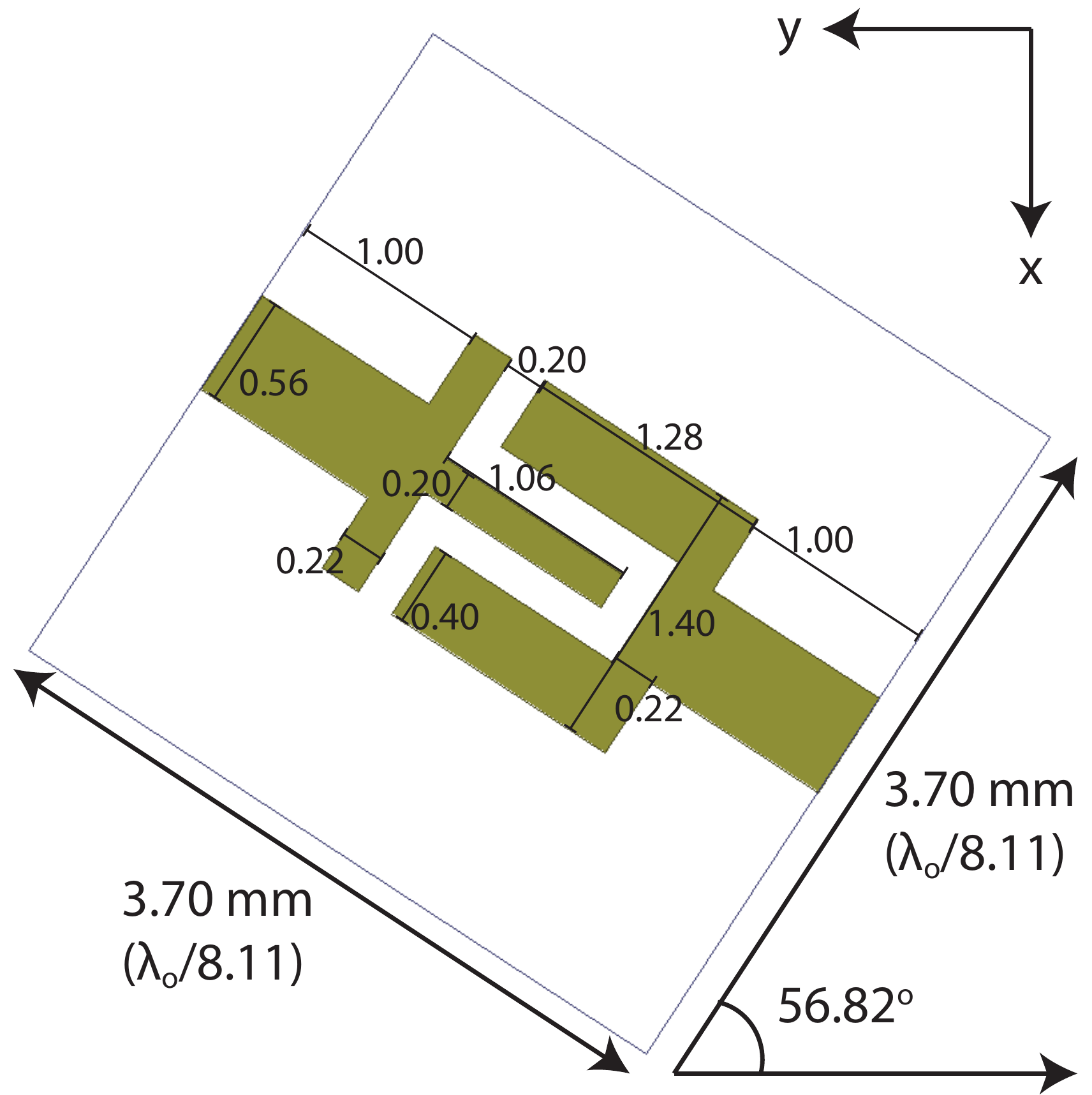}
    \label{fig:PolRotatorYs1}}
     \subfigure[]{
    \includegraphics[width=1.55in]{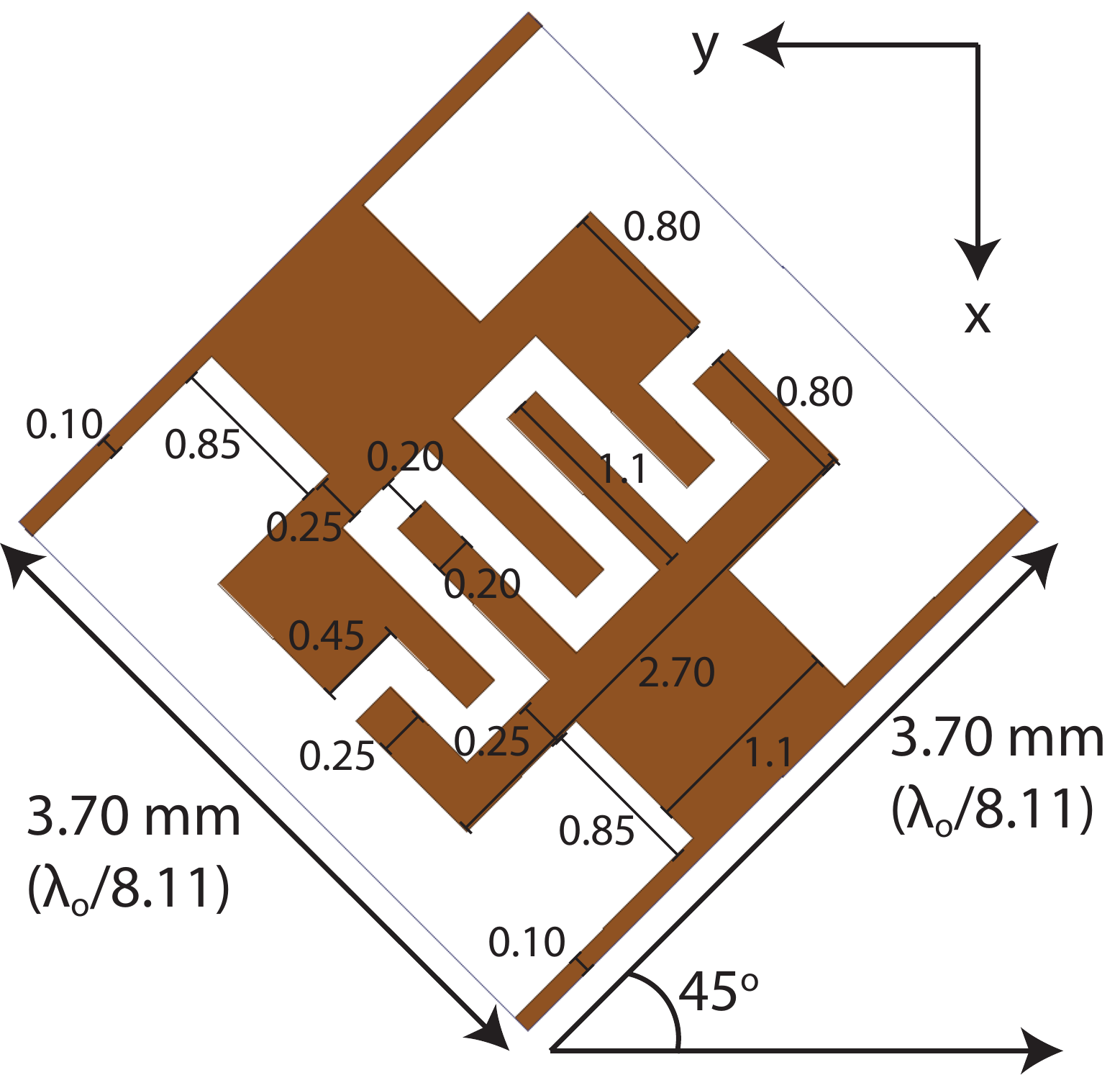}
    \label{fig:PolRotatorYs2}}
    \subfigure[]{
    \includegraphics[width=1.55in]{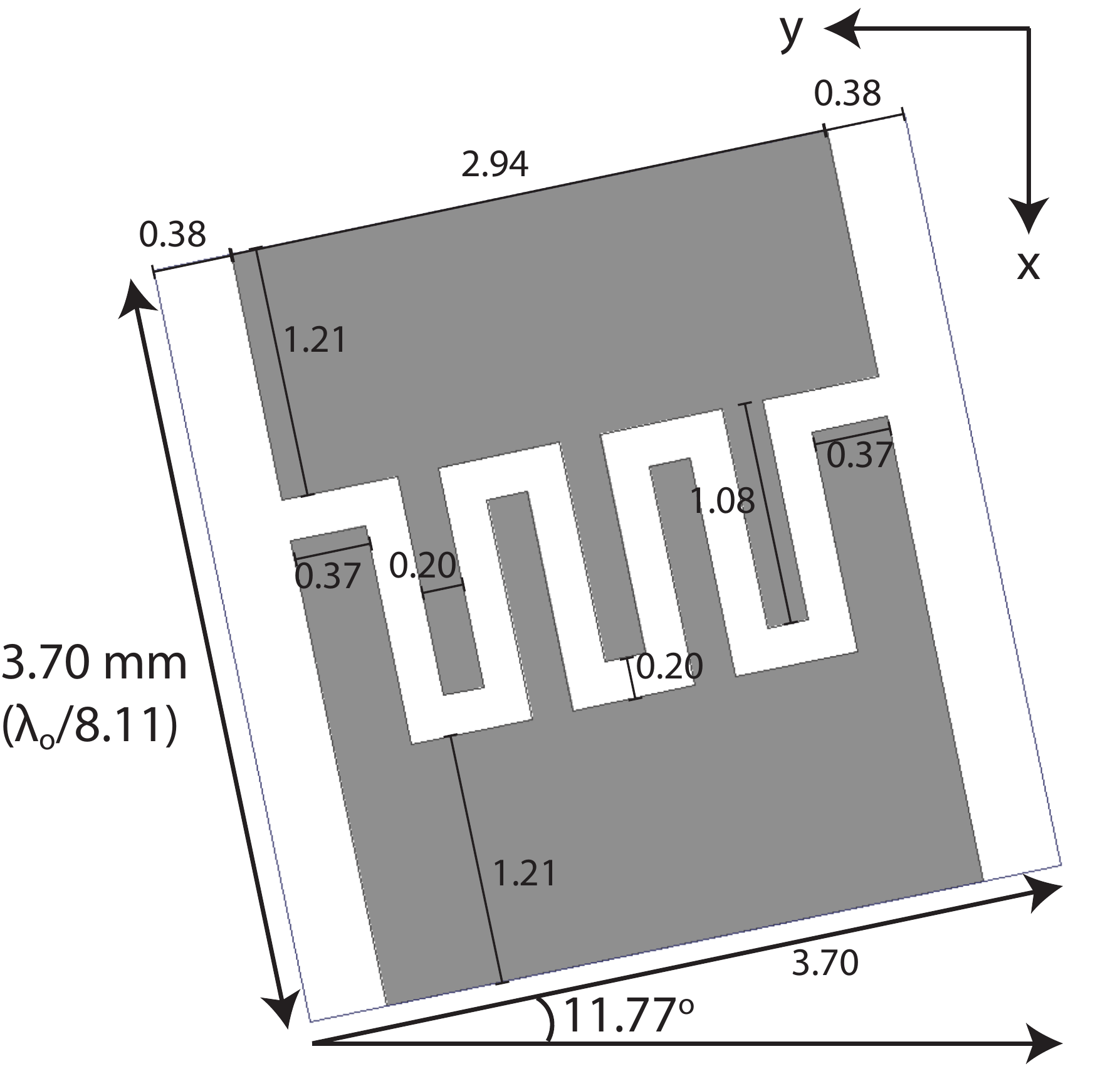}
    \label{fig:PolRotatorYs3}}
    \subfigure[]{
    \includegraphics[width=1.55in]{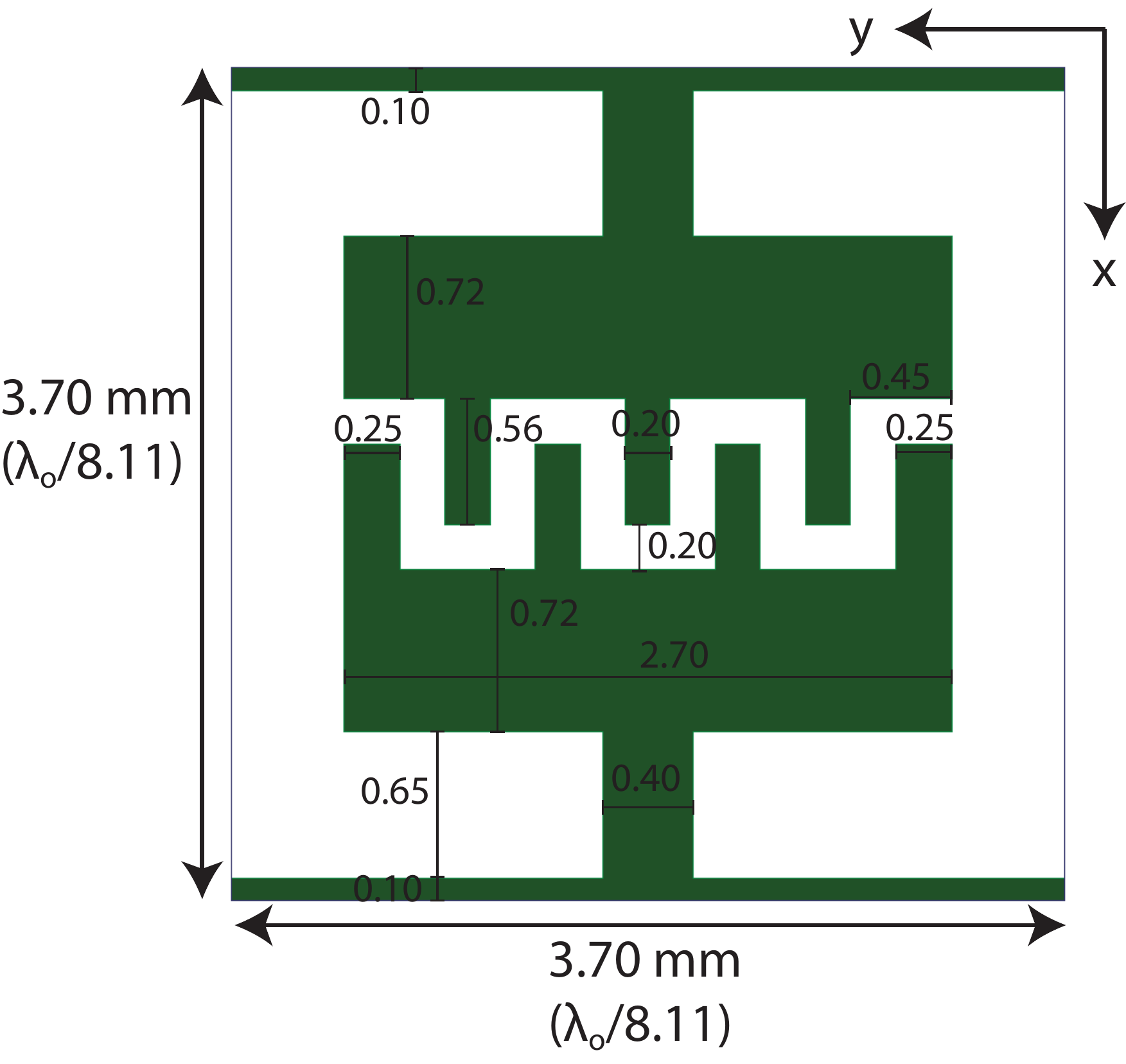}
    \label{fig:PolRotatorYs4}}
    \subfigure[]{
    \includegraphics[width=1.55in]{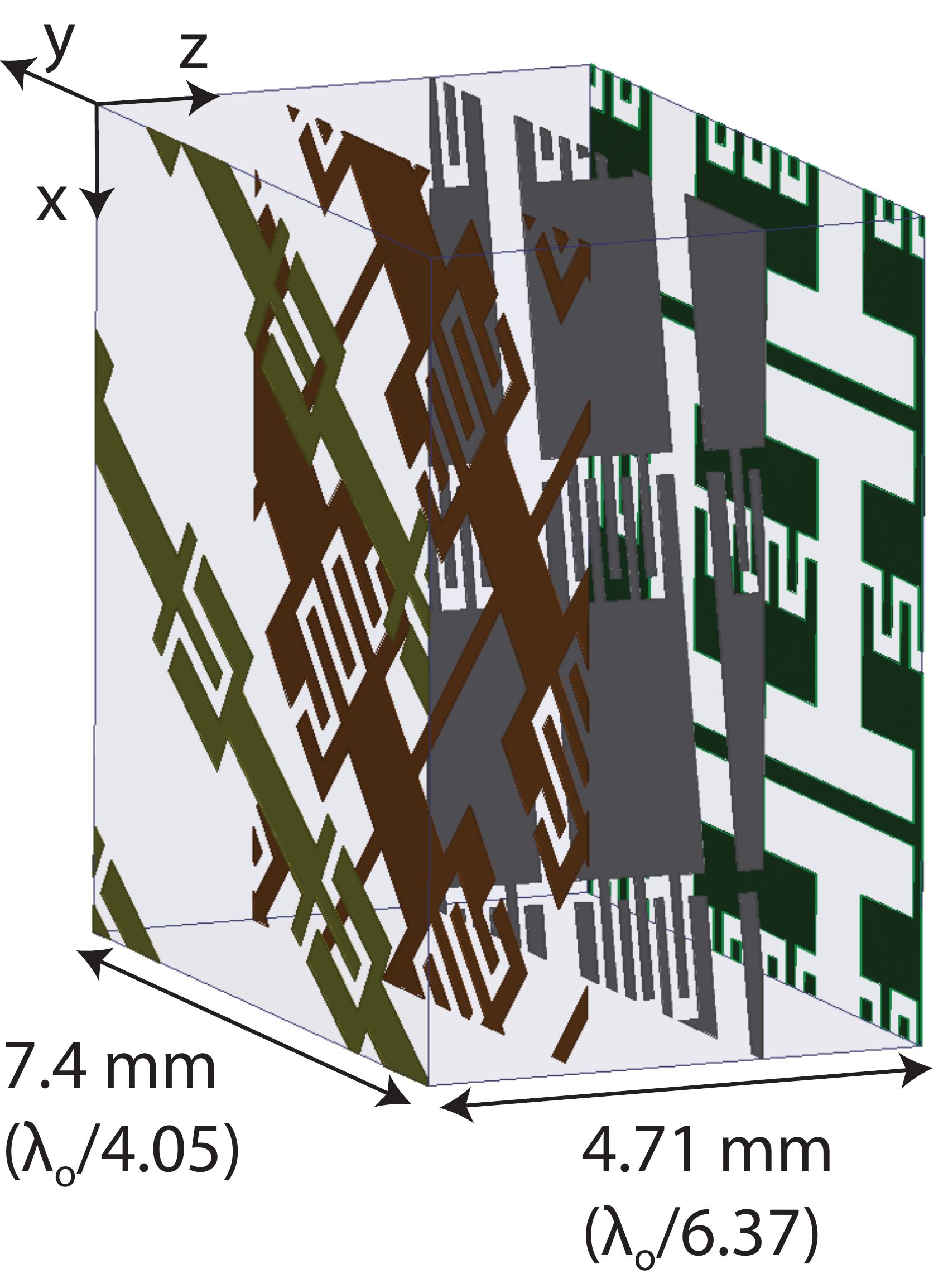}
    \label{fig:PolRotatorTotalDim}}
  \caption{Dimensions of the polarization rotator. \textbf{(a)-(d)} Dimensions (mm) of the first, second, third, and fourth sheets, respectively. \textbf{(e)} Perspective view of a section of the polarization rotator.} \label{fig:PolRotatorDimm}
\end{figure}

To realize the sheet admittances, copper is patterned on 1.52 mm thick, Rogers 4003 substrates $(\epsilon_r=3.55$, tan $\delta=0.0027)$. The patterns that realize the desired sheet admittances are shown in Fig. \ref{fig:PolRotatorDimm}. Each sheet has a periodicity of 3.7 mm x 3.7 mm ($\lambda_{\circ}/8.11$ x $\lambda_{\circ}/8.11$). It can be seen that the sheet admittance of the first sheet ($\textbf{Y}_{s1}$) has a small capacitance along $0.838 \hat{x}+0.545 \hat{y}$ and a larger capacitance along $0.547 \hat{x}-0.837 \hat{y}$, which are its principle axes. The large capacitance along the $0.547 \hat{x}-0.837 \hat{y}$ direction is realized with interdigitated capacitors. For the second sheet ($\textbf{Y}_{s2}$), the sheet admittance is inductive along the $(\hat{x}+\hat{y})/\sqrt{2}$ direction and capacitive along the $(\hat{x}-\hat{y})/\sqrt{2}$ direction. For the third sheet ($\textbf{Y}_{s3}$), the sheet admittance is capacitive along both principle axes, $0.204 \hat{x}+0.979 \hat{y}$ and $0.979 \hat{x}-0.205 \hat{y}$. The fourth sheet is similar to the second in that it is inductive along one principle axis $(\hat{y})$ and capacitive along the other $(\hat{x})$.

\begin{figure}[ht]
    \centering
    \subfigure[]{
    \includegraphics[width=1.55in]{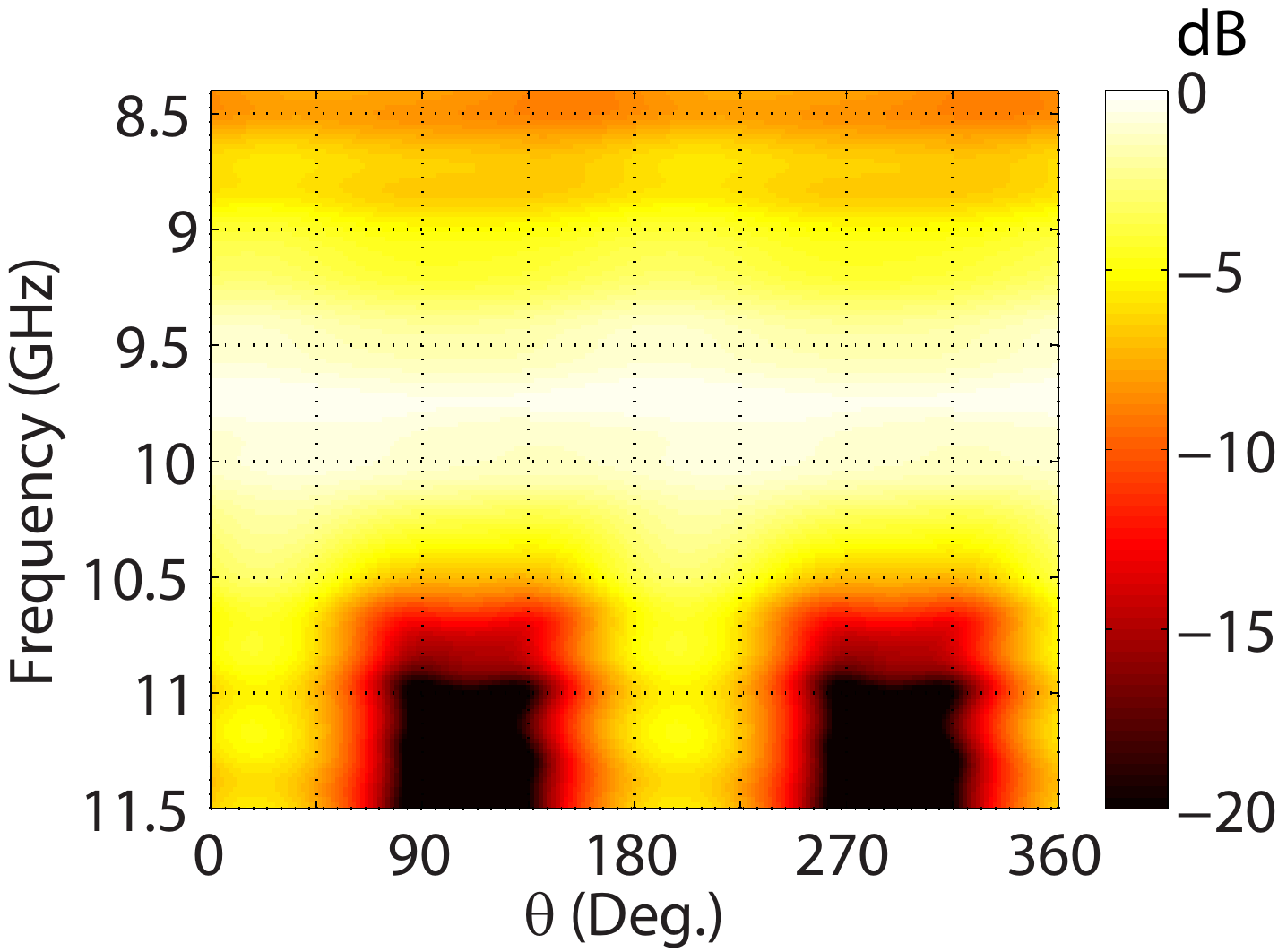}
    \label{fig:PolRotatorAngles_Myx}}
    \subfigure[]{
    \includegraphics[width=1.55in]{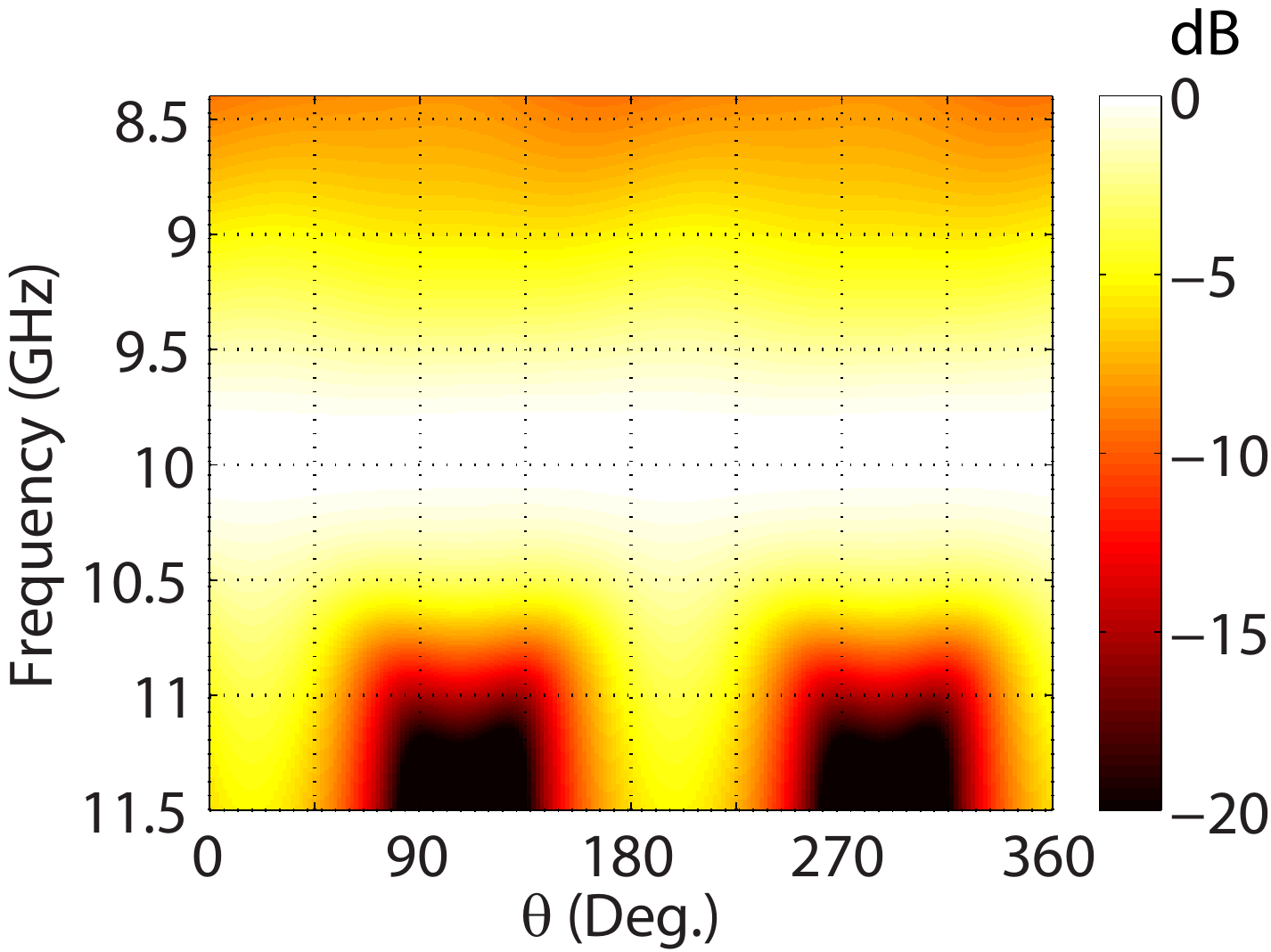}
    \label{fig:PolRotatorAngles_Syx}}
    \subfigure[]{
    \includegraphics[width=1.55in]{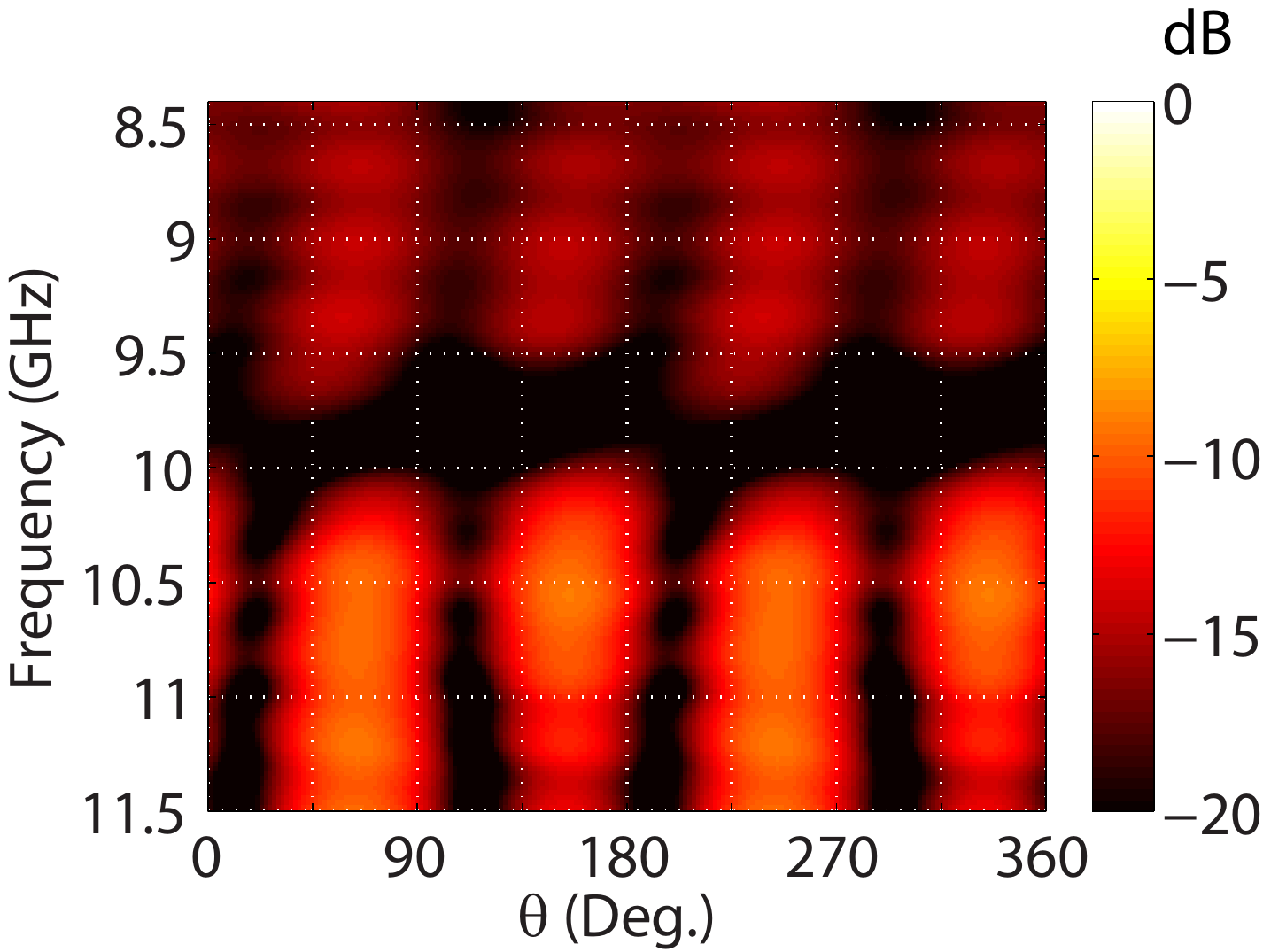}
    \label{fig:PolRotatorAngles_Mxx}}
    \subfigure[]{
    \includegraphics[width=1.55in]{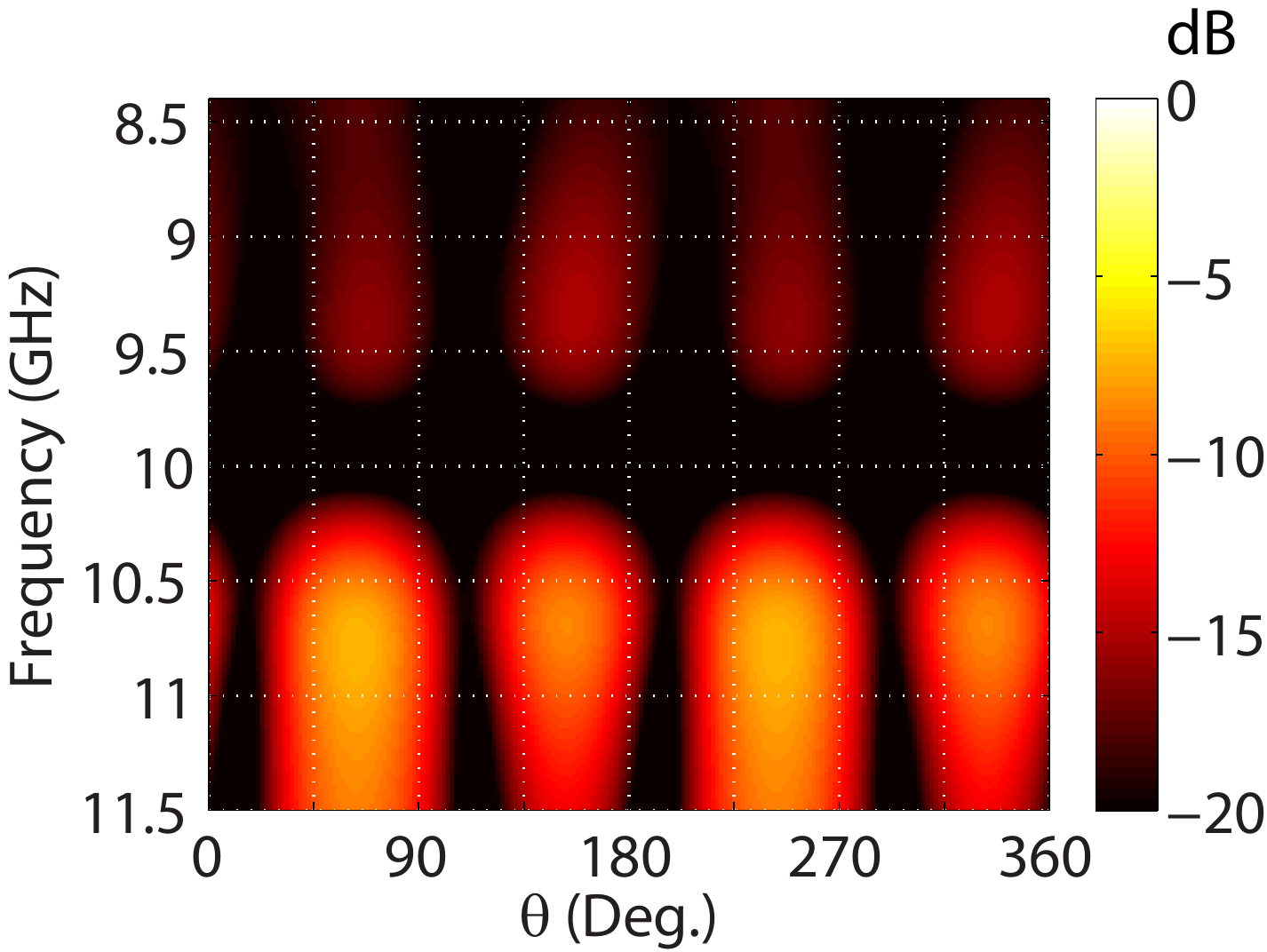}
    \label{fig:PolRotatorAngles_Sxx}}
      \subfigure[]{
    \includegraphics[width=1.55in]{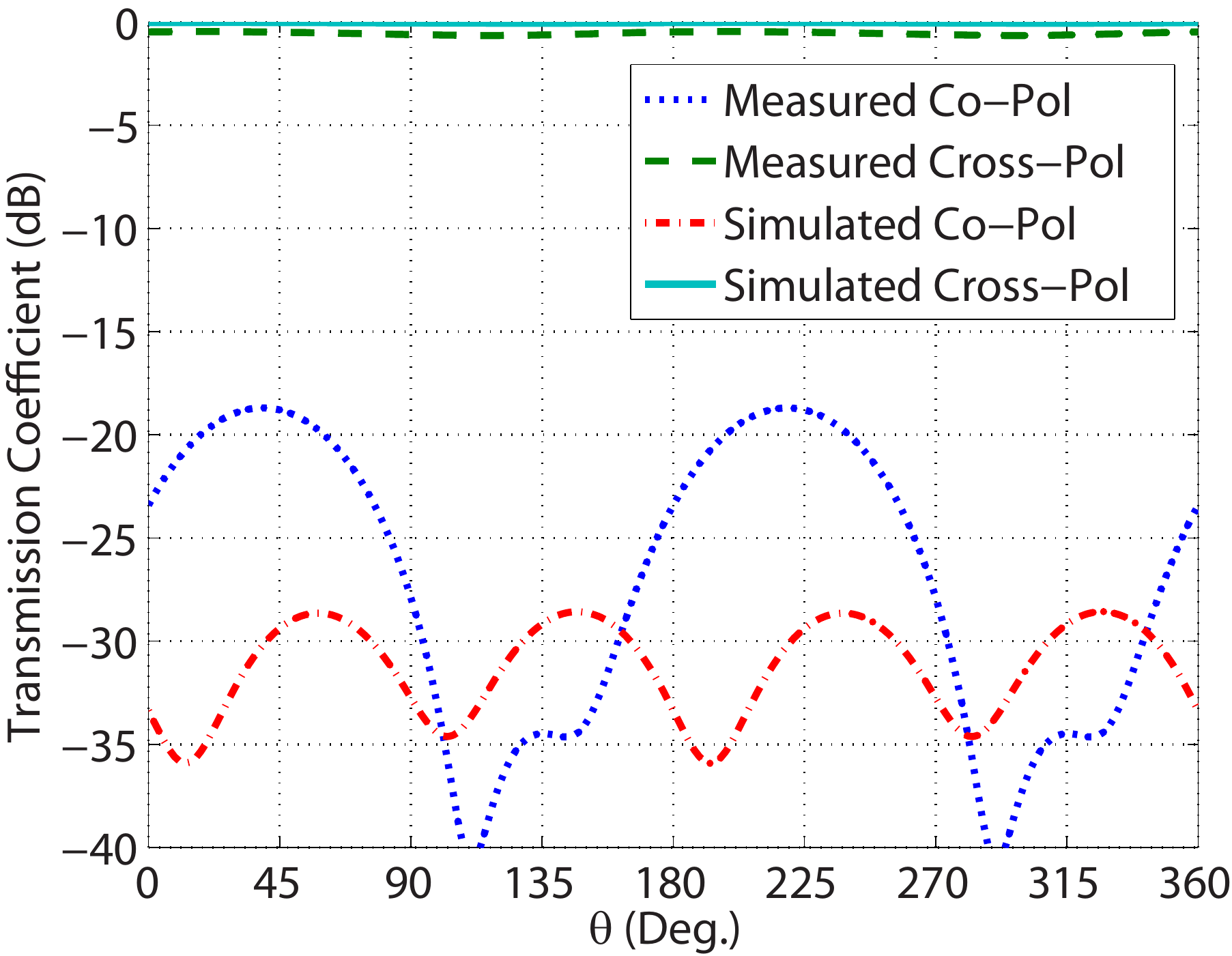}
    \label{fig:PolRotatorAngles_S21at9p9}}
          \subfigure[]{
    \includegraphics[width=1.55in]{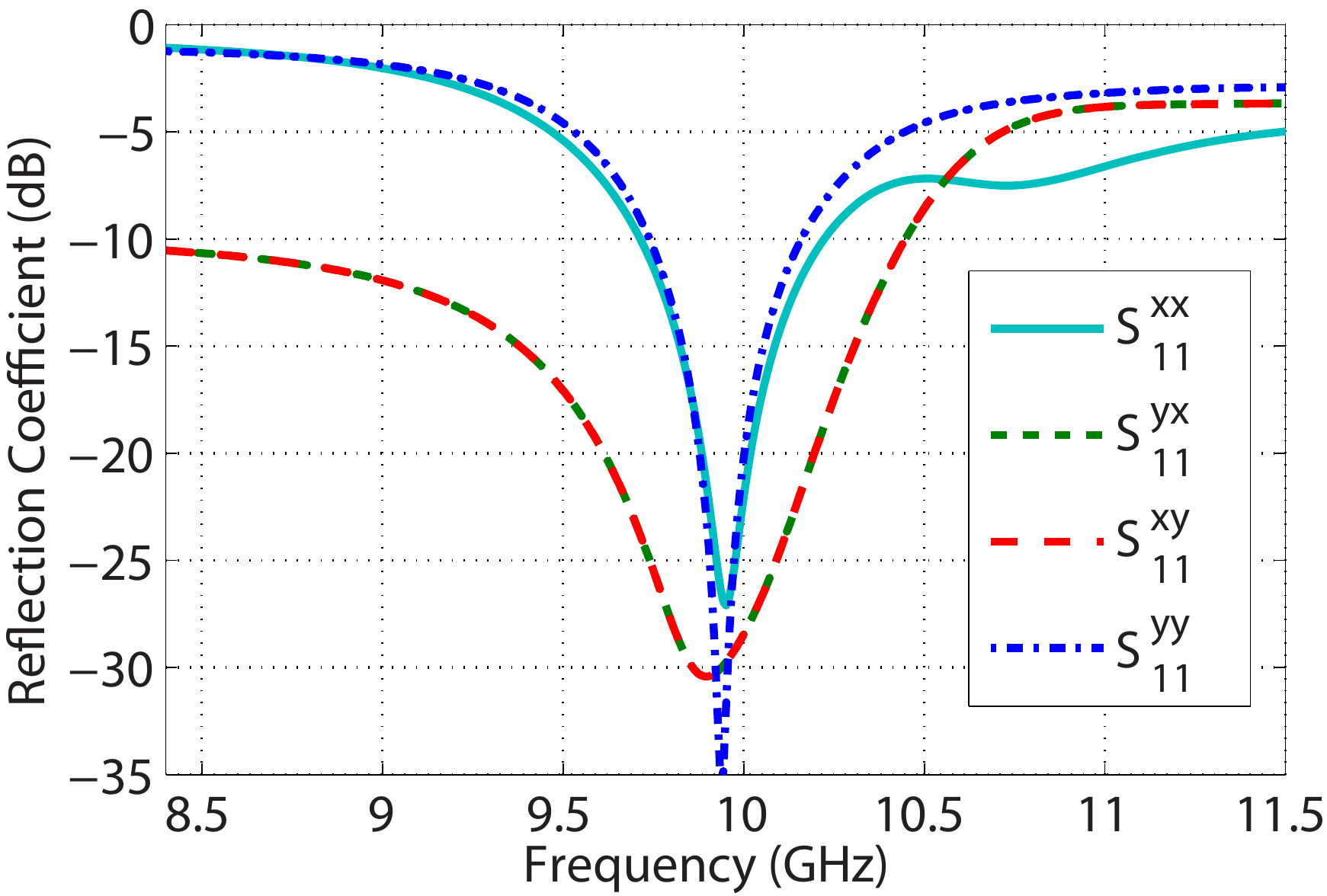}
    \label{fig:PolRotatorS11}}
  \caption{Additional performance metrics of the polarization rotator. \textbf{(a)} Measured cross-polarized transmission $(S_{21}^{yx})$ as a function of frequency and input linear polarization. The angle $\theta$ refers to the angle between the $x$ and $y$ axes of the input linear polarization. \textbf{(b)} Simulated cross-polarized transmission $(S_{21}^{yx})$ as a function of frequency and input linear polarization. \textbf{(c)} Measured co-polarized transmission $(S_{21}^{xx})$ as a function of frequency and input linear polarization. \textbf{(d)} Simulated co-polarized transmission $(S_{21}^{xx})$ as a function of frequency and input linear polarization. \textbf{(e)} Co-polarized and cross-polarized transmission as a function of the input linear polarization at the measured (9.78 GHz) and simulated (10.00 GHz) operating frequencies. \textbf{(f)} Simulated reflection coefficient.} \label{fig:PolRotatorAngles}
\end{figure}

Unlike the other metasurfaces presented here, the sheet admittances comprising the polarization rotator are not periodic with respect to a single coordinate system. Thus Fig. 2(a) of the main text and Fig. S\ref{fig:PolRotatorTotalDim} are not unit cells of the structure, but rather a section of the structure. Since the polarization rotator cannot be discretized into a single unit cell, its performance cannot be verified using a full wave electromagnetics simulator \cite{ma2005analysis}. Instead, the simulated responses shown in Fig. 2(c) of the main text and Fig. \ref{fig:PolRotatorAngles} are found by simulating each sheet admittance individually, and calculating the overall cascaded response analytically. Specifically, each sheet admittance is extracted using (\ref{eqn:SheetExtraction}) and their values are inserted into (\ref{eqn:ABCD2}) to find the overall $\textbf{ABCD}$ matrix of the entire structure. The S-parameters are then evaluated by inserting the $\textbf{ABCD}$ matrix into (\ref{eqn:SparamsFromABCD}). In contrast, all the other structures presented here (e.g. asymmetric circular polarizer, asymmetric linear polarizer, symmetric circular polarizer) can be discretized into unit cells, and their performances are simulated using Ansys HFSS.

It is important to note that (\ref{eqn:LambdaPolRotatorS}) dictates that the polarization rotator be isotropic at the operating frequency even though each sheet admittance is not. This can be verified by rotating the incident linear polarization by an angle $\theta$ about the $z$-axis and noting the co-polarized and cross-polarized transmission coefficients. As shown in Figs. \ref{fig:PolRotatorAngles} (a)-(e), the cross-polarized transmission is near 0 dB and co-polarized transmission is near or below -20 dB, for all angles $\theta$ around 10 GHz. A slight frequency shift of 2\% can be seen between the measured (Figs. \ref{fig:PolRotatorAngles} (a) and (c)) and simulated (Figs. \ref{fig:PolRotatorAngles} (b) and (d)) transmission coefficients. The simulated metasurface is well matched at the operating frequency, as shown in Fig. S\ref{fig:PolRotatorS11}.

\begin{figure}[ht]
    \centering
    \includegraphics[width=3.1in]{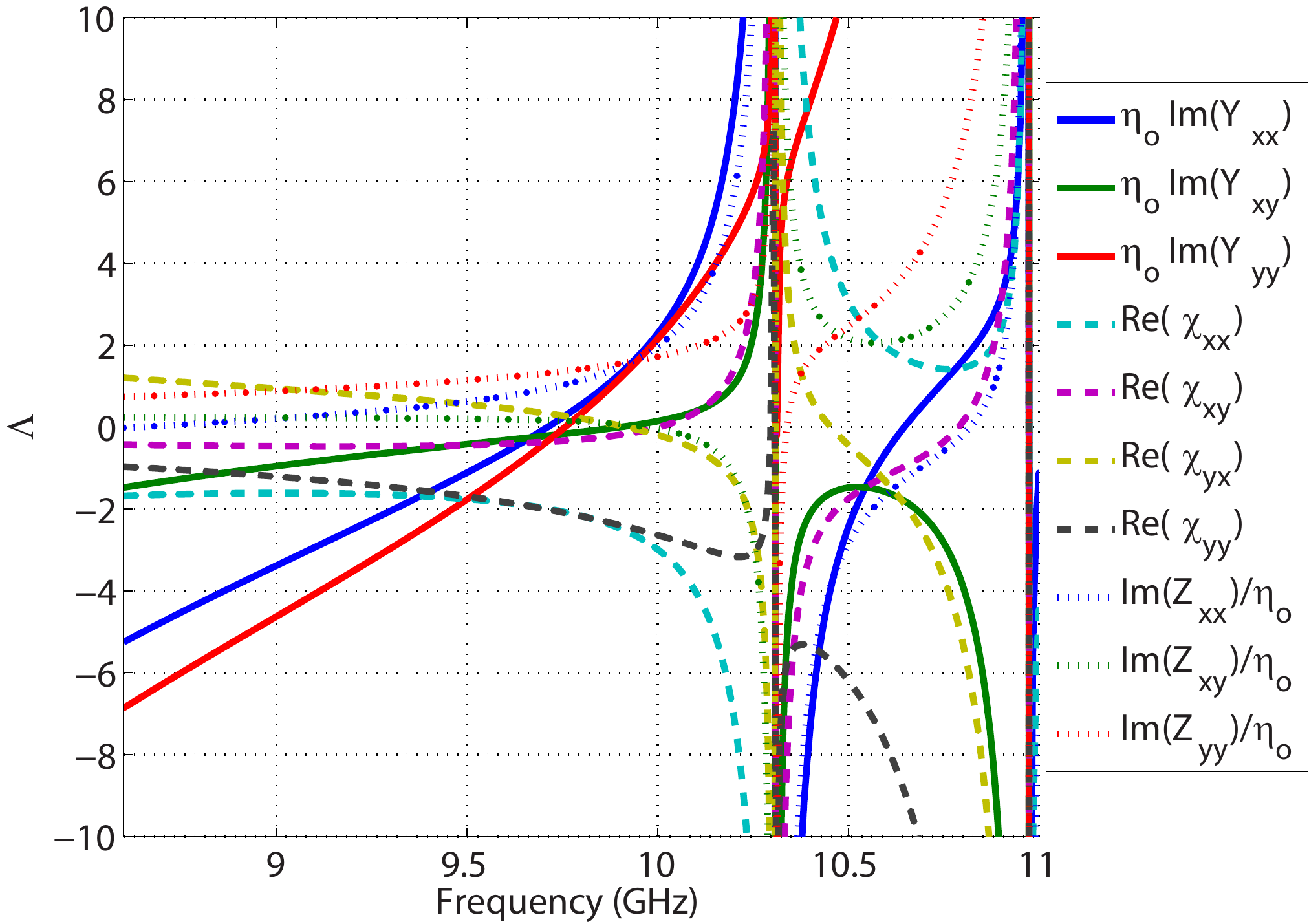}
  \caption{Constituent surface parameters of the simulated polarization rotator. The terms leading to loss (Re($\textbf{Y}$), Re($\textbf{Z}$), Im($\boldsymbol{\chi}$)) are low and are not plotted. All other terms that are not plotted can be inferred by noting that the structure is reciprocal.}\label{fig:PolRotatorLambda}
\end{figure}

The constituent surface parameters of the polarization rotator can be determined from simulation by inserting the S-parameters of the structure into (\ref{eqn:LambdafromSparam}). They are shown in Fig. \ref{fig:PolRotatorLambda}. It can be seen that at the operating frequency of 10 GHz, $Y_{xy}\eta_{\circ}=Z_{xy}/\eta_{\circ}=2$, and $\chi_{xy}=\chi_{yx}=-2.8$. In addition, all the off diagonal constituent parameters are zero (e.g. $Y_{xy}=Z_{xy}=\chi_{xy}=\chi_{yx}=0$). This is consistent with (\ref{eqn:LambdaPolRotatorS}) when $\phi=-45^{\circ}$. The terms leading to loss (Re($\textbf{Y}$), Re($\textbf{Z}$), Im($\boldsymbol{\chi}$)) are low and are not plotted. All other terms that are not plotted can be inferred by noting that the structure is reciprocal.

It should be noted that the polarization rotator is the only structure presented here that utilizes four patterned sheets. Initially, three sheets were used to realize a polarization rotator. However, the simulated bandwidth was narrow (0.6\%) and the loss was high (S$_{21}$=-1.7 dB). This led to a structure that was extremely sensitive to fabrication tolerances. Therefore, a four layer structure was considered since this structure has additional degrees of freedom that can be varied to increase bandwidth and reduce loss.

\section{Asymmetric circular polarizer}
Additional control over the constituent surface parameters is demonstrated with a metasurface that provides asymmetric transmission for circularly polarized waves at millimeter-wave frequencies. This effect exhibits some similarity to Faraday rotation, but requires no magnetic field or nonreciprocal materials \cite{fedotov2006asymmetric}. In addition, this metasurface acts as a circular polarizer since it transmits circular polarization of one handedness and reflects the other. Demonstrating this effect at millimeter-wave frequencies is particularly useful. Millimeter wavelengths are long enough to easily propagate through visibly opaque media, while short enough to realize large operating bandwidths and millimeter resolution \cite{wells2009faster}. This enables high resolution radar and imaging systems, as well as high bandwidth communication. In particular, 77 GHz is relevant to automotive radar systems \cite{schneider2005automotive}.

To date, the most common method to realize asymmetric circular transmission is to utilize a purely electric response by printing two-dimensional chiral patterns on a single sheet \cite{fedotov2006asymmetric}. However, the asymmetric response is significant only when the eigenvectors of the sheet admittance are complex, which requires high loss. Therefore, the efficiency of these structures is fundamentally limited. The asymmetric response is often defined as the difference between the transmittance of a given handedness of circular polarization, propagating in the $+z$ and $-z$ directions. It is typically low for single sheets geometries (not exceeding 0.25) \cite{wu2013giant}. Alternatively, it was recently shown that a bi-layered metasurface realized by cascading two-dimensional chiral patterns can achieve a larger asymmetric response of 0.6 \cite{wu2013giant}. However, the design procedure and description of the physics are vague, and the transmittance and asymmetric response are still too low for most applications. In contrast, the metasurface presented here achieves a near-optimal asymmetric response (0.99) at the design frequency, and a thorough analysis and systematic design procedure are provided.

As mentioned in the main text (see Eqs. (10) and (11)) the asymmetric circular polarizer that is considered has the following transmission coefficient,
\begin{equation}\label{eqn:asymmetriccircularS}
\textbf{S}_{21}=\frac{e^{j\phi}}{2}\begin{pmatrix}1 & j\\j&-1\end{pmatrix}
\end{equation}
This device has constituent surface parameters given by,

\small
\begin{equation}\label{eqn:LambdaAsymmetricCircularTransmissionS}
\boldsymbol{\Lambda}= \begin{pmatrix}
      \frac{-2j\textrm{ tan}(\phi/2)}{\eta_{\circ}} & 0 & 0 & 0 \\
       0 & \frac{2j\textrm{ cot}(\phi/2)}{\eta_{\circ}} & 0 & 0 \\
       0 & 0 & -2j\eta_{\circ}\textrm{ tan}(\phi) & 2j\eta_{\circ}\textrm{ sec}(\phi) \\
       0 & 0 & 2j\eta_{\circ}\textrm{ sec}(\phi) & -2j\eta_{\circ}\textrm{ tan}(\phi)
   \end{pmatrix}
\end{equation}
\normalsize

The asymmetric circular polarizer is designed by setting $\phi=175^{\circ}$, $\beta d=2\pi/6.37$, $\eta_d=\eta_{\circ}/1.483$, and $\eta_1=\eta_2=\eta_{\circ}$. The necessary sheet admittances are then numerically solved for, $\textbf{Y}_{s1}=\frac{j}{\eta_{\circ}}\begin{pmatrix}1.01 & -1.00\\-1.00 & 1.01\end{pmatrix}$, $\textbf{Y}_{s2}=\frac{j}{\eta_{\circ}}\begin{pmatrix}2.19 & 0\\0 & -200\end{pmatrix}$, and $\textbf{Y}_{s3}=\frac{j}{\eta_{\circ}}\begin{pmatrix}1.01 & -1.00\\-1.00 & 1.01\end{pmatrix}$.

\begin{figure}[ht]
    \centering
    \subfigure[]{
    \includegraphics[width=1.55in]{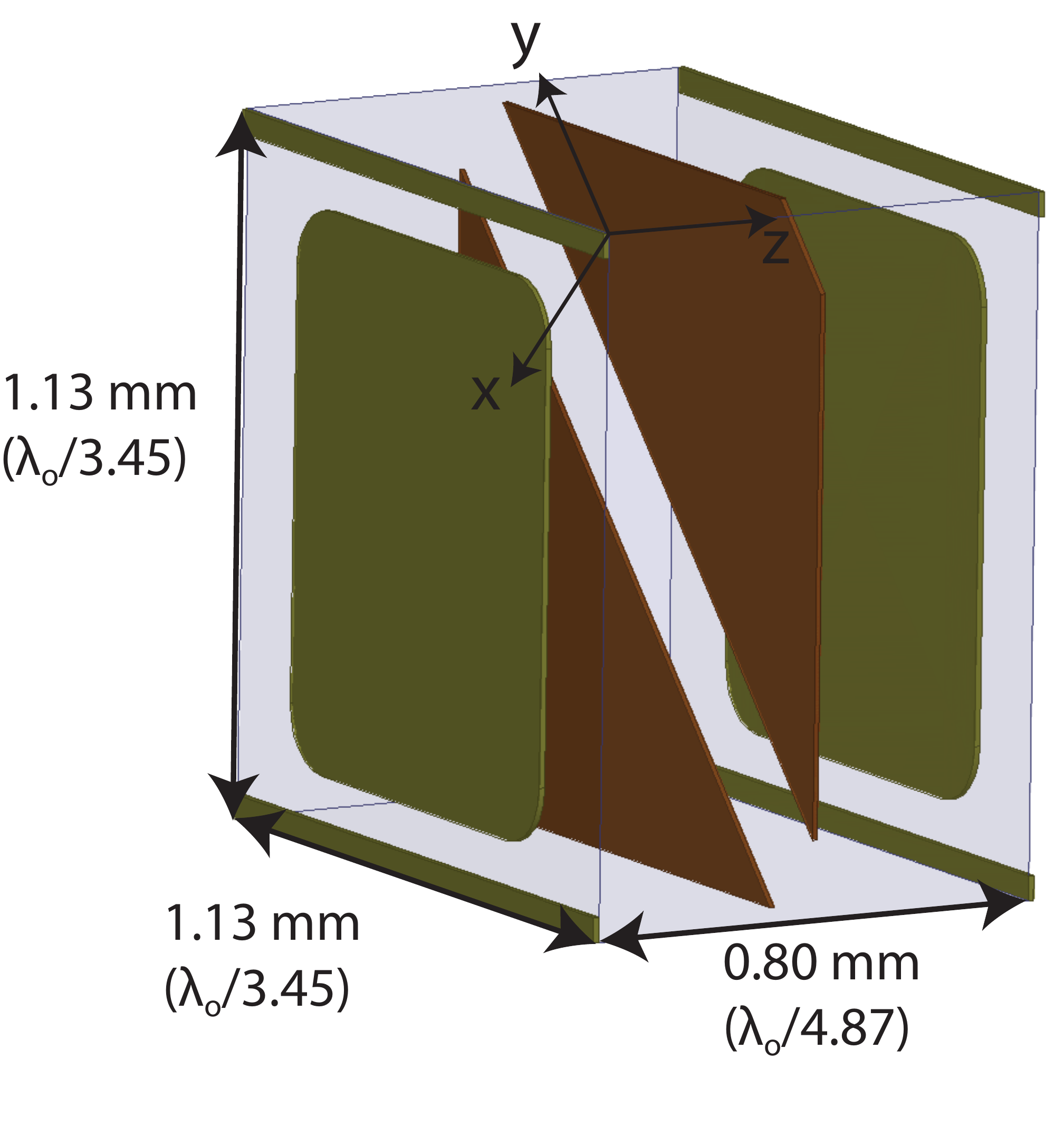}
    \label{fig:AsymmetricCircularTotalDim}}
    \subfigure[]{
    \includegraphics[width=1.55in]{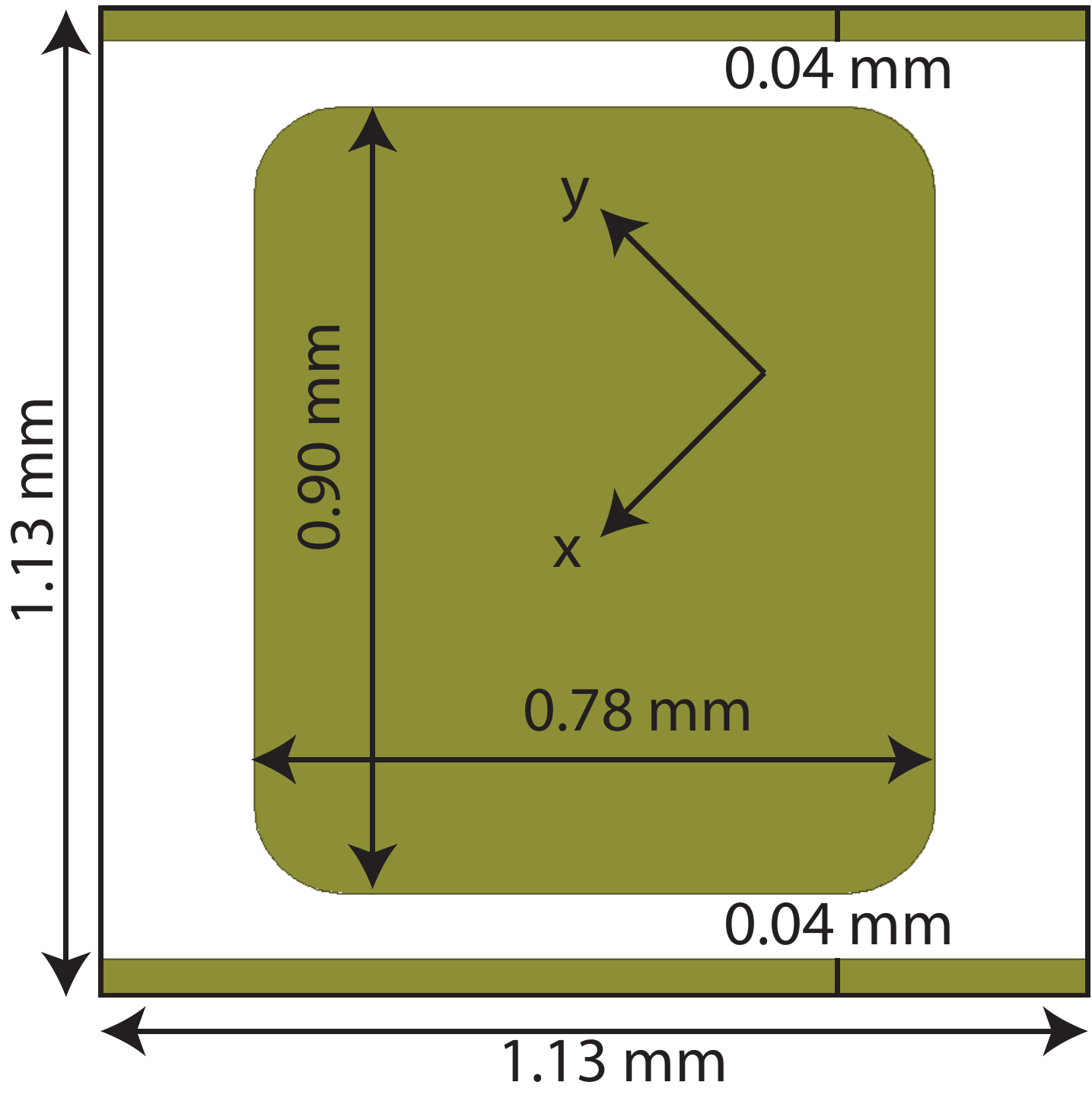}
    \label{fig:AsymmetricCircularYs1}}
     \subfigure[]{
    \includegraphics[width=1.55in]{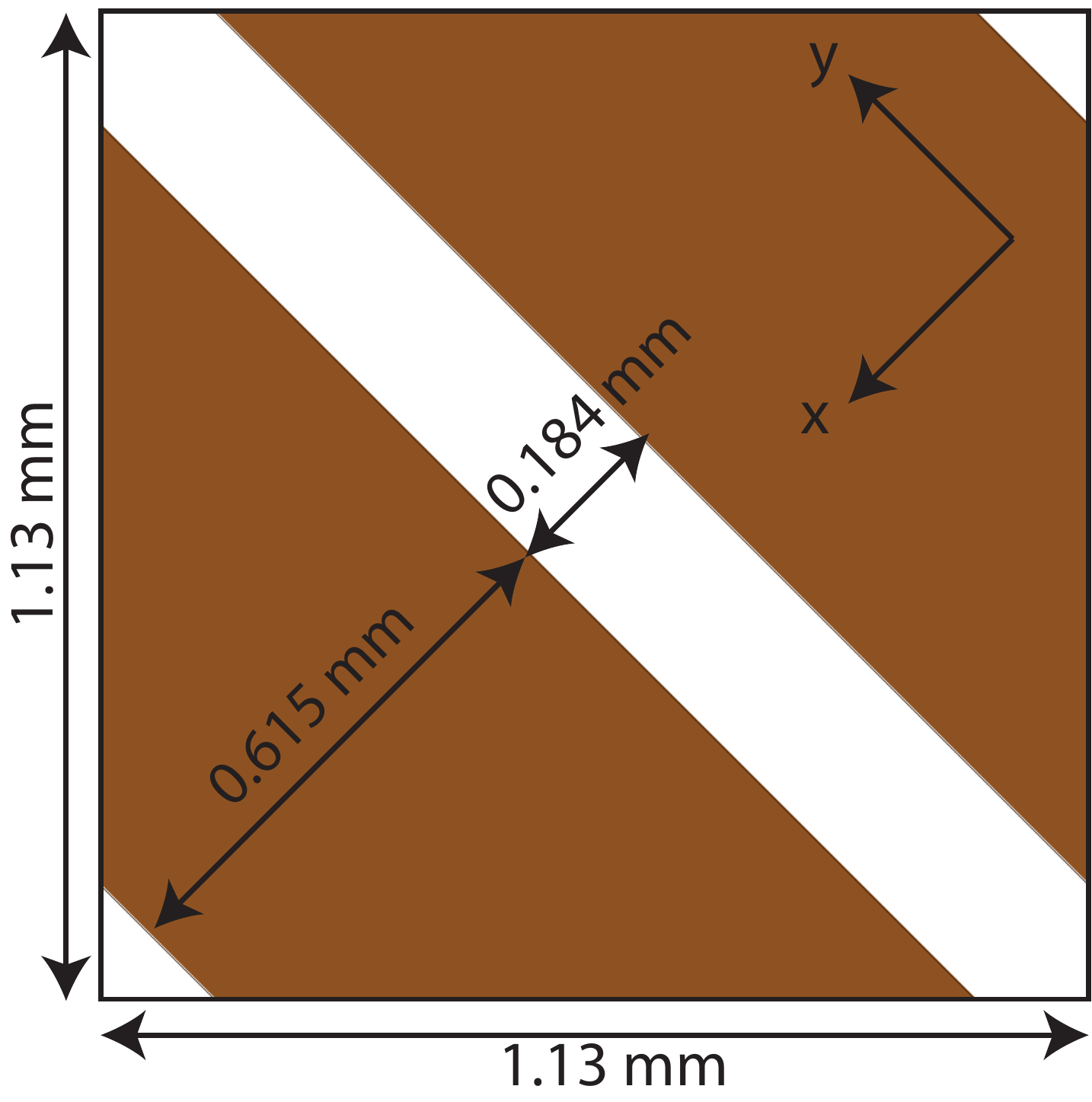}
    \label{fig:AsymmetricCircularYs2}}
  \caption{Dimensions of the asymmetric circular polarizer. \textbf{(a)} Perspective view with the overall thickness and coordinate system shown. \textbf{(b)} Dimensions of the first and third sheets. \textbf{(c)} Dimensions of the second sheet.} \label{fig:AsymmetricCircularDimm}
\end{figure}

To realize the sheet admittances, copper is patterned on 380 $\mu$m thick, Rogers 5880 Duroid substrates $(\epsilon_r=2.2$, tan $\delta=0.0009)$. The detailed patterns of the sheets are shown in Fig. \ref{fig:AsymmetricCircularDimm}. The structure is reflection symmetric along the $z=0$ plane $(\textbf{Y}_{s1}=\textbf{Y}_{s3})$, which causes all chiral terms to reduce to zero \cite{menzel2010advanced}. All three sheets are capacitive along one principle axis and inductive along the other. It should be noted that although the second sheet looks very similar to a wire grid polarizer, there are some differences. Wire grid polarizers are generally designed to minimize the inductance along the $y$ axis, while also minimizing the capacitance along the $x$ axis to provide high reflection and transmission, respectively. Here, the cell size and patterned copper are chosen to realize a specific capacitance $(Y_{s2}^{xx}=2.19j/\eta_{\circ})$ along the $x$ axis in order to achieve an optimal performance at the design frequency.

\begin{figure}[ht]
    \centering
    \includegraphics[width=3.1in]{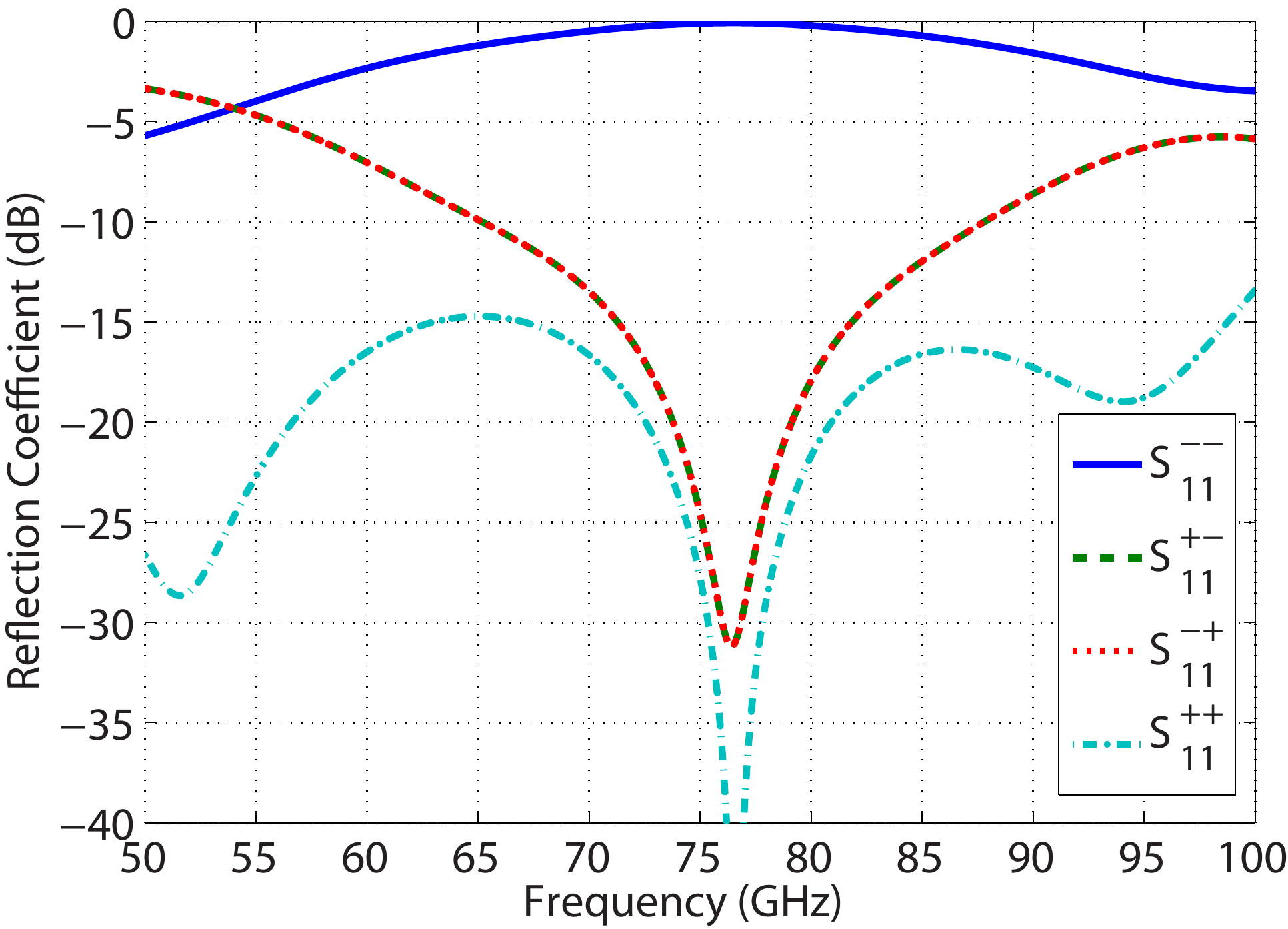}
  \caption{Reflection coefficient of the asymmetric circular polarizer.}\label{fig:AssymetricCircularS11}
\end{figure}

It should be noted that the constituent surface parameters of the ideal asymmetric circular polarizer (Eq. (\ref{eqn:LambdaAsymmetricCircularTransmissionS})) are a function of the stipulated reflection coefficients in addition to the transmission coefficient. Eq. (\ref{eqn:LambdaAsymmetricCircularTransmissionS}) assumes the reflection coefficients are equal to,
\begin{equation}\label{eqn:asymmetricCicularS11}
\textbf{S}_{11}=\textbf{S}_{22}=\frac{e^{j\phi}}{2}\begin{pmatrix}1&-j\\-j&-1\end{pmatrix}
\end{equation}
When left-handed-circular is incident in the $+z$ direction, all of the power is reflected to left-handed-circular \cite{Roy1996Reciprocal}. Similarly, when right-handed-circular is incident in the $-z$ direction, all of the power is reflected to right-handed-circular, as shown in Fig. \ref{fig:AssymetricCircularS11}. It should be emphasized that, in general, the stipulated reflection coefficients are not unique. In all the examples presented here, the reflection coefficients are assumed to be identical to those of the fabricated structure.

\begin{figure}[ht]
    \centering
    \includegraphics[width=3.1in]{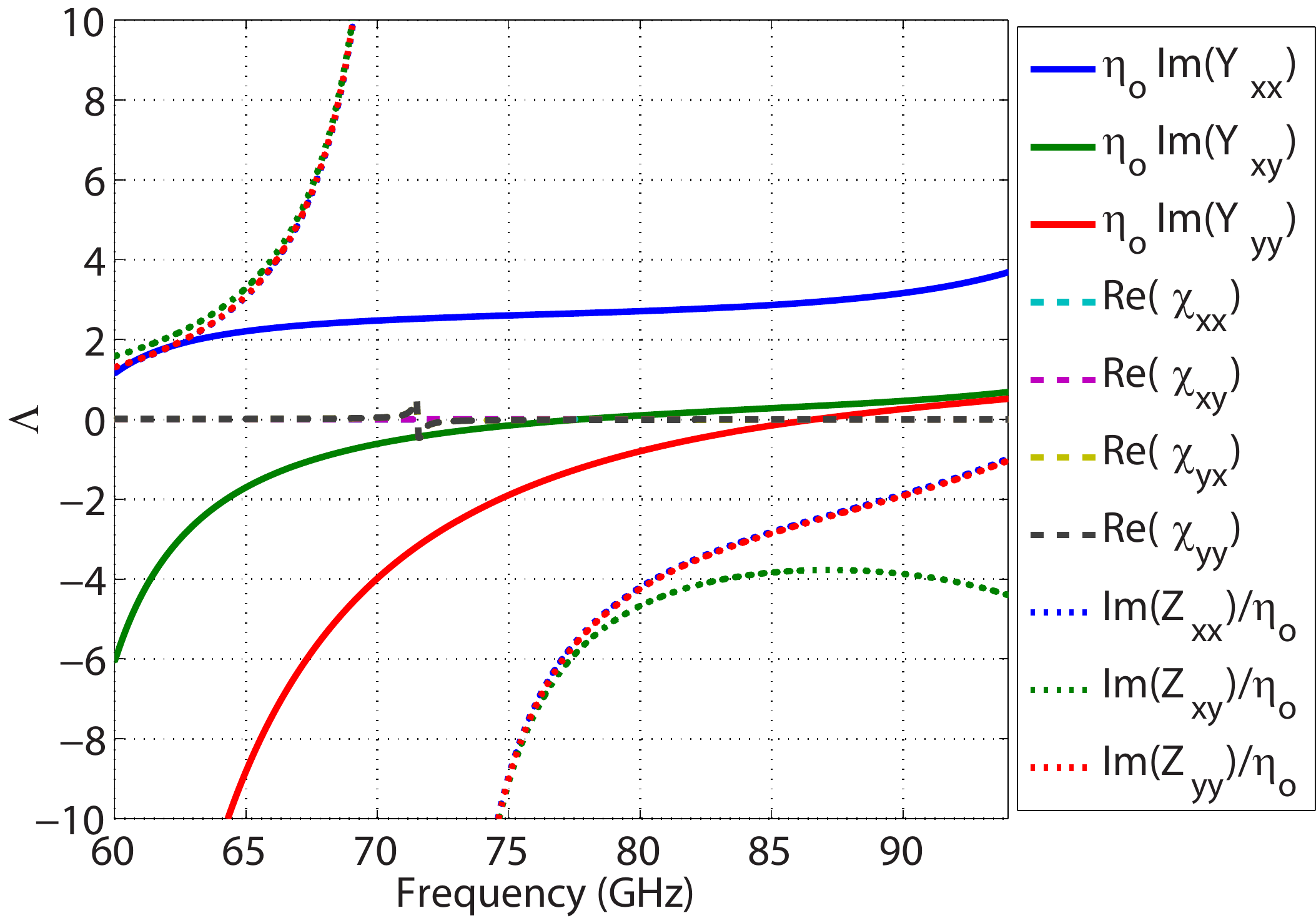}
  \caption{Constituent surface parameters of the simulated asymmetric circular polarizer. The terms leading to loss (Re($\textbf{Y}$), Re($\textbf{Z}$), Im($\boldsymbol{\chi}$)) are low and are not plotted. All other terms that are not plotted can be inferred by noting that the structure is reciprocal.}\label{fig:AsymmetricCircularLambda}
\end{figure}

The constituent parameters of the asymmetric circular polarizer are shown in Fig. \ref{fig:AsymmetricCircularLambda}. It can be seen that all chiral terms are zero since the structure is reflection symmetric about the $z=0$ plane. At the design frequency of 77 GHz, the principle axes of the electric susceptibility are aligned along the $\hat{x}$ and $\hat{y}$ axes since $Y_{xy}=0$, and the principle axes of the magnetic susceptibility are aligned along $(\hat{x}+\hat{y})/\sqrt{2}$ and $(\hat{x}-\hat{y})/\sqrt{2}$, since $Z_{xx}=Z_{yy}$ and $Z_{xy}\ne0$. In other words, the principle axes of the electric and magnetic response are rotated by $45^{\circ}$ from each other to achieve an optimal performance. It should be noted that, some degree of asymmetric transmission is present whenever the principle axes of the electric and magnetic susceptibilities are not aligned. This can be shown using (\ref{eqn:SparamfromLambda}).

\section{Asymmetric linear polarizer}
Asymmetric transmission for linearly polarized waves requires geometries that do not exhibit any mirror or rotational symmetry \cite{menzel2010advanced}. These metasurfaces can be used to increase the polarization diversity of microwave and optical devices. Previously, devices exhibiting asymmetric linear transmission were realized with two and three layered chiral meta-atoms \cite{menzel2010asymmetric,mutlu2012diodelike} or asymmetric helical geometries \cite{zhang2013Interference}. Although a near-optimal performance was achieved, the response was narrowband \cite{mutlu2012diodelike}. A more straightforward design procedure based on Fabry-Perot resonances was reported in \cite{grady2013terahertz}. This approach led to an enhanced bandwidth, but at the expense of an increased electrical thickness.

Here, an analytic approach is used to systematically design arbitrarily thin metasurfaces. The asymmetric linear polarizer considered here has the transmission coefficient given by,
\begin{equation}\label{eqn:asymmetricLinear}
\textbf{S}_{21}=e^{j\phi}\begin{pmatrix}0 & 0\\1&0\end{pmatrix}
\end{equation}
When an $x$-polarized plane wave is incident from Region 1, all the power is transmitted into the $y$-component of Region 2. However, if an $x$-polarized plane wave is incident from Region 2, all the power is reflected. Hence, the structure exhibits asymmetric transmission for linear polarization. Assuming the reflection coefficients are equal to,
\begin{equation}\label{eqn:asymmetricLinearS11}
\textbf{S}_{11}=e^{j\phi}\begin{pmatrix}0&0\\0&-1\end{pmatrix},\quad
\textbf{S}_{22}=e^{j\phi}\begin{pmatrix}-1&0\\0&0\end{pmatrix}
\end{equation}
this metasurface has the constituent surface parameters given by,
\begin{equation}\label{eqn:asymmetricLinearLambda}
 \boldsymbol{\Lambda}=\begin{pmatrix}
      4j\eta_{\circ}^{-1}\textrm{ cot}(\phi) & -4j\eta_{\circ}^{-1}\textrm{ csc}(\phi) & 0 & -2 \\
       -4j\eta_{\circ}^{-1}\textrm{ csc}(\phi) & 4j\eta_{\circ}^{-1}\textrm{ cot}(\phi) & -2 & 0 \\
       0 & 2 & 0 & 0 \\
       2 & 0 & 0 & 0
   \end{pmatrix}
\end{equation}

\begin{figure}[ht]
      \centering
    \subfigure[]{
    \includegraphics[width=2in]{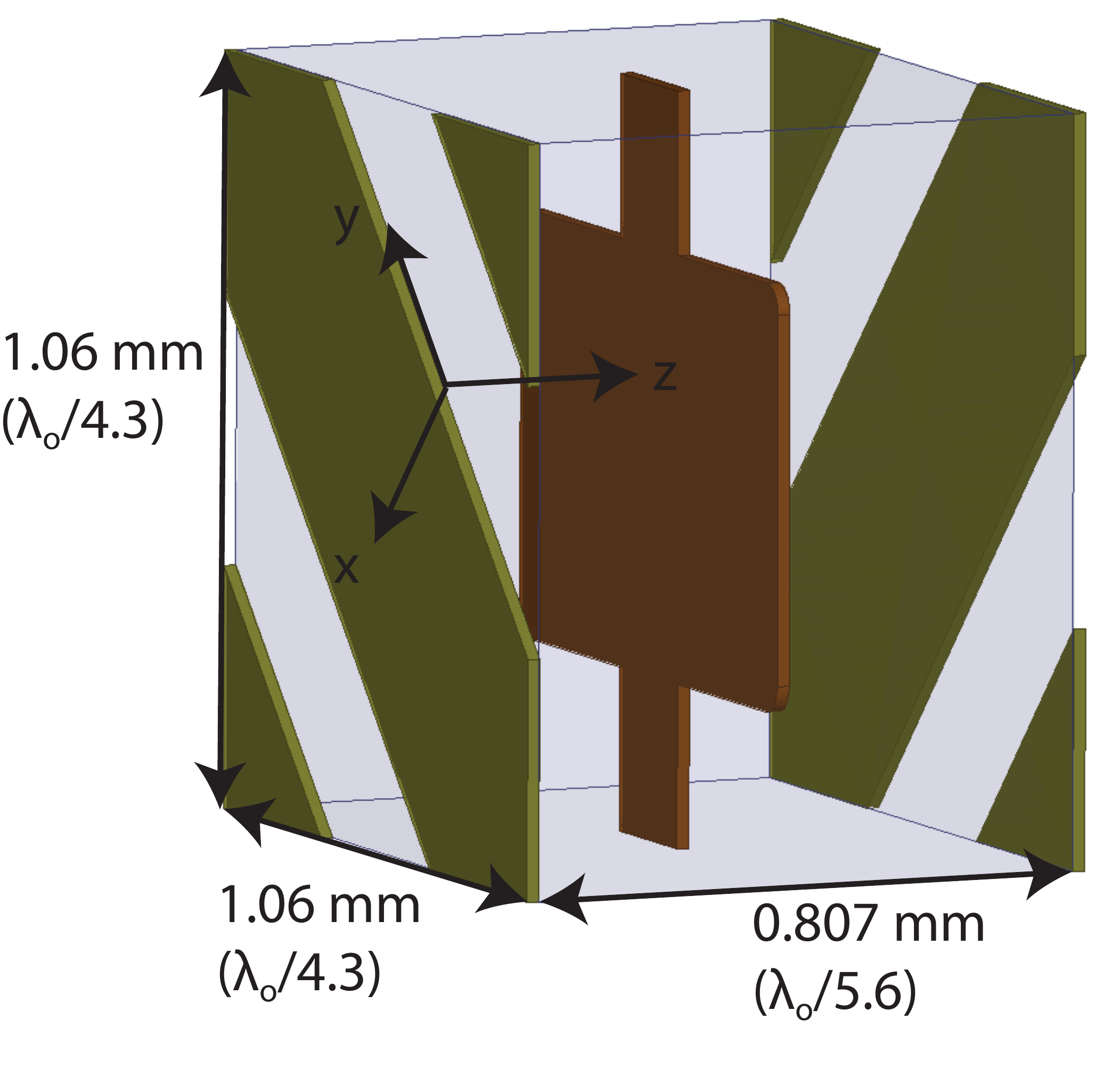}
    \label{fig:AsymmetricLinearPic}}
    \subfigure[]{
    \includegraphics[width=2in]{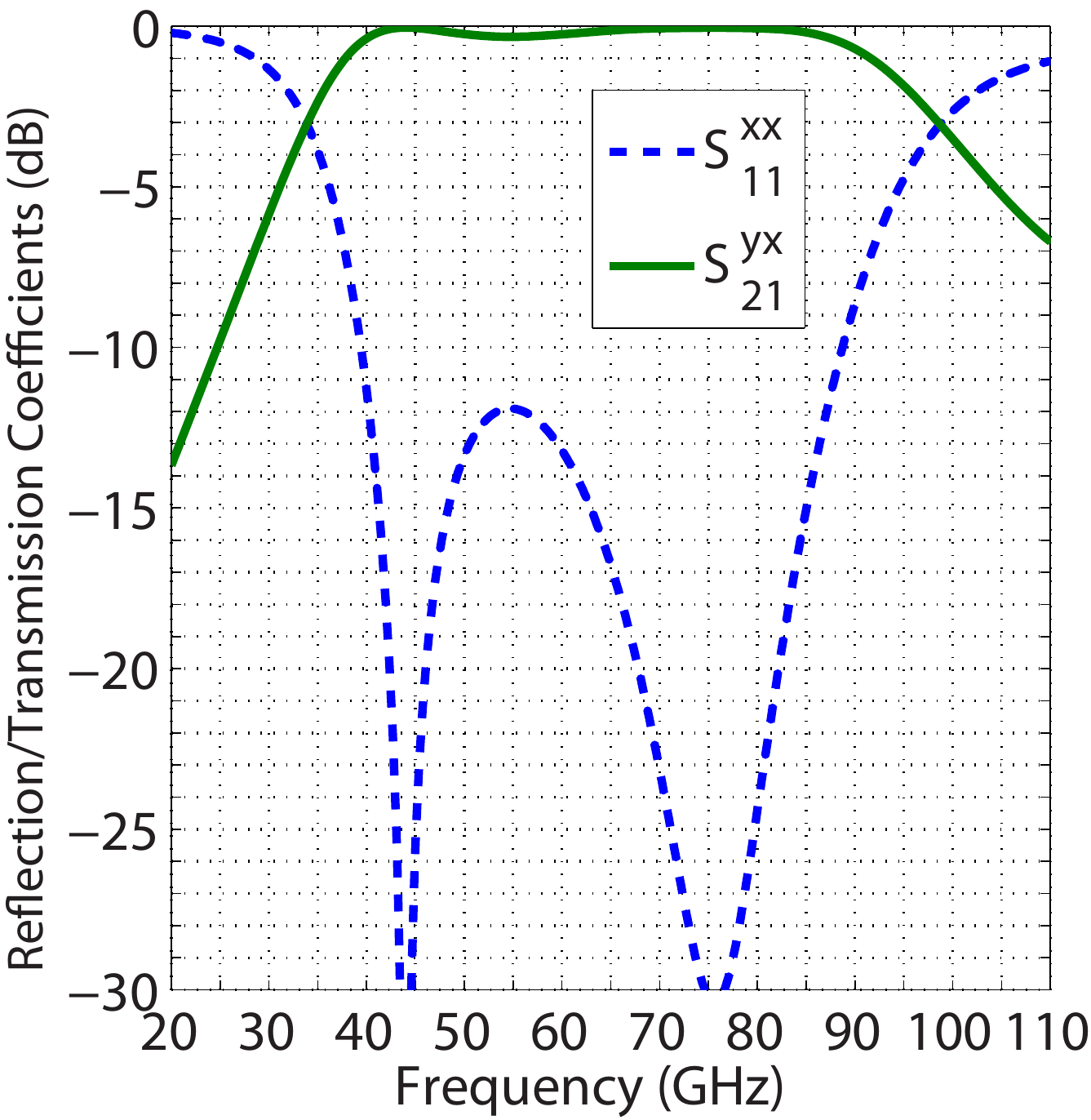}
    \label{fig:AsymmetricLinearPerformance}}
  \caption{Metasurface exhibiting asymmetric linear transmission. \textbf{(a)} Schematic of the unit cell. \textbf{(b)} Simulated co-polarized reflection coefficient $(S_{11}^{xx})$ and cross-polarized transmission coefficient $(S_{21}^{yx})$ for an incident plane wave traveling in the +$z$ direction. $S_{11}^{yy}$ is greater than -0.01 dB, and all other S-parameters are less than -30 dB over the entire frequency range, and are not shown.}\label{fig:AsymmetricLinear}
\end{figure}

It can be seen that both an anisotropic electric susceptibility and an anisotropic chiral response are required. Again, the necessary sheet admittances are solved by inserting (\ref{eqn:asymmetricLinear}) and (\ref{eqn:asymmetricLinearS11}) into (\ref{eqn:SparamsFromABCD}), and combining the result with (\ref{eqn:ABCD3layer}). Upon setting $\phi=135^{\circ}$, $\beta d=2\pi/6.37$, $\eta_d=\eta_{\circ}/1.483$, and $\eta_1=\eta_2=\eta_{\circ}$, the necessary sheet admittances are given by $\textbf{Y}_{s1}=\frac{j}{\eta_{\circ}}\begin{pmatrix}0.88 & 0\\0 & -77.0\end{pmatrix}$, $\textbf{Y}_{s2}=\frac{j}{\eta_{\circ}}\begin{pmatrix}-0.70 & 4.15\\4.15 & -0.70\end{pmatrix}$, and $\textbf{Y}_{s3}=\frac{j}{\eta_{\circ}}\begin{pmatrix}-77.0 & 0\\0 & 0.88\end{pmatrix}$. A unit cell of this metasurface is shown in Fig. S\ref{fig:AsymmetricLinearPic}. The simulated cross-polarized and co-polarized transmission coefficients are shown in Fig. S\ref{fig:AsymmetricLinearPerformance}. $S_{11}^{yy}$ is greater than -0.01 dB, and all other S-parameters are less than -30 dB over the entire frequency range, and are not shown. Although the structure is designed for an operating frequency of 77 GHz, the performance is quite broadband. A 1 dB transmission bandwidth of 2.43:1 for the desired polarization is achieved. The rejection of the unwanted polarization is greater than 30 dB in this band.

\begin{figure}[ht]
    \centering
    \subfigure[]{
    \includegraphics[width=1.55in]{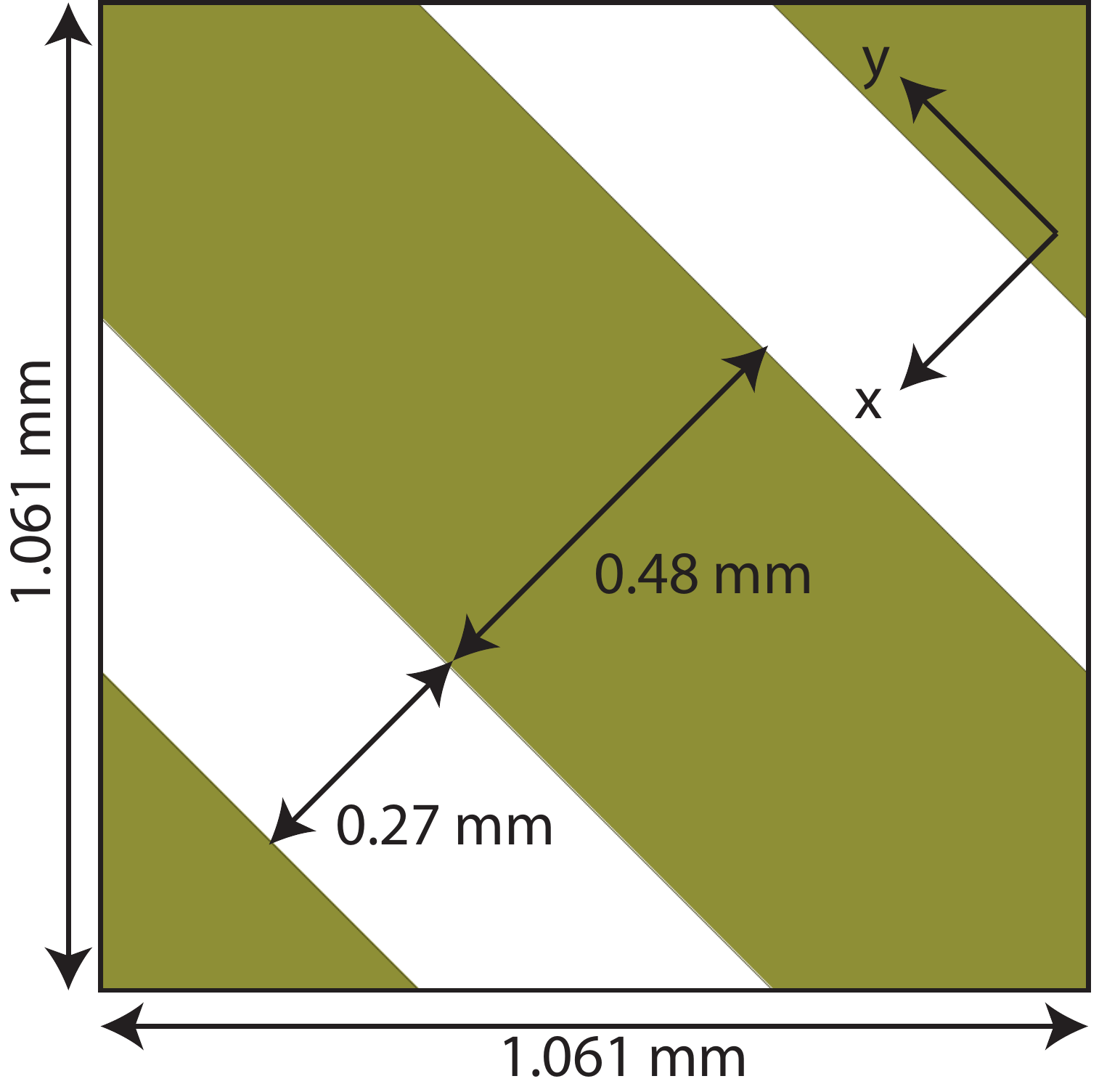}
    \label{fig:AsymmetricLinearYs1}}
     \subfigure[]{
    \includegraphics[width=1.55in]{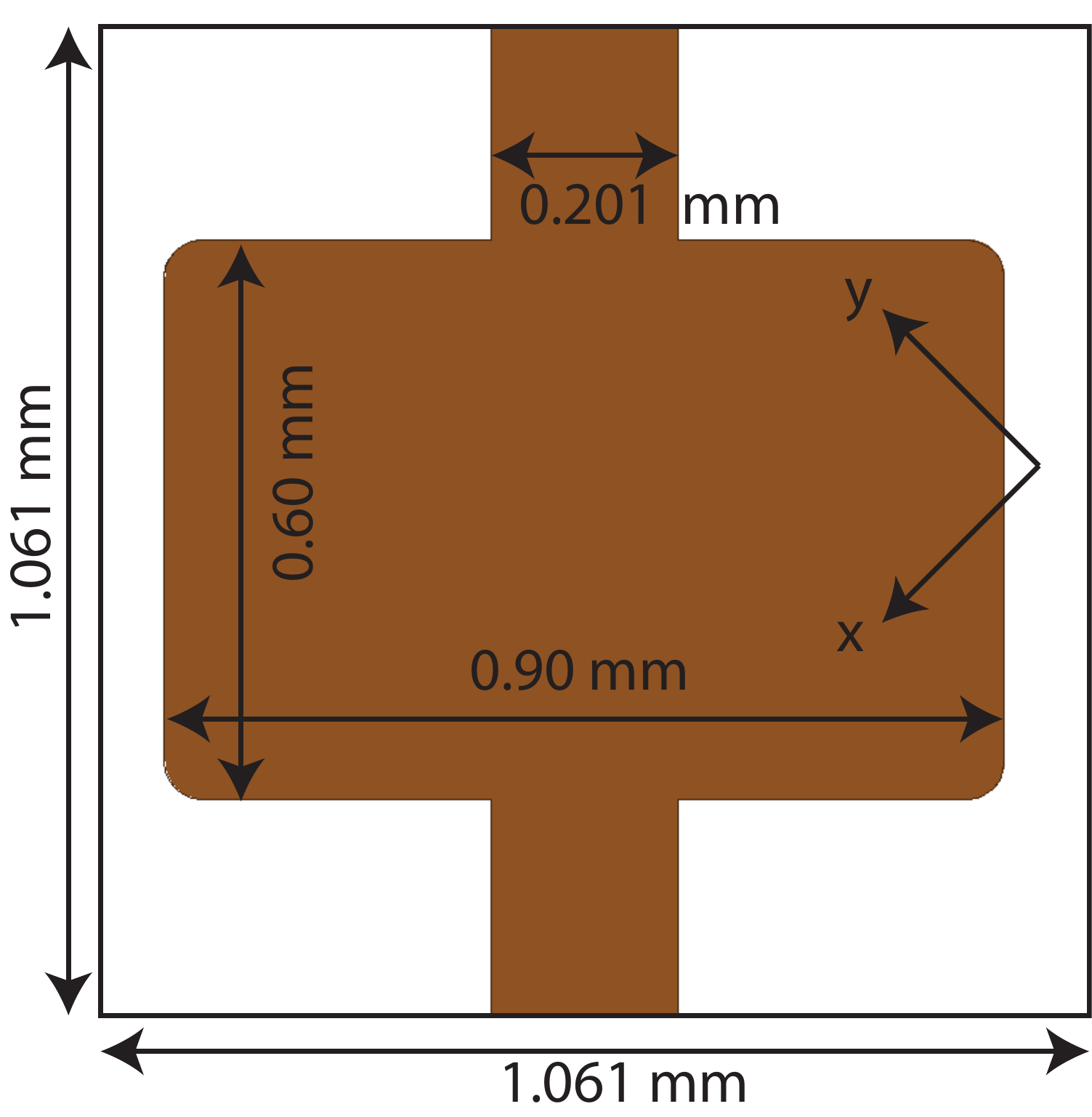}
    \label{fig:AsymmetricLinearYs2}}
    \subfigure[]{
    \includegraphics[width=1.55in]{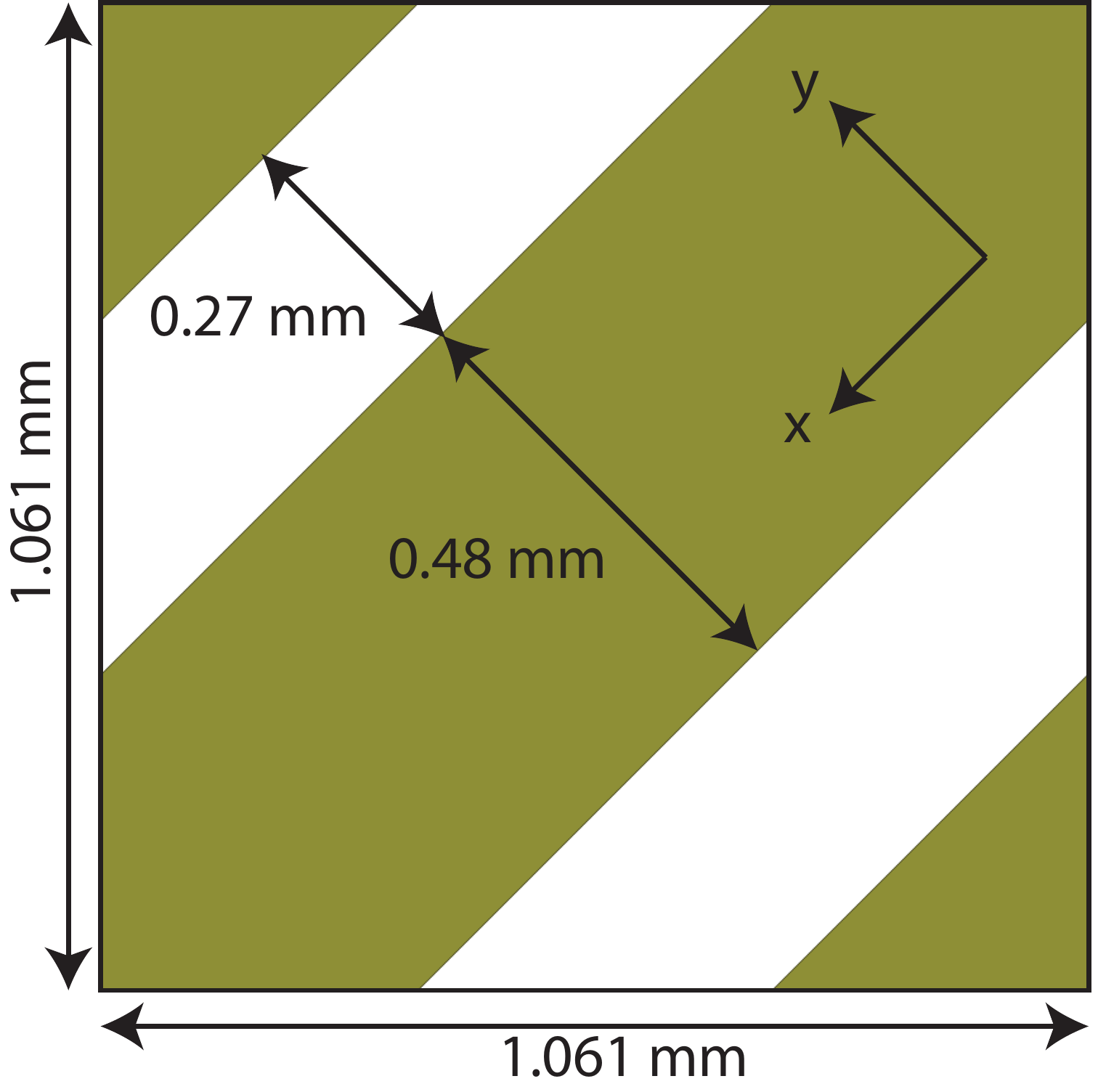}
    \label{fig:AsymmetricLinearYs3}}
  \caption{ \textbf{(a)-(c)}Dimensions first, second and third sheets of the asymmetric linear polarizer, respectively.} \label{fig:AsymmetricLinearDim}
\end{figure}

The patterns that realize the desired sheet admittances are shown in Fig. \ref{fig:AsymmetricLinearDim}. It can be seen that the sheet admittance of the first layer ($\textbf{Y}_{s1}$) is capacitive along $\hat{x}$ and inductive along $\hat{y}$. The second sheet is inductive along the $(\hat{x}-\hat{y})/\sqrt{2}$ direction and capacitive along the $(\hat{x}+\hat{y})/\sqrt{2}$ direction. The third sheet is identical to the first, except rotated by 90$^{\circ}$.

\begin{figure}[ht]
    \centering
    \includegraphics[width=3.1in]{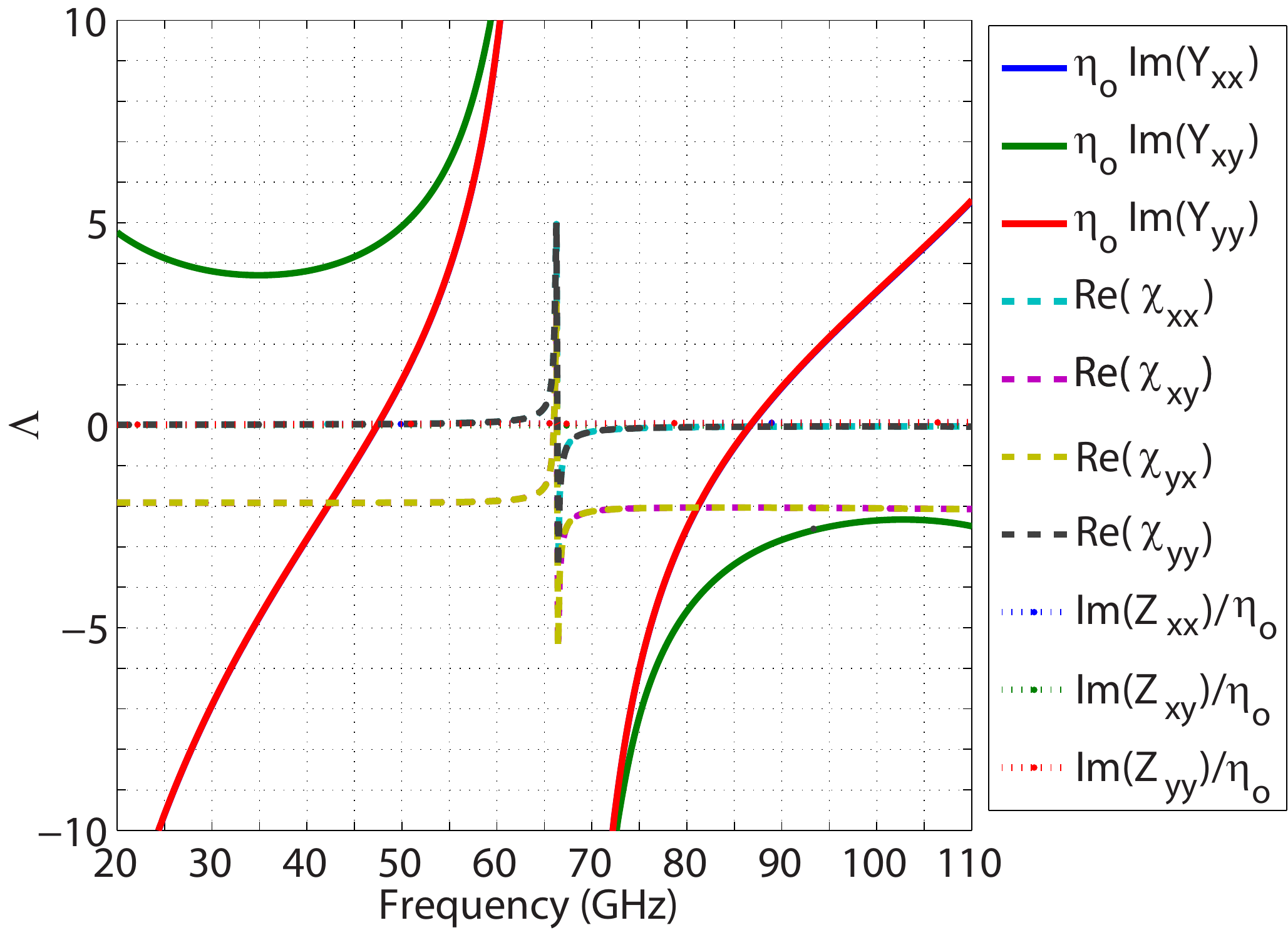}
  \caption{Constituent surface parameters of the simulated asymmetric linear polarizer. Note that $Y_{xx}=Y_{yy}$ for all plotted frequencies. The terms leading to loss (Re($\textbf{Y}$), Re($\textbf{Z}$), Im($\boldsymbol{\chi}$)) are low and are not plotted. All other terms that are not plotted can be inferred by noting that the structure is reciprocal.}\label{fig:AsymmetricLinearLambda}
\end{figure}

The constituent surface parameters of the metasurface providing asymmetric linear transmission can be determined from simulation by inserting the S-parameters of the structure into (\ref{eqn:LambdafromSparam}), as shown in Fig. \ref{fig:AsymmetricLinearLambda}. It can be seen that over the entire frequency range, the magnetic susceptibility is near zero, the crystal axes of the electric susceptibility are located in the $(\hat{x}+\hat{y})/\sqrt{2}$ and $(\hat{x}-\hat{y})/\sqrt{2}$ directions, and the chirality term is $\boldsymbol{\chi}=\begin{pmatrix}0 & -2\\-2&0\end{pmatrix}$, which are consistent with (\ref{eqn:asymmetricLinearLambda}). Note that $Y_{xx}=Y_{yy}$ for all plotted frequencies.

\section{Symmetric circular polarizer}
As a final demonstration of the versatility of the design process, a symmetric circular polarizer is demonstrated at optical frequencies. Analogous to conventional linear polarizers, these structures transmit one handedness of circular polarization but reflect the other, independent of the propagation direction. At optical frequencies, circular polarizers are attractive for color displays, microscopy, and photography \cite{yang2010ultrabroadband}. These devices are most commonly realized by combining quarter wave plates and linear polarizers \cite{Roy1996Reciprocal}. However, this leads to bulky structures that do not lend themselves to system integration. Alternatively, helical structures have demonstrated a reduced thickness and much broader bandwidth \cite{gansel2009gold,yang2010ultrabroadband}. However, they require an involved fabrication process. This motivated cascading patterned metallic sheets with rotated principle axes \cite{zhao2012twisted}.

Most reflectarray antennas at microwave frequencies are linearly polarized. These antennas could also be made to operate with circularly polarized radiation by incorporating symmetric circular polarizers, which would be beneficial for applications such as satellite communication, remote sensing, and radar \cite{tarn2007new,Roy1996Reciprocal}. Conventional circular polarizers at microwave frequencies are realized with a helical geometry \cite{Roy1996Reciprocal,pierrot1966patent,morin1990simple,tilston1988polarization}. However, these designs suffer from a complex fabrication process that is prohibitive at higher frequencies. This motivated two and three layer structures that can be fabricated using standard printed-circuit-board processes \cite{ye2011homogeneous,tarn2007new}. However these designs suffered from a narrow bandwidth, relatively low extinction ratio, and high insertion loss.

First, a symmetric circular polarizer is considered at optical frequencies. It transmits right-handed-circularly polarized light but reflects the left-handed-circular polarization, regardless of the propagation direction. The transmission coefficient is given by,
\begin{equation}\label{eqn:symmetriccircular}
\textbf{S}_{21}=\frac{e^{j\phi}}{2}\begin{pmatrix}1 & j\\-j&1\end{pmatrix}
\end{equation}
The constituent surface parameters are given by,

\small
\begin{equation}\label{eqn:LambdaSymmetricCircular}
 \boldsymbol{\Lambda}=\begin{pmatrix}
      \frac{-2j\textrm{ tan}(\phi/2)}{\eta_{\circ}} & 0 & 0 & 0 \\
       0 & \frac{2j\textrm{ cot}(\phi)}{\eta_{\circ}} & 0 & -2\textrm{ csc}(\phi) \\
       0 & 0 & -2j\eta_{\circ}\textrm{ tan}(\phi/2) & 0 \\
       0 & -2\textrm{ csc}(\phi) & 0 & 2j\eta_{\circ}\textrm{ cot}(\phi)
   \end{pmatrix}
\end{equation}
\normalsize
This metasurface exhibits anisotropic electric and magnetic susceptibilities, as well as anisotropic chirality. Eq. (\ref{eqn:LambdaSymmetricCircular}) assumes that both Regions 1 and 2 are composed of free space and that the reflection coefficients are given by,
\begin{equation}\label{eqn:symmetricCicularS11}
\textbf{S}_{11}=\frac{e^{j\phi}}{2}\begin{pmatrix}1&-j\\-j&-1\end{pmatrix},\quad
\textbf{S}_{22}=\frac{e^{j\phi}}{2}\begin{pmatrix}1&j\\j&-1\end{pmatrix}
\end{equation}
Left-handed-circularly polarized light is completely reflected from both the front and the back of the metasurface, whereas right-handed-circularly polarized light has zero reflection.

Since optical metasurfaces are typically fabricated on a bulk substrate, it is assumed that Region 2 is SiO$_2$ ($n=1.444$). The dielectric spacer between the sheets is assumed to be SU-8 ($n=1.572$), and the permittivity of the Au is described by a Drude model, $\epsilon_{Au}=\epsilon_{\infty}-\omega_p^2/(\omega^2+j\omega\omega_c)$, with $\epsilon_{\infty}=9.0$, $\omega_p=1.363\textrm{x}10^{16}$ rad/s (8.97 eV), and collision frequency $\omega_c=3.60\textrm{x}10^{14}$ rad/s (0.24 eV). This assumes a collision frequency that is three times greater than that of bulk Au \cite{johnson1972optical}, in order to account for thin film scattering and grain boundary effects \cite{chen2010drude}. This large loss does present some limitations on achieving extreme values of the sheet admittance. Nevertheless, a high performance is still achievable. This structure can be fabricated by sequential patterning of each metallic layer using standard lithography and lift-off processes \cite{liu2007three,zhao2012twisted}.

\begin{figure*}[hb]
      \centering
    \subfigure[]{
    \includegraphics[width=2.5in]{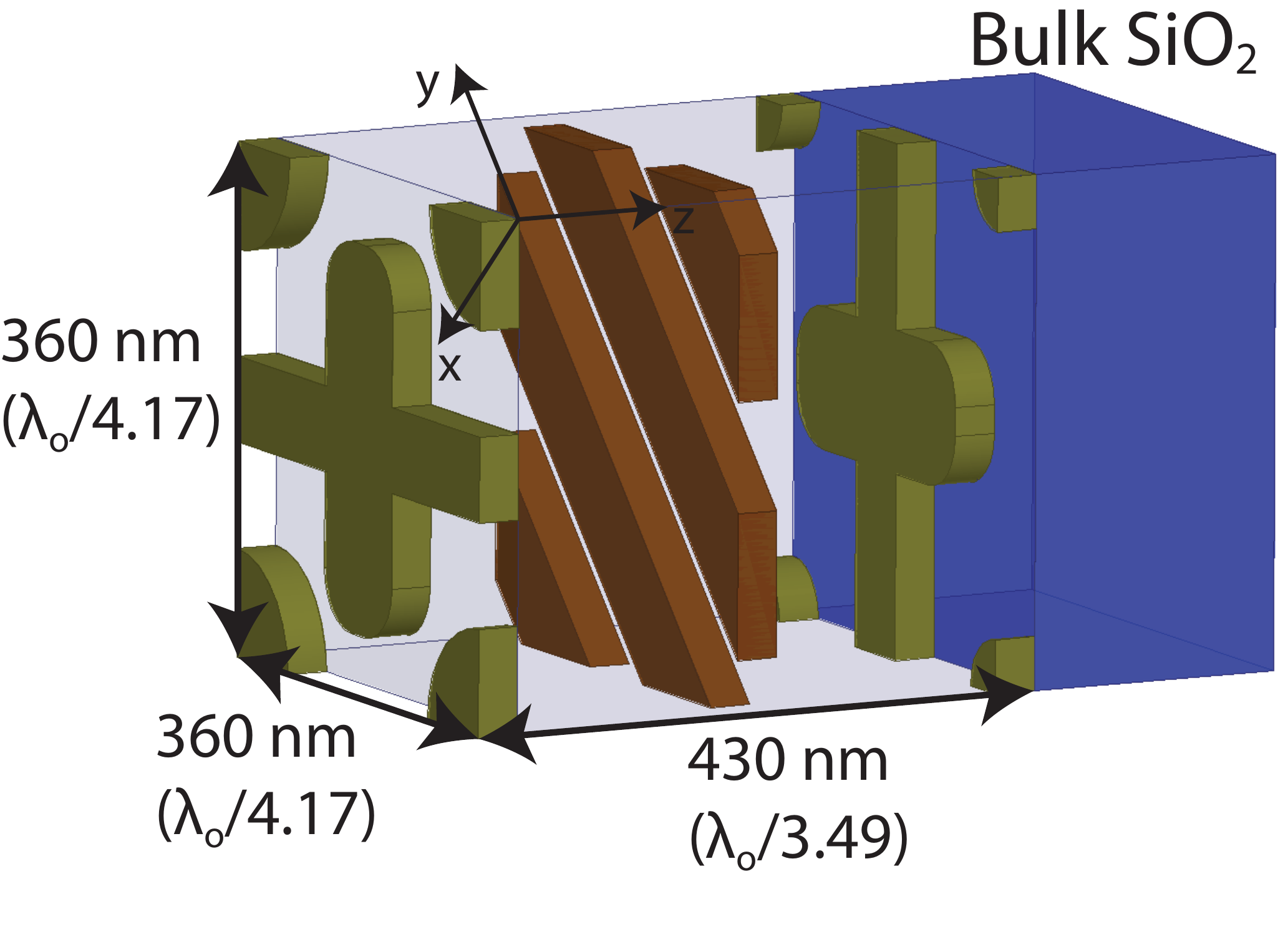}
    \label{fig:SymmetricCircularPic}}
    \subfigure[]{
    \includegraphics[width=2in]{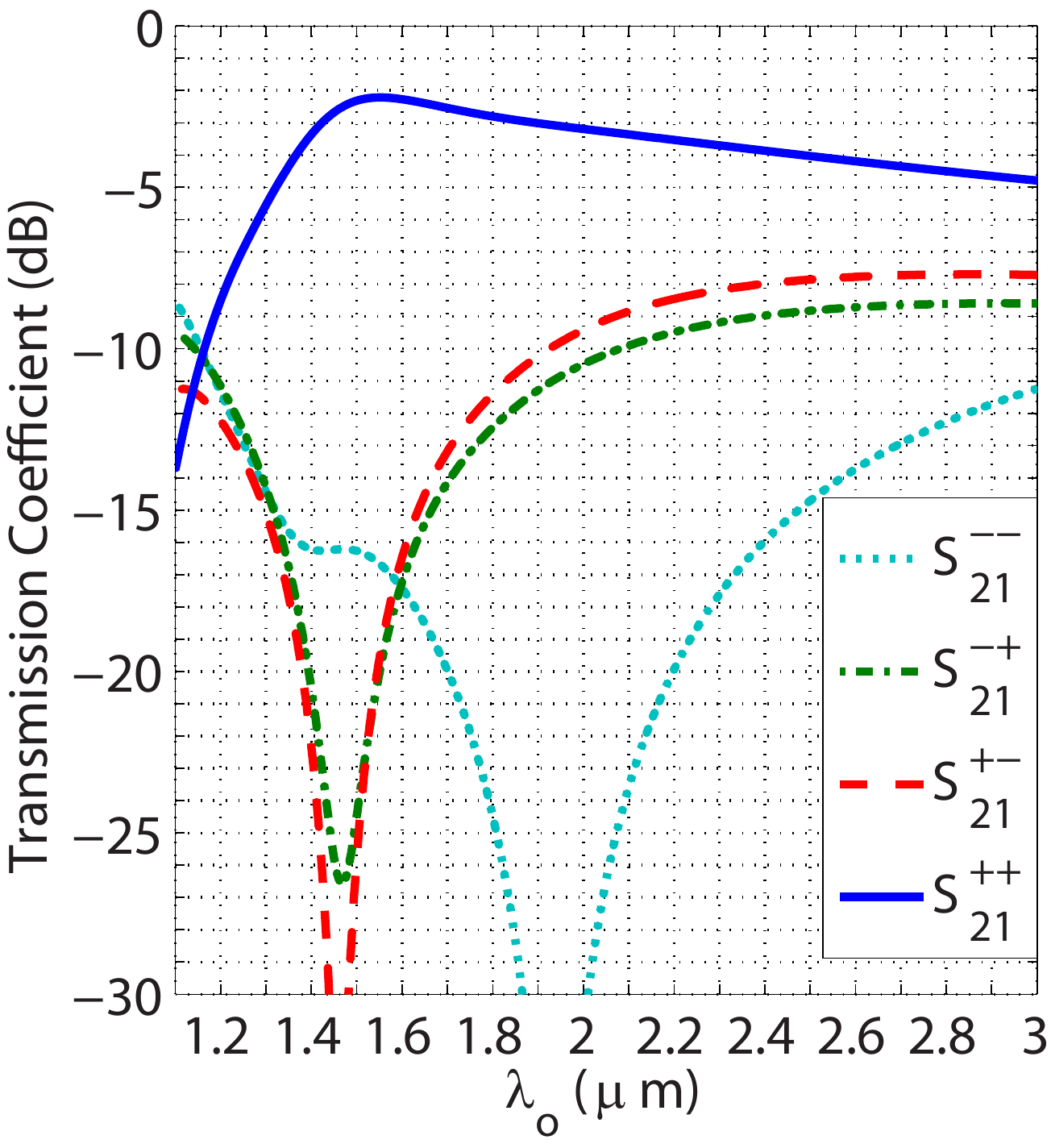}
    \label{fig:SymmetricCircularPerformance}}
    \subfigure[]{
    \includegraphics[width=2in]{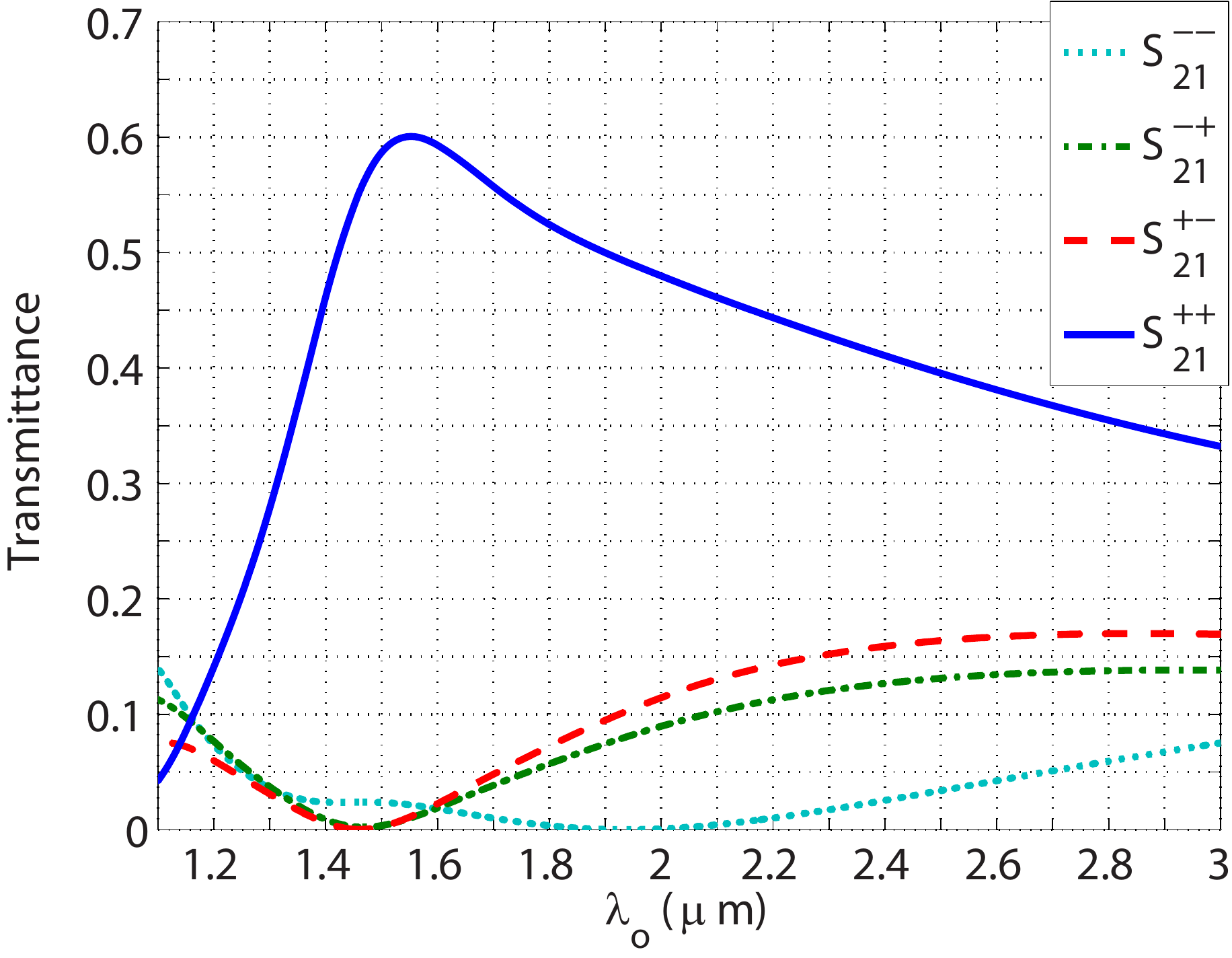}
    \label{fig:SymmetricCircularS21Linear}}
    \subfigure[]{
    \includegraphics[width=2.5in]{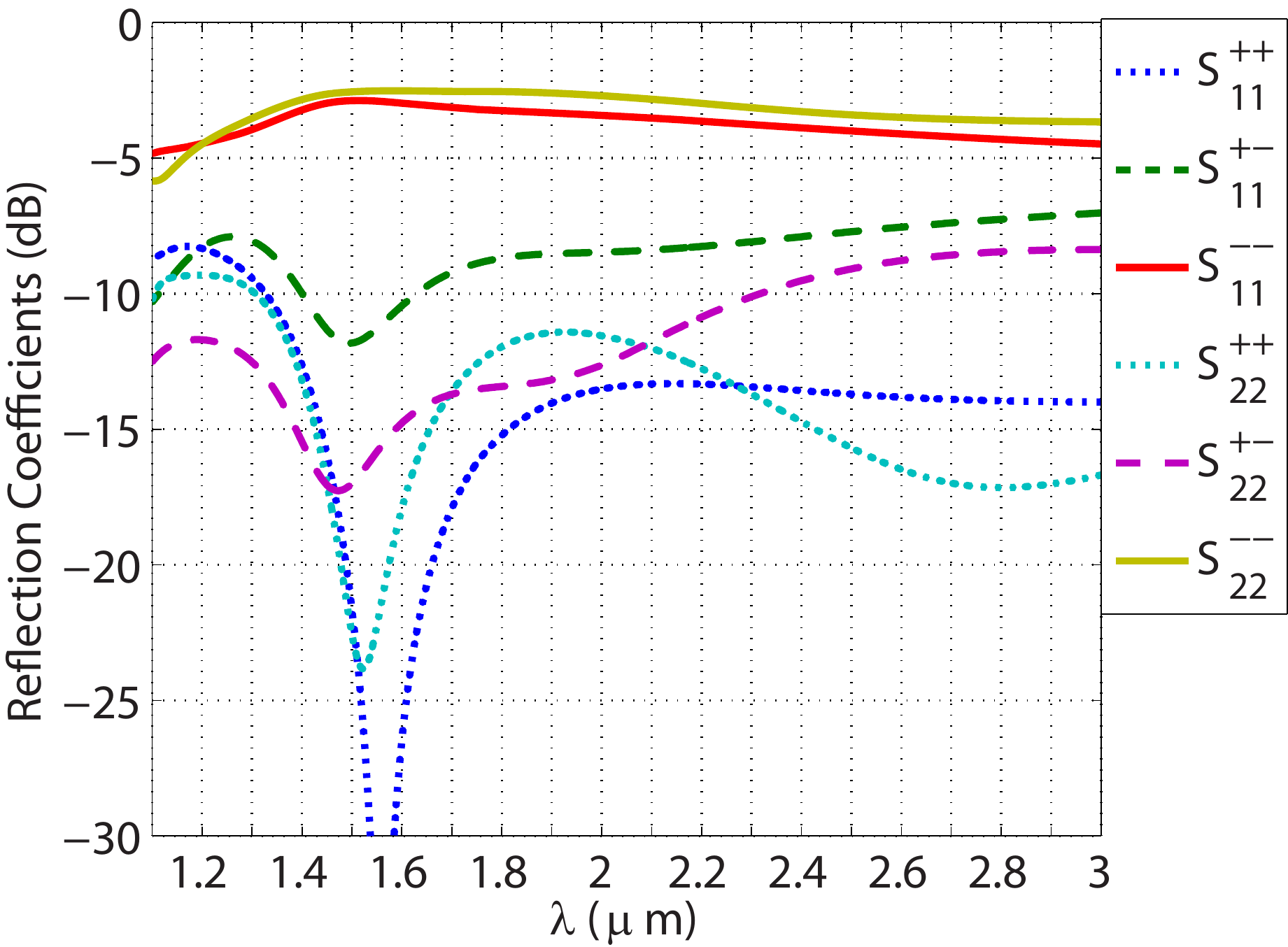}
    \label{fig:SymmetricCircularS11}}
  \caption{Symmetric circular polarizer at near-infrared wavelengths. The surface is designed to operate at a wavelength of 1.5 $\mu$m. \textbf{(a)} Schematic of the unit cell. \textbf{(b)} Transmission coefficient, where the superscript `+' denotes right-handed-circular and `-' denotes left-handed-circular. \textbf{(c)} Transmittance $(|\textbf{S}_{21}|^2)$ on a linear scale. \textbf{(d)} Reflection coefficient.}\label{fig:SymmetricCircular}
\end{figure*}

Again looking to (\ref{eqn:SparamsFromABCD}) and (\ref{eqn:ABCD3layer}), and setting $\phi=170^{\circ}$, $\beta d=2\pi/4.77$, $\eta_d=\eta_{\circ}/1.572$, $\eta_1=\eta_{\circ}$, and $\eta_2=\eta_{\circ}/1.444$, the necessary sheet admittances to realize a symmetric circular polarizer are given by, $\textbf{Y}_{s1}=\frac{j}{\eta_{\circ}}\begin{pmatrix}0.34 & -1.11\\-1.11 & 0.34\end{pmatrix}$, $\textbf{Y}_{s2}=\frac{j}{\eta_{\circ}}\begin{pmatrix}1.10 & 0\\0 & -9.00\end{pmatrix}$, and $\textbf{Y}_{s3}=\frac{j}{\eta_{\circ}}\begin{pmatrix}0.57 & 1.57\\1.57 & 0.57\end{pmatrix}$. When solving for the cascaded sheet admittances, their maximum magnitude was required to be less than $9.0/\eta_{\circ}$ along any given principle axis. This is because the operating wavelength imposes limitations in achieving extremely small feature sizes and the plasmonic nature of Au presents an added inductance. The designed unit cell is shown in Fig. S\ref{fig:SymmetricCircularPic}. Its simulated transmission coefficient is shown in Fig. S\ref{fig:SymmetricCircularPerformance}. The superscript `+' denotes right-handed-circular polarization and `-' denotes left-handed-circular polarization. It can be seen that at the design frequency of 1.5 $\mu$m, the metasurface achieves low loss for right-handed-circularly polarized light, and it provides greater than 15 dB rejection for left-handed-circularly polarized light. For completeness, the transmittance is also plotted on a linear scale in Fig. S\ref{fig:SymmetricCircularS21Linear}, so that its performance can be easily compared to earlier reported structures \cite{gansel2009gold,zhao2012twisted}. This metasurface is impedance matched at the design frequency, as shown by the reflection coefficients in Fig. S\ref{fig:SymmetricCircularS11}. The power absorbed by the metasurface can be calculated by subtracting the incident power from the transmitted and reflected power $1-|S_{11}|^2-|S_{21}|^2$, which is $\sim$40\%.

\begin{figure*}[ht]
    \centering
    \subfigure[]{
    \includegraphics[width=2in]{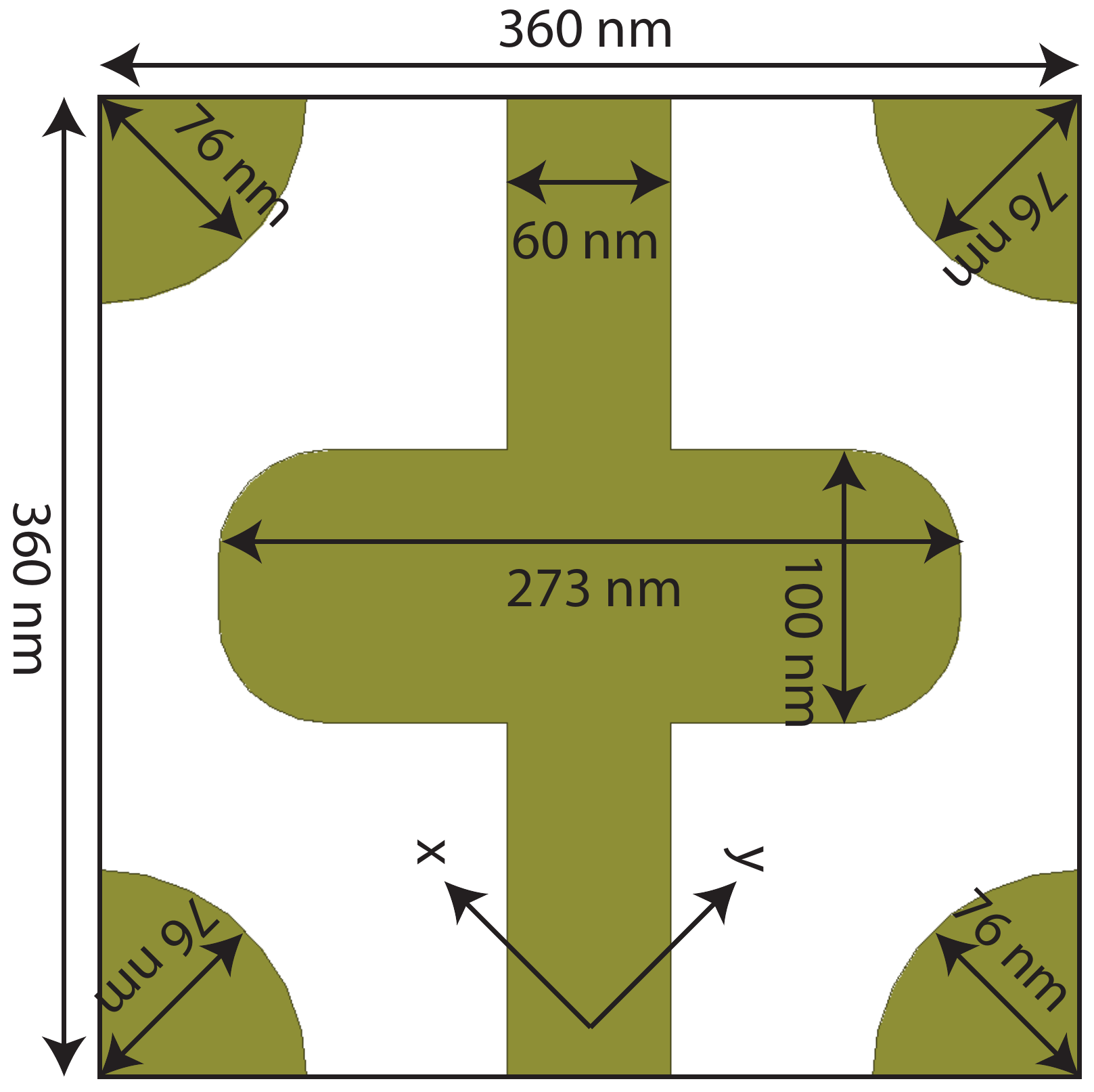}
    \label{fig:SymmetricCircularYs1}}
     \subfigure[]{
    \includegraphics[width=2in]{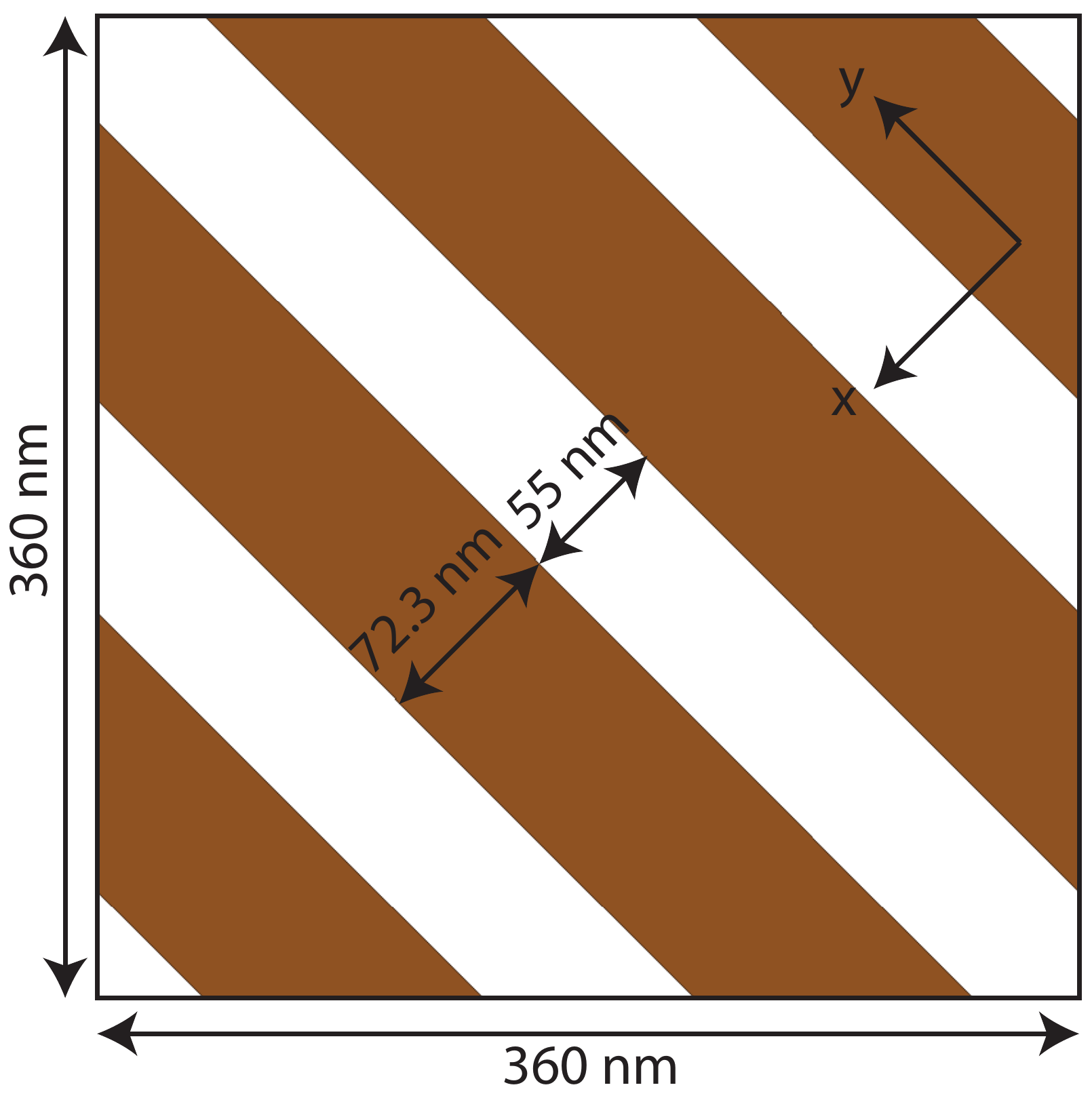}
    \label{fig:SymmetricCircularYs2}}
    \subfigure[]{
    \includegraphics[width=2in]{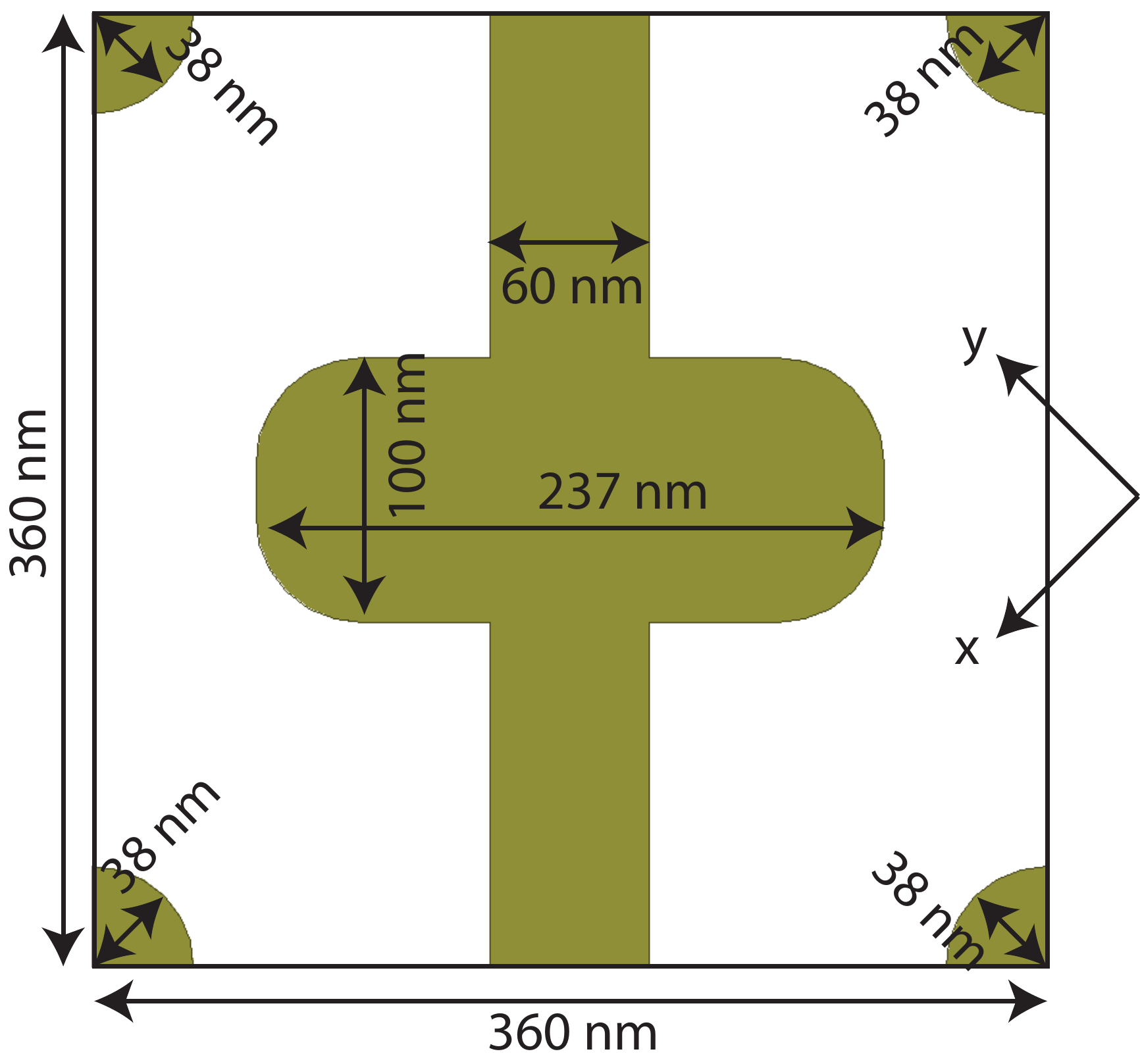}
    \label{fig:SymmetricCircularYs3}}
  \caption{\textbf{(a)-(c)} Dimensions of the first, second, and third sheets of the symmetric circular polarizer, respectively.} \label{fig:SymmetricCircularDimm}
\end{figure*}

The patterns that realize the desired sheet admittances are shown in Fig. \ref{fig:SymmetricCircularDimm}. All corners are rounded with a radius of curvature of 40 nm. As in the previous metasurfaces, each sheet admittance is individually designed such that its imaginary part is identical to the desired sheet admittances. However, the analytic model assumes the sheets are lossless, which is somewhat of an approximation. Therefore, the performance can be improved further by using the optimizer provided by Ansys HFSS. The dimensions are varied to minimize the real part of the admittance while also approaching an ideal imaginary part. The optimization process is relatively quick since the initial structure represents a good starting point. The dimensions shown in Fig. \ref{fig:SymmetricCircularDimm} correspond to the optimized structure. Each sheet is capacitive along one principle axis and inductive along the other.

\begin{figure*}[ht]
      \centering
    \subfigure[]{
    \includegraphics[width=3in]{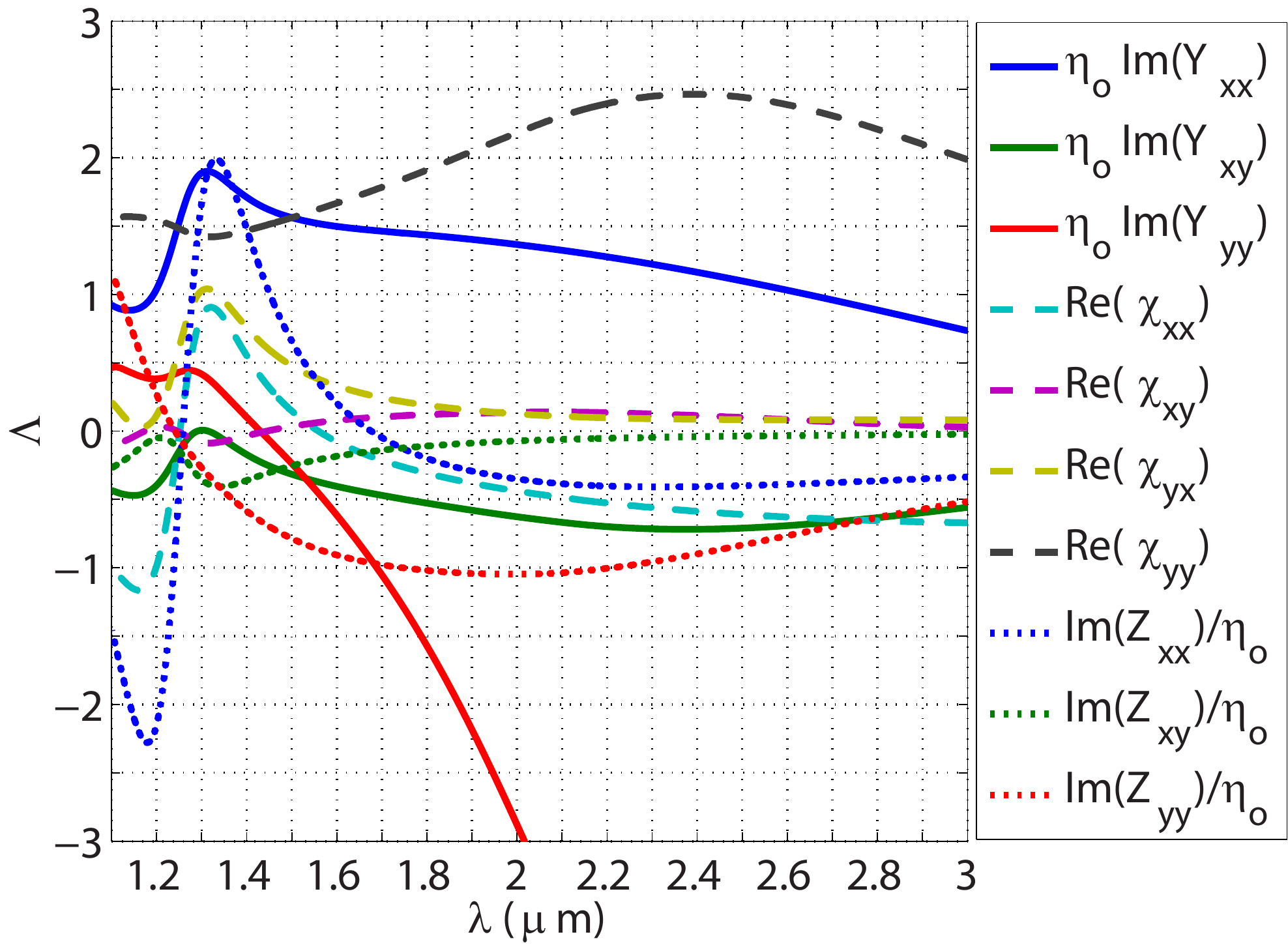}
    \label{fig:SymmetricCircularLambdaOptimized}}
     \subfigure[]{
    \includegraphics[width=3in]{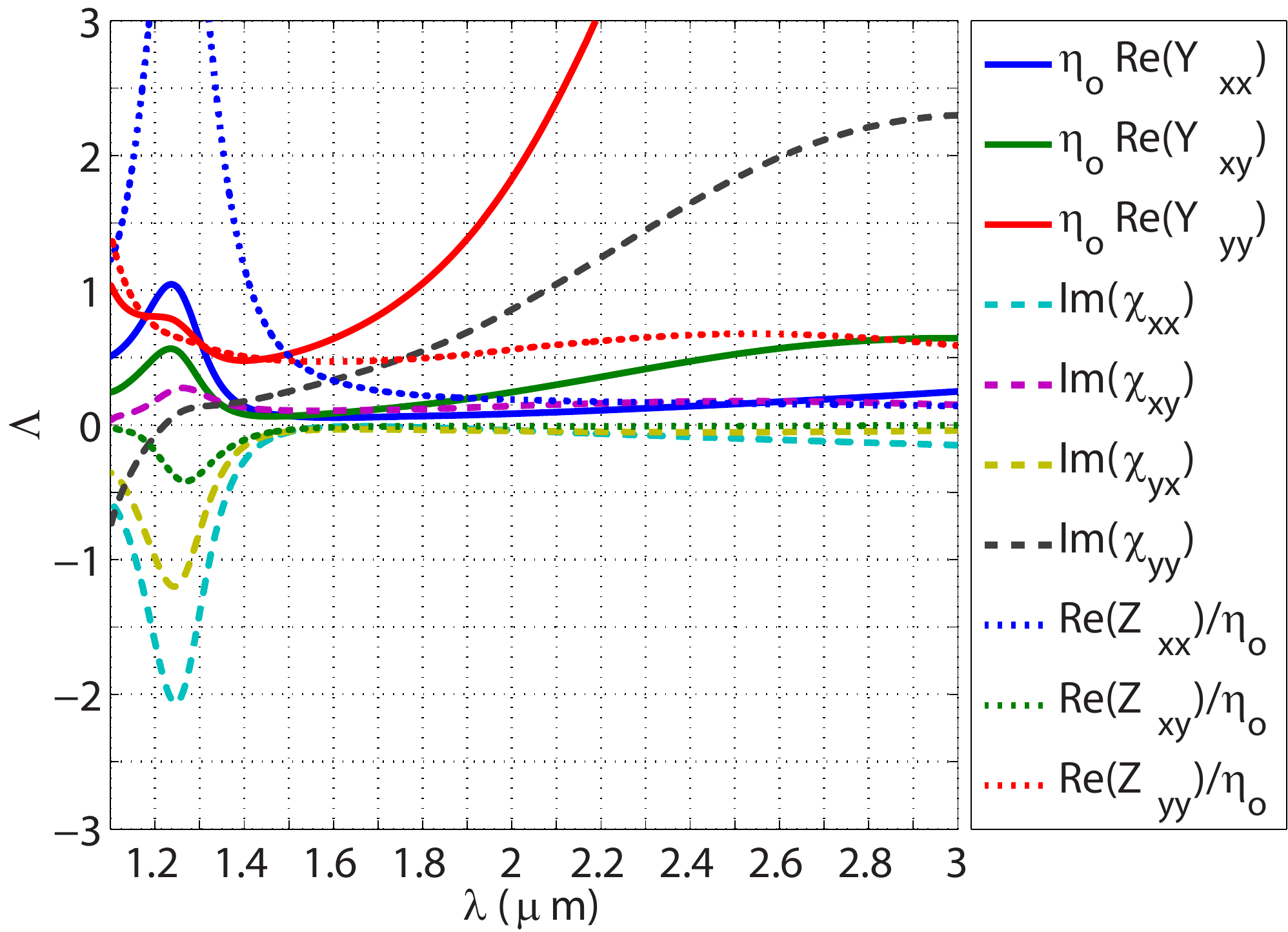}
    \label{fig:SymmetricCircularLambdaOptimizedReal}}
  \caption{\textbf{(a)} Constituent surface parameters of the symmetric circular polarizer. \textbf{(b)} The terms leading to loss (Re($\textbf{Y}$), Re($\textbf{Z}$), Im($\boldsymbol{\chi}$)). All other terms that are not plotted can be inferred by noting that the structure is reciprocal.}\label{fig:SymmetricCircularLambda}
\end{figure*}

The constituent surface parameters of the symmetric circular polarizer can be determined from simulation by inserting the S-parameters of the structure into (\ref{eqn:LambdafromSparam}). They are shown in Figs. \ref{fig:SymmetricCircularLambda} (a) and (b). The terms representing loss (Re($\textbf{Y}$), Re($\textbf{Z}$), Re($\boldsymbol{\chi}$)) cannot be neglected. It can be seen that at the operating wavelength of 1.5 $\mu$m, Re($\chi_{yy}$) is much larger than the other chiral terms. In addition Im($Y_{xx}$) $\sim$ Im($Z_{xx}$) and Im($Y_{yy}$) $\sim$ Im($Z_{yy}$), which are all necessary conditions for symmetric circular polarization as given by (\ref{eqn:LambdaSymmetricCircular}). Although (\ref{eqn:LambdaSymmetricCircular}) does assume that $\eta_1=\eta_2$, which is not the case here, it is still valuable at providing some physical insight since the wave impedance of SiO$_2$ is similar to that of free space.

The performance of this structure exceeds that of previous metasurfaces in a few respects. At the operating frequency, the structure presented here achieves a polarization rejection of 15 dB. This is comparable to the rejection levels of the Au helix metamaterial \cite{gansel2009gold}, which achieved a larger bandwidth at the expense of increased fabrication complexity and overall thickness. This polarization rejection is also an order of magnitude higher than the previous three layer structure that cascaded identical electric dipoles with a rotation between the sheets \cite{zhao2012twisted}.

\begin{figure*}[ht]
      \centering
     \subfigure[]{
    \includegraphics[width=1.7in]{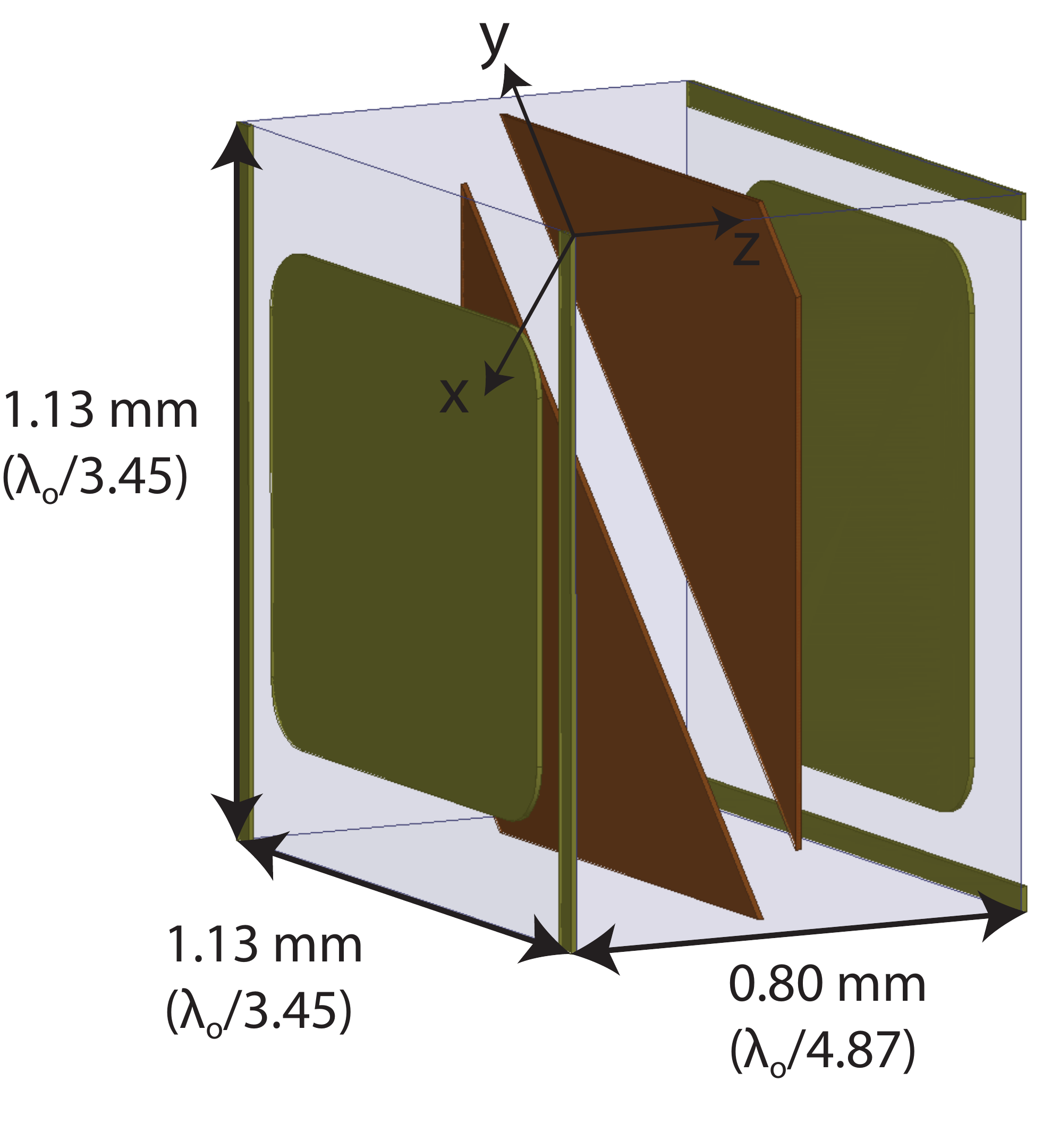}
    \label{fig:SymmetricCircularPicmmWave}}
    \subfigure[]{
    \includegraphics[width=2in]{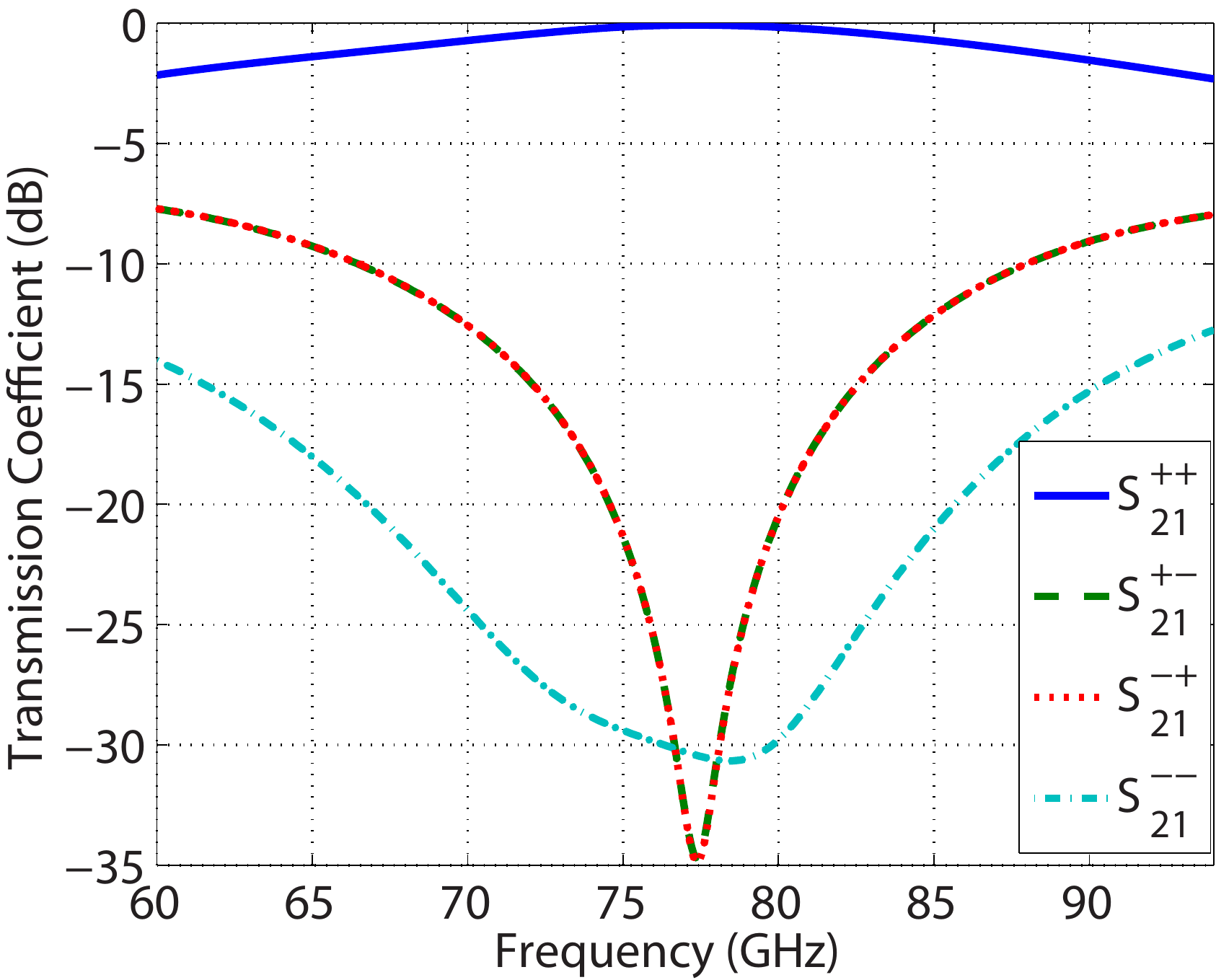}
    \label{fig:SymmetricCircularS21mmWave}}
    \subfigure[]{
    \includegraphics[width=2.3in]{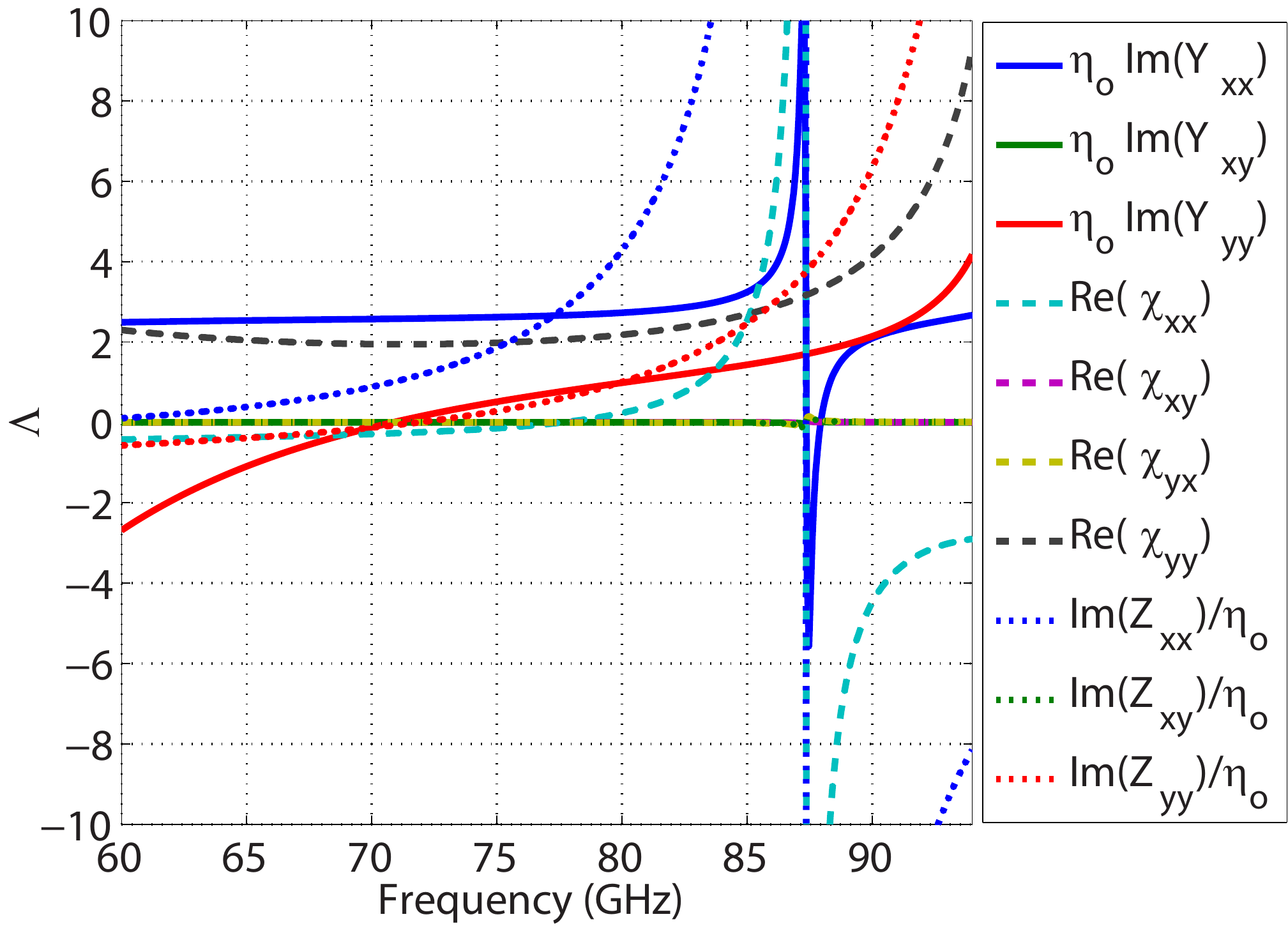}
    \label{fig:SymmetricCircularmmWaveLambda}}
  \caption{Symmetric circular polarizer at mm-wave frequencies. \textbf{(a)} Perspective view of the symmetric circular polarizer. \textbf{(b)} Transmission coefficient, where the superscript `+' denotes right-handed-circular and `-' denotes left-handed-circular. \textbf{(c)} Constituent surface parameters.}\label{fig:SymmetricCircularmmWave}
\end{figure*}

As with many optical designs, the analysis of this structure is relatively approximate, which does provide a reduced performance as compared to the other metasurfaces presented at lower frequencies. To demonstrate that this reduced performance is only a result of high loss at near-infrared frequencies and not a result of the stipulated polarization transformation, a design is also developed at 77 GHz, as shown in Fig. \ref{fig:SymmetricCircularmmWave}. Using the same design procedure as before, it can be shown that the necessary sheets of this structure are identical to those of the metasurface providing asymmetric circular transmission, except the first sheet is rotated by $90^{\circ}$. The transmission coefficient for the desired polarization is near unity, while the rejection of the undesired polarization is greater than 30 dB at the design frequency (see Fig. S\ref{fig:SymmetricCircularS21mmWave}. In addition, the constituent surface parameters are very close to the ideal values that achieve a symmetric polarizer at the design frequency of 77 GHz. This can be verified by looking to (\ref{eqn:LambdaSymmetricCircular}) and letting $\phi=-103^{\circ}$, and then comparing the result with Fig. S\ref{fig:SymmetricCircularmmWaveLambda}.

\end{bibunit}
\end{document}